\documentclass[12pt]{article}
\pdfoutput=1
\usepackage{amsmath,amsfonts,graphicx,yfonts,mathabx}
\usepackage[nosort]{cite}

\usepackage{amssymb}
\usepackage{epsfig, float}

\def\baselinestretch{1.1}
\usepackage{tikz}
\usetikzlibrary{decorations.pathmorphing}
\usepackage{capt-of}
\usepackage{caption}
\usepackage{subcaption}

\newcommand{\be}{\begin{equation}}
\newcommand{\ee}{\end{equation}}
\newcommand{\bea}{\begin{eqnarray}}
\newcommand{\eea}{\end{eqnarray}}

\newcommand{\mt}[1]{\textrm{\tiny #1}}

\def\rh {r_\mt{H}}

\renewcommand{\title}[1]{\vbox{\center\LARGE{#1}}\vspace{3mm}}
\renewcommand{\author}[1]{\vbox{\center#1}\vspace{3mm}}
\newcommand{\address}[1]{\vbox{\center\em#1}}

\begin{document}
%\begin{titlepage}
\begin{center}
\rightline{\tt}
%\vskip 2.5cm
\title{Chaos, Diffusivity, and Spreading of Entanglement\\ in Magnetic Branes, and\\ the strengthening of the internal interaction}
%\vskip .6cm
\author{Daniel \'Avila$^{a}$, Viktor Jahnke$^{b}$, Leonardo Pati\~no$^{a}$}
\vskip -.5cm 
\address{
$^a$Departamento de F\'isica, Facultad de Ciencias, Universidad Nacional Aut\'onoma de M\'exico,
Apartado Postal 70-542, CDMX 04510, M\'exico

$^b$Departamento de F\'isica de Altas Energias, Instituto de Ciencias Nucleares, Universidad Nacional Aut\'onoma de M\'exico\\ Apartado Postal 70-543, CDMX 04510, M\'exico}
%\vskip -.1cm

%\email{viktor.jahnke@correo.nucleares.unam.mx}
\end{center}
%\vskip 3cm

%%

\abstract{
We use holographic methods to study several chaotic properties of a super Yang-Mills theory at temperature $T$ in the presence of a background magnetic field of constant strength $\mathcal{B}$. The field theory we work on has a renormalization flow between a fixed point in the ultraviolet and another in the infrared, occurring in such a way that the energy at which the crossover takes place is a monotonically increasing function of the dimensionless ratio $\mathcal{B}/T^2$.
By considering shock waves in the bulk of the dual gravitational theory, and varying $\mathcal{B}/T^2$, we study how several chaos-related properties of the system behave while the theory they live in follows the renormalization flow. In particular, we show that the entanglement and butterfly velocities generically increase in the infrared theory, violating the previously suggested upper bounds but never surpassing the speed of light.
We also investigate the recent proposal relating the butterfly velocity with diffusion coefficients. We find that electric diffusion constants respect the lower bound proposed by Blake.
All our results seem to consistently indicate that the global effect of the magnetic field is to strengthen the internal interaction of the system.

%To describe this system we use a gravity dual that smoothly interpolates between a near horizon BTZ $\times R^2$ geometry and an asymptotic region that approaches $AdS_5$ near the holographic boundary. The radius at which the crossover occurs is , where $T$ is the temperature due to the presence of a horizon. In the field theory side, this transition is perceived as renormalization flow

%upper bounds proposed in  1311.1200 and 1612.00082

}

%\vfill
%\end{titlepage}
%%\keywords{Gauge-gravity correspondence}  

\tableofcontents

\section{Introduction} \label{sec-intro}
Recently, there has been a growing interest in the study of chaotic properties of many-body quantum systems, especially in the context of gauge/gravity correspondence \cite{duality1,duality2,duality3}. It turned out that the chaotic phenomena of the boundary theory have a rather simple description in the dual gravitational theory in terms of shock waves traveling in the vicinity of the black hole horizon \cite{BHchaos1,BHchaos2,BHchaos3,BHchaos4}. Since the properties of black hole horizons are connected to transport properties of strongly interacting systems, this suggests a connection between chaos and diffusion phenomena \cite{Blake1,Blake2}. Moreover, the chaotic properties of the boundary theory have also shed light on the inner working mechanisms of gauge/gravity correspondence. For example, a maximum Lyapunov coefficient seems to be a necessary condition for a quantum system to have a description in terms of Einstein's gravity \cite{bound-chaos,Kitaev-2014}, whereas the butterfly velocity seems to play an important role in determining the bulk causal structure \cite{Qi-2017}.

The characterization of chaos in a quantum many-body system can be done by considering how much an early perturbation ${\cal{O}}_1$ is correlated with a later measurement of some other operator\footnote{It is customary to refer to these operators as $V$ and $W$, but we brake this tradition to avoid confusion with the metric functions} ${\cal{O}}_2$. This can be conveniently quantified by
\be
C(t,\vec{x})= - \langle [{\cal{O}}_2(t,\vec{x}),{\cal{O}}_1(0,0)]^2 \rangle_{\beta},
\ee
where $\langle \cdot \rangle_{\beta} = Z^{-1} \text{tr}( \cdot)$ denotes a thermal expectation value at temperature $\beta^{-1}$. We assume ${\cal{O}}_1$ and ${\cal{O}}_2$ to be hermitian operators normalized such that $\langle {\cal{O}}_1 {\cal{O}}_1 \rangle=\langle {\cal{O}}_2{\cal{O}}_2 \rangle=1$. For simplicity, let us first consider the case where the two operators are not separated in space, i.e. $\vec{x}=0$. For a sufficiently chaotic system, $C(t,0)$ approaches a first order constant value at large times \cite{AMPSS}. The time scale $t_*$ at which this occurs is the so-called scrambling time \cite{scrambling1,scrambling2}. Just before saturation, $C(t,\vec{x})$ is expected to grow exponentially with time as
\be
C(t,\vec{x}) \sim \frac{1}{\mathcal{N}} \exp \left[\lambda_L\left(t-\frac{|\vec{x}|}{v_B} \right) \right]\,,\,\,\,\,\text{for}\,\,t_d << t << t_*\,,
\label{eq-C(t,x)}
\ee
where $\mathcal{N}$ denotes the total number of degrees of freedom and $t_d$ is the dissipation time of the system, which characterizes the time decay of two point functions $\langle {\cal{O}}_1(t) {\cal{O}}_1(0)\rangle \sim e^{-t/t_d}$. The growth of $C(t,\vec{x})$ with time is characterized by the Lyapunov exponent $\lambda_L$. For systems with a large number of degrees of freedom and with a large hierarchy between $t_d$ and $t_*$, the Lyapunov exponent was shown to be bounded by the temperature $\lambda_L \leq 2 \pi T$ \cite{bound-chaos}. The butterfly velocity characterizes the rate at which the information about the operator ${\cal{O}}_1$ spreads in space. When ${\cal{O}}_1$ and ${\cal{O}}_2$ are separated in space, there is a delay in scrambling. This delay is controlled by the butterfly velocity. This velocity defines a {\it butterfly effect cone} as $t-t_* =|\vec{x}|/v_B$. Inside the cone, for $t-t_* \geq |\vec{x}|/v_B$, we expect $C(t,\vec{x}) \sim \mathcal{O}(1)$, while outside the cone, for $t-t_* < |\vec{x}|/v_B$, we expect $C(t,\vec{x}) \sim 1/\mathcal{N}<<1$.

In the context of the gauge/gravity correspondence, the chaotic properties of the boundary theory can be extracted from shock waves in the bulk\footnote{Some works in this direction include, for instance, \cite{Leichenauer-2014,Roberts-2016,Ling-2016,Huang-2016,Sircar-2016,Mezei-2016,Cai-2017,
Giataganas-2017,Jahnke-2017,Qaemmaqami-2017,Qaemmaqami-2017-2,deBoer-2017,Murata-2017,Wu-2017,Huang-2017,Baggioli-2018,Huang-2018}.}. To do this it is convenient to consider a thermofield double state made out of two copies of the boundary theory. Let us call them $L$ and $R$ boundary theories, respectively.  At the $t=0$ slice, this state can be schematically written as
\be
|TFD \rangle = Z^{-1/2} \sum_n e^{-\frac{\beta E_n}{2}} | E_n\rangle_L |E_n \rangle_R ,
\ee
where $\beta$ is the inverse temperature of the system. From the gravitational point of view this state is represented in a two-sided black hole geometry \cite{eternalBH}. The two asymptotic theories live at the two asymptotic boundaries of the geometry and do not interact with each other, which is consistent with the fact that the wormhole is not traversable.

In order to diagnose chaos we perturb the $L$ part of the system by acting with an operator ${\cal{O}}_2(t_0)$ at time $t_0$ in the past. From the bulk perspective, this creates a `particle' near the boundary, which then falls into the black hole and generates a shock wave geometry, as illustrated in figure \ref{fig-PenroseShockWave}.

The profile of this shock wave $\alpha(t,\vec{x})$ turns out to be related to $C(t,\vec{x})$ in a simple way and we can extract the Lyapunov exponent $\lambda_L$, the scrambling time $t_*$ and the butterfly velocity $v_B$ from it. In systems that can be described by a black hole geometry, the Lyapunov exponent is always maximal $\lambda_L = 2 \pi/\beta$, while the leading order contribution to the scrambling time always scales as $t_* \approx \frac{\beta}{2 \pi} \log \mathcal{N}$. The only chaotic property that turns out to be more interesting is the butterfly velocity, because it depends on more specific characteristics of the system.

If we consider a homogeneous perturbation, such that $\alpha(t,\vec{x}) = \alpha(t)$, we can also diagnose the chaos in the boundary by studying the mutual information $I(A,B)$ between subsystems $A$ and $B$ of the $L$ and $R$ systems, respectively. This way of characterizing chaos is interesting because it has some connections with spreading of entanglement.

The basic idea is that at $t=0$, the TFD state has a very particular pattern of entanglement between the $L$ and $R$ systems and this can be diagnosed by a non-zero mutual information $I(A,B)$ between large subsystems $A \subset L$ and $B \subset R$. When we perturb the $L$ system at a time $t_0$ in the past, the perturbation scrambles the left-side Hilbert space and destroys the pattern of entanglement that was present in the unperturbed system at $t=0$. Indeed, an initially positive mutual information smoothly drops to zero as we move the perturbation further into the past. As we will explain in section \ref{sec-mutual}, this phenomenon has a very simple description in terms of Ryu-Takayanagi surfaces in the bulk.

From the Lyapunov exponent we can define a time scale known as {\it Lyapunov time}, which is given by $t_L =1/\lambda_L$. The upper bound in $\lambda_L$ implies a lower bound in the Lyapunov time $t_L \geq \frac{\beta}{2 \pi}.$\footnote{Here we are using units such that Planck and Boltzmann constants $\hbar$ and $k_B$ are both equal to unity. If we reintroduce $\hbar$ and $k_B$ in our formulas we obtain $\lambda_L \leq \frac{2\pi k_B}{ \hbar \beta}$ or $t_L \geq \frac{\hbar \beta}{2 \pi k_B}$.} For convenience, let us call $\tau_L$ the lower bound on the Lyapunov time.

In \cite{Sachdev-97,Sachdev-99} it was proposed that $\tau_L =\hbar/(2 \pi k_B T)$ provides a fundamental dissipative time scale that controls the transport in strongly coupled systems. Such a universal time scale would be responsible for the universal properties of several strongly coupled systems that do not a have a description in terms of quasiparticle excitations. Working on these ideas and aiming to explain the linear-T resistivity behavior of strange metals, Hartnoll \cite{Hartnoll-2014} proposed the existence of a universal bound on the diffusion constants related to the collective diffusion of charge and energy $D \gtrsim \hbar v^2 /(k_B T)$, where $v$ is some characteristic velocity of the theory. As the thermoeletric diffusion constant $D$ is proportional to the conductivity $\sigma$, the saturation of the lower bound on $D$ implies the scaling $\sigma \sim 1/T$, that results in a linear-T resistivity behavior.

In an holographic treatment, both the transport and the chaotic properties of the gauge theory are determined by the dynamics close to the black hole horizon in the gravitational dual. It is then natural to question if there is any connection between chaos and diffusion phenomena. With this in mind, Blake proposed in \cite{Blake1,Blake2} that, for particle-hole symmetric theories, the eletric diffusivity $D_{c}$ should be controlled by $v_B$ and $\tau_L$ as
\begin{equation}
D_{c}\geq C_{c}v_{B}^{2}\tau_{L},
\label{bound1}
\end{equation}
where $C_{c}$ is a constant that depended on the universality class of theory. The above proposal works well for system where energy and charge diffuse independently, but it is not valid in more general situations. See, for instance \cite{Lucas-2016,Davison-2016,Baggioli-2016,Kim-2017}.

In \cite{Blake3} it was proposed that, for a general family of holographic  Q-lattice  models, the thermal diffusivity $D_T$ should be generically related to chaos exponents at infrared fixed points through 
\begin{equation}
D_{T}\geq C_{T}v_{B}^{2}\tau_{L},
\end{equation} 
where $C_{T}$ is another universality constant different from $C_{c}$ (this was latter generalized to theories with an spatial anisotropy in \cite{Ahn-2017}). This Q-lattice models do not have translational symmetry and features a finite charge density, which makes $D_{T}$ finite.

In this work we use holographic techniques to study chaos, diffusivity and spreading of entanglement in a gauge theory at a finite temperature $T$ in the presence of a background magnetic field of strength $\mathcal{B}$. The theory flows between two fixed points of the renormalization group, one in the ultraviolet corresponding to a four dimensional $\mathcal{N}=4$ super Yang-Mills theory, and the other in the infrared where the theory is also conformal but, due to the magnetic field, reduces to 1+1 dimensions. The gravity dual of this theory was presented in \cite{DHoker-2009} and has been used to investigate the effects of an external magnetic field in several physical observables, of which a comprehensive list would be difficult to achieve, but some studies relevant to our current topic are \cite{Arciniega:2013dqa,Arean:2016het,Martinez-y-Romero:2017awl,Ammon:2016szz,Ammon:2017ded,Rahimi:2018ica}.

In \cite{Martinez-y-Romero:2017awl} holographic methods were used to show that the energy scale at which the crossover from one fixed point to the other occurs is a monotonically increasing function of the dimensionless parameter $\mathcal{B}/T^2$. It is this last result that allow us to investigate how the chaotic properties of the theory are changed by the RG flow, because it indicates we can explore it by varying $\mathcal{B}/T^2$, since at a fixed energy scale large values of this ratio will pull the theory closer to the IR limit and small values will move it towards the UV one.

%\footnote{Note that, for fixed $B$, the ratio $B/T^2$ is small for high temperatures and large for small temperatures, and this limits correspond to the ultraviolet (UV) and infrared (IR) limits, respectively. So, the geometry is given by a 5-dimensional black D3-brane solution at the UV and by 3-dimensional BTZ black hole at the IR.}

The paper is organized as follows. In section \ref{sec-gravity} we review the gravity dual of the gauge theory we work with and how the renormalization flow is realized in it. We show how to extract the chaotic properties of the boundary theory in section \ref{sec-vb1}. In section \ref{sec-mutual} we study the disruption of the two-sided mutual information in shock wave geometries and show how this is connected to spreading of entanglement. We discuss the connection between chaos and diffusion phenomena in section \ref{sec-diffusion}. Finally, we discuss our results in section \ref{sec-discussion}. We relegate some technical details to the appendices \ref{InteriorNumeric}, \ref{AppD} and \ref{AppEigen}.

\section{Gravity setup} \label{sec-gravity}

The gravitational theory we will be working with is the consistent truncation \cite{Cvetic:1999xp} of type IIB supergravity that will leave us with the action
\be
S=-\frac{1}{16\pi G_5}\int d^5x\sqrt{-g}(R+F^{\mu\nu}F_{\mu\nu}-\frac{12}{L^2}), \label{EMact}
\ee
describing Einstein-Maxwell gravity with a negative cosmological constant.

Following \cite{DHoker-2009}, to obtain a gravitational background that accommodates a constant magnetic field, we consider solutions to the theory governed by (\ref{EMact}) that are of the form
\be
ds^2=-U(r)dt^2+V(r)(dx^2+dy^2)+W(r)dz^2+\frac{dr^2}{U(r)}, \label{5dmet}
\ee
where $t, x, y$ and $z$ are the directions of the holographic boundary.

Given that $U, V$ and $W$ depend exclusively on the radial coordinate, a field strength of the form  $F=\mathcal{B}\, dx\wedge dy$ will identically satisfy Maxwell equations, and to find a solution only Einstein equations will have to be solved to determine the specific shape of the metric functions.

We have not been able to solve this system analytically, so we have resorted to a numerical construction, of which the particulars have been previously discussed and can be found, for instance, in \cite{Arean:2016het}, so here we will only mention the properties relevant to the present work while a minor necessary extension will be discussed in appendix \ref{InteriorNumeric}.

The solutions we construct have an event horizon at a value of the radial coordinate that we will denote as $\rh$, and in the region close to it, the geometry approaches BTZ$\times R^2$. From the details in \cite{Arean:2016het}, it is easy to see that the temperature associated with the near horizon geometry is determined to be $T=\frac{3\rh}{2\pi}$ for the Euclidean continuation of the background to be regular. As we move away from the horizon, all solutions start looking like a five dimensional black brane and as $r\rightarrow\infty$, they transits to another asymptotic region where their geometry approaches $AdS_5$. This backgrounds group into a one parameter family of solutions where each physically different member\footnote{The only scale parameter in the background is $\rh$, and since Faraday tensor is a 2-form, the intensity of the magnetic field can only be measured in multiples of $\rh^2$, or equivalently, $T^2$. A solution with a given value for $\frac{\mathcal{B}}{T^2}=\frac{\mathcal{B}}{(\frac{3\rh}{2\pi})^2}$ and arbitrary values of $\rh$ and $\mathcal{B}$ can be brought to have $\rh=1$ through scaling the radial coordinate by a constant, so, as long as $\mathcal{B}$ is adjusted to keep the actual intensity of the magnetic field $\mathcal{B}/T^2$ fixed, the two backgrounds will be physically equivalent. This is confirmed to be consistent with the gravitational equations presented in \cite{Martinez-y-Romero:2017awl}.} is solely characterized by the dimensionless ratio $\mathcal{B}/T^2$, and as this quantity increases, so it does the dimensionless radial position $\tilde{r}=r/\rh$ at which the background undergoes the crossover between the two asymptotic geometries. When $\mathcal{B}/T^2\rightarrow 0$ the background becomes the five dimensional black brane solution all the way to the horizon, and as $\mathcal{B}/T^2\rightarrow \infty$, the BTZ$\times R^2$ geometry keeps on extending farther towards the boundary.

On the gauge side of the correspondence, the field theory is at temperature $T$ and subject to an external magnetic field of intensity $\mathcal{B}$, while the behavior over the radial coordinate is perceived as a renormalization flow between two fixed points corresponding to the infrared and ultraviolet theories. The dimensionless radial position $r/\rh$ is roughly dual to the energy scale\cite{Martinez-y-Romero:2017awl}, so, from the behavior of the gravitational background described in the previous paragraph, we see that an increment of the ratio $\mathcal{B}/T^2$ will increase the amount of energy require to access the ultraviolet degrees of freedom \cite{Martinez-y-Romero:2017awl}. Conversely, if we work at fixed energy, moving from small to large values of $\mathcal{B}/T^2$ will take us from the ultraviolet theory to the infrared fixed point, which is the way that the results will be presented in the following sections.

To compute the entanglement velocity we will need to know how the geometry extends across the horizon. Given that the equations of motion degenerate at $\rh$, in the past we had only constructed the exterior solutions, as explained for instance in \cite{Arean:2016het}. The extension is simple and we show how to do it in appendix \ref{InteriorNumeric}, where it can also be seen that the BTZ$\times R^2$ geometry extends to the interior of the horizon only down to a given radial coordinate, below which, the solution again approaches that of the black brane close to the singularity. Just as in the exterior, the size of the region where the geometry is approximately BTZ$\times R^2$ grows with $\mathcal{B}/T^2$, and the radial position at which the geometry transits to the black brane solution gets closer to the singularity as $\mathcal{B}/T^2$ increases.

\section{Shock wave geometry} \label{sec-vb1}

\begin{figure}[h!]
\centering
\captionsetup{justification=centering}

\begin{tikzpicture}[scale=1.5]

\draw [thick]  (0,0) -- (0,3);
\draw [thick]  (3,0) -- (3,3);
\draw [thick,dashed]  (0,0) -- (3,3);
\draw [thick,dashed]  (0,3) -- (3,0);
\draw [thick,decorate,decoration={zigzag,segment length=2mm, amplitude=0.5mm}]  (0,3) -- (3,3);
\draw [thick,decorate,decoration={zigzag,segment length=2mm,amplitude=.5mm}]  (0,0) -- (3,0);
\draw [thick,<->] (1,2.15) -- (1.5,1.65) -- (2,2.15);
\node [above] at (0.95,2.13) {$u$};
\node [above] at (2.07,2.13) {$v$};

\node[scale=0.9, align=center] at (1.5,2.65) {Future\\ Interior};
\node[scale=0.9,align=center] at (1.5,.6) {Past\\ Interior};
\node[scale=0.9,align=center] at (0.6,1.6) {Left\\ Exterior};
\node[scale=0.9,align=center] at (2.4,1.6) {Right\\ Exterior};
\end{tikzpicture}
\vspace{0.1cm}
\caption{ \small Penrose diagram for the two-sided black branes we consider. This geometry is dual to a thermofield double state $|TFD \rangle$ made by entangling two copies of the boundary theory.}
\label{fig-PenroseAdS}
\end{figure}
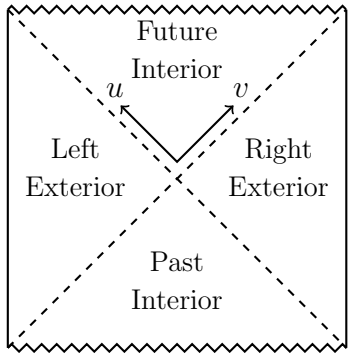

In this section we explain how to extract some chaotic properties of the boundary theory from shock waves in the bulk. We start with a generic black hole metric of the form
\be
ds^2 = G_{mn}dx^m dx^n= - G_{tt} dt^2+G_{rr}dr^2+G_{ij}dx^i dx^j\,,
\ee
that in particular can accommodate (\ref{5dmet}), and in agreement with the previous sections, the metric potentials depend solely on the radial coordinate $r$, the boundary is at $r=\infty$ and the black hole horizon at $r=\rh$.\footnote{We emphasize that the shock wave solutions that we construct here do not assume that the geometry is asymptoically AdS.} In the near-horizon region, we assume that
\be
G_{tt}= c_0 (r-\rh)\,,\,\,\,\,G_{rr}=\frac{c_1}{r-\rh}\,,\,\,\,\, G_{ij}(\rh)= \text{finite}\,,
\ee
where $c_0$ and $c_1$ are constants. The black hole Hawking temperature can be written as
\be
T=\frac{1}{4 \pi}\sqrt{\frac{c_0}{c_1}}\,.
\ee
We consider a maximally extended black hole solution that represents a wormhole geometry. In this case it is more convenient to work with Kruskal coordinates that cover smoothly the two sides of the geometry. We first define the Tortoise coordinate
\be
r_* = -\int_{r}^{\infty}\sqrt{\frac{G_{rr}(r')}{G_{tt}(r')}}dr'\,,
\ee
and then we define the Kruskal coordinates as\footnote{These are actually the Kruskal coordinates for the left-exterior region. See figure \ref{fig-PenroseAdS}.}
\be
uv = -e^{\frac{4\pi}{\beta} r_*}\,,\,\,\,\,u/v =-e^{-\frac{4\pi}{\beta} t}.
\ee
In terms of these coordinates the metric takes the form
\be
ds^2 = 2 A(u,v)du dv+G_{ij}dx^idx^j\,,
\ee
where
\be
A(u,v)=\frac{1}{8 \pi^2 T^2} \frac{G_{tt}}{uv}\,,
\ee
so the boundary is located at $uv=-1$, the horizon at $u=0$ or $v=0$, and the singularity at $uv =1$.

Figure \ref{fig-PenroseAdS} shows the Penrose diagram of this geometry, which is dual to a thermofield double state made by entangling two copies of the boundary theory. We now want to know how this background changes when we perturb it a very long time in the past.

Let us say that we act with some operator ${\cal{O}}_2(t_0)$ on the left-side boundary theory. In the bulk description, this creates a `particle' near the boundary of AdS, which then falls into the black hole. If the perturbation is done early enough, the particle will follow an almost null trajectory very close to the past horizon, as we will now see. Let $(u,v)=(u_0,v_0)$ be the initial position of the perturbation in Kruskal coordinates. Under time evolution these coordinates change as $(u_0,v_0) \rightarrow (e^{-\frac{2\pi}{\beta}t}u_0,e^{\frac{2\pi}{\beta}t}v_0)$, and this means that, as time passes, the perturbation gets more and more localized at $u=0$, and stretched along the $v$-direction. Besides that, from the point of view of the $t=0$ frame, the energy of the perturbation increases exponentially as $t_0$\footnote{In our convention the Killing time coordinate $t$ runs forward on the right boundary and backwards on the left. Hence, a perturbation on the left boundary at the time $t_0>0$ is in the past of the $t=0$ slice of the geometry.} moves farther into the past, {\it{i.e.}} $E=E_0 e^{\frac{2 \pi}{\beta}t_0}$. As a result, for $t_0$ far enough into the past, the energy-momentum tensor of the perturbation can be very well approximated by
\be
T^\text{shock}_{uu}= E_0\,e^{\frac{2 \pi}{\beta}t_0} \delta(u)\,a(x^i)\,,
\ee
where $E_0$ is the asymptotic energy of the perturbation and $a(x^i)$ is some function representing the localization of the operator ${\cal{O}}_2(t_0)$. Note that the shock wave divides the geometry into two halves: the causal future of the shock wave (the region $u>0$), and its causal past (the region $u<0$).

\begin{figure}[h!]
\centering
\captionsetup{justification=centering}

\begin{tikzpicture}[scale=1.]

\node[scale=1.5] at (.5,1.5) {${\cal{O}}_2(t_0)|TFD \rangle $};

\draw [line width=2,<->] (3.5,1.5) -- (4.5,1.5);

%%% singularities
\draw [thick,decorate,decoration={zigzag,segment length=2mm, amplitude=0.5mm}]  (5.5,3) -- (9.5,3);

\draw [thick,decorate,decoration={zigzag,segment length=2mm, amplitude=0.5mm}]  (6.6,0) -- (10.6,0);

%%% bottom left - top right diagonal
\draw [thick] (6.5,0) -- (9.5,3);
\draw [thick] (6.6,0) -- (9.6,3);

% left boundary
\draw [thick] (5.5,3) -- (6.5,0);

% right boundary
\draw [thick] (9.6,3) -- (10.6,0);

% top left - bottom right diagonal
\draw [thick,dashed] (5.5,3) -- (7.5,1);
\draw [thick,dashed] (8.6,2) -- (10.6,0);

\draw [thick,<->] (8.43,2.13) -- (7.43,1.13);

\node [scale=1.2] at (7.7,1.8) {$\alpha$};

\end{tikzpicture}
\vspace{0.1cm}
\caption{ \small Penrose diagram for the shock wave geometry. This geometry is dual\\ to a thermofield double state perturbed at a time $t_0$ in the far past.}
\label{fig-PenroseShockWave}
\end{figure}
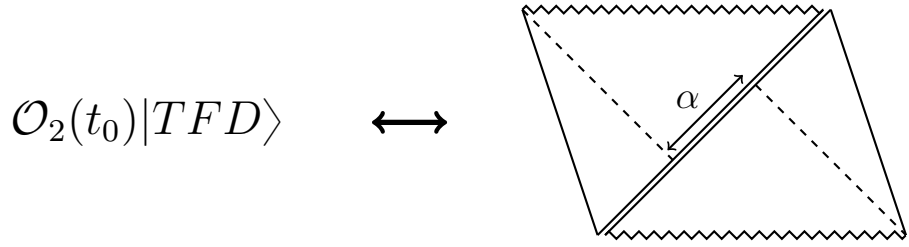

The backreaction to this perturbation on the geometry is actually very simple and can be described by a shift $v \rightarrow v + \alpha(t,x^i)$ in the causal future of the shock wave, while the causal past is unaffected\footnote{This was first done in \cite{Dray-85} for Minkowski spacetime, and then generalized for generic curved spacetimes in \cite{Sfetsos-94}. More details about the case of anisotropic metrics can be found, for instance, in \cite{Jahnke-2017}.}. This is illustrated in figure \ref{fig-PenroseShockWave}. 

The shock wave metric is simply given by \cite{Dray-85,Sfetsos-94,Jahnke-2017}
\be
ds^2= 2 A(u,v)du dv+G_{ij}dx^idx^j-2 A(u,v)\alpha(t,x^i) \delta(u) du^2\,,
\ee
where the shock wave profile $\alpha(t,x^i)$ has to obey
\be
\left[ G^{ii}\partial_i \partial_i-\left( \frac{2\pi}{\beta} \right)^2 \frac{G^{ii}(\rh)G_{ii}'(\rh)}{G_{tt}'(\rh)} \right]\alpha(t,x^i)= \frac{8 \pi G_{N} E}{A(\rh)} e^{2 \pi t / \beta} a(x^i)\,,
\ee
and $G_{ij}$ has been considered to be diagonal. Taking $a(x^i)\sim \delta(x^i)$ and assuming that $|\vec{x}|>>1$, the above equation has a solution of the form
\begin{equation}
\alpha(t,x^i) \sim \text{exp}\, \left[ \frac{2\pi}{\beta}\left(t-t_* - \frac{|x^i|}{v_B^i} \right) \right]\,,
\label{eq-h(x)}
\end{equation}
where $t_* \sim \frac{\beta}{2 \pi} \log \frac{1}{G_N}\sim \frac{\beta}{2 \pi} \log N^2$ is the scrambling time and
\begin{equation}
v_{B}^i=\frac{1}{\sqrt{G_{ii}(\rh)}}\frac{\sqrt{G_{tt}'(\rh)}}{\sqrt{G^{jj}(\rh)G_{jj}'(\rh)}}\,,
\end{equation}
is the butterfly velocity along the $x^i$-direction. By comparing (\ref{eq-h(x)}) and (\ref{eq-C(t,x)}) we can also extract the Lyapunov exponent of the system as $\lambda_L = 2\pi/\beta$.

We now specialize our formula for the butterfly velocity of the magnetic brane solution described in section \ref{sec-gravity}. 

Along the direction of the magnetic field, the butterfly velocity reads
\be
v_{B,\parallel}^2 =\frac{U'}{W \left(2\frac{V'}{V}+\frac{W'}{W} \right)} \Big|_{r=\rh}\,,
\label{eq-VBz}
\ee
while the butterfly velocity along any direction perpendicular to the magnetic field is given by
\be
v_{B,\perp}^2 =\frac{U'}{V \left(2\frac{V'}{V}+\frac{W'}{W} \right)}\Big|_{r=\rh}\,.
\label{eq-VBxy}
\ee
Both $v_{B,\parallel}^2$ and $v_{B,\perp}^2$ are functions of the ratio $\mathcal{B}/T^2$, which controls the strength of the magnetic field on the system. Figure \ref{fig-VB2} shows how these velocities are affected by the presence of the external magnetic field.

\subsubsection*{$v_B$ at the UV fixed point ($\mathcal{B}/T^2 \rightarrow 0$)}
The gravitational dual to the UV fixed point reads \footnote{As explained in appendix \ref{InteriorNumeric}, the coordinate $r$ in \eqref{eq-metric-UV} has been shifted so that the Hawking temperature is the same across the family of solutions.}
\bea
U(r)&=&(r+\rh/2)^2\left(1-\frac{(3\rh/2)^2}{(r+\rh/2)^2} \right)\,,\nonumber\\
V(r)&=&W(r)=(r+\rh/2)^2\,,
\label{eq-metric-UV}
\eea
with Hawking temperature $T=3 \rh /(2 \pi)$. In this case, as the system is isotropic, so is the butterfly velocity, which is given by
\be
v_B^\mt{UV}=  \sqrt{\frac{2}{3}}\,,
\label{eq-vB-UV}
\ee
This result is consistent with the result for a $d$-dimensional CFT, which is $v_B^2 =\frac{d}{2(d-1)}$ \cite{BHchaos3}. This is expected, since the theory is effectively described by a 4-dimensional CFT at the UV.
\subsubsection*{$v_B$ at the IR fixed point ($\mathcal{B}/T^2 \rightarrow \infty$)}
The gravitational dual to the IR fixed point reads \footnote{Although the numerical solutions need to be rescaled in order to approach $AdS_{5}$ for $r\rightarrow\infty$ (see appendix \ref{InteriorNumeric} for details), they still go to \eqref{eq-metric-IR} as $\mathcal{B}/T^{2}\rightarrow\infty$. In the notation of appendix \ref{InteriorNumeric}, this is because the ratio $V(r)/\mathcal{B}$ is unchanged under the rescaling and $w_{\infty}\rightarrow 1$ as $\mathcal{B}/T^{2}\rightarrow\infty$.}
\bea
U(r)&=&\left(3 r^2 - 3\rh^2 \right)\,,\nonumber\\
V(r)&=& \mathcal{B}/\sqrt{3}\,,\nonumber\\
W(r)&=&3 r^2\,,
\label{eq-metric-IR}
\eea
with Hawking temperature $T=3 \rh /(2 \pi)$. The butterfly velocity along the direction of the magnetic field is 
\be
v_{B,\parallel}^\mt{IR}=1\,,
\label{eq-vBz-IR}
\ee
which is the expected result for a BTZ black hole. The butterfly velocity along any direction perpendicular to the magnetic field is 
\be
v_{B,\perp}^\mt{IR}= \frac{2 \pi}{3^{1/4}} \frac{T}{\sqrt{\mathcal{B}}} << 1\,,
\label{eq-vBxy-IR}
\ee
which, consistently with the dimensional reduction suffer by the theory at the IR fixed point\cite{Martinez-y-Romero:2017awl}, vanishes as $\mathcal{B}/T^2 \rightarrow \infty$ at any set temperature.

\subsubsection*{$v_B$ at intermediate values of $\mathcal{B}/T^2$ }
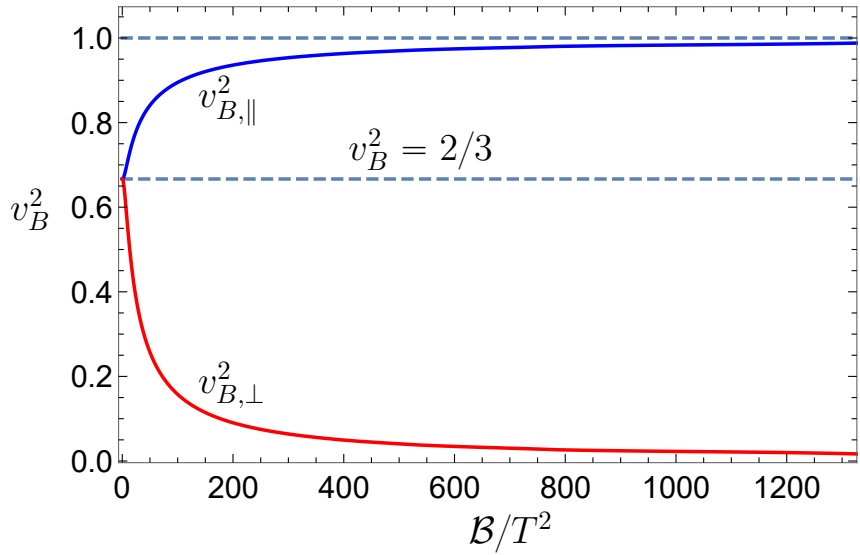
\begin{figure}[H]
\begin{center}
\setlength{\unitlength}{1cm}
\includegraphics[width=0.6\linewidth]{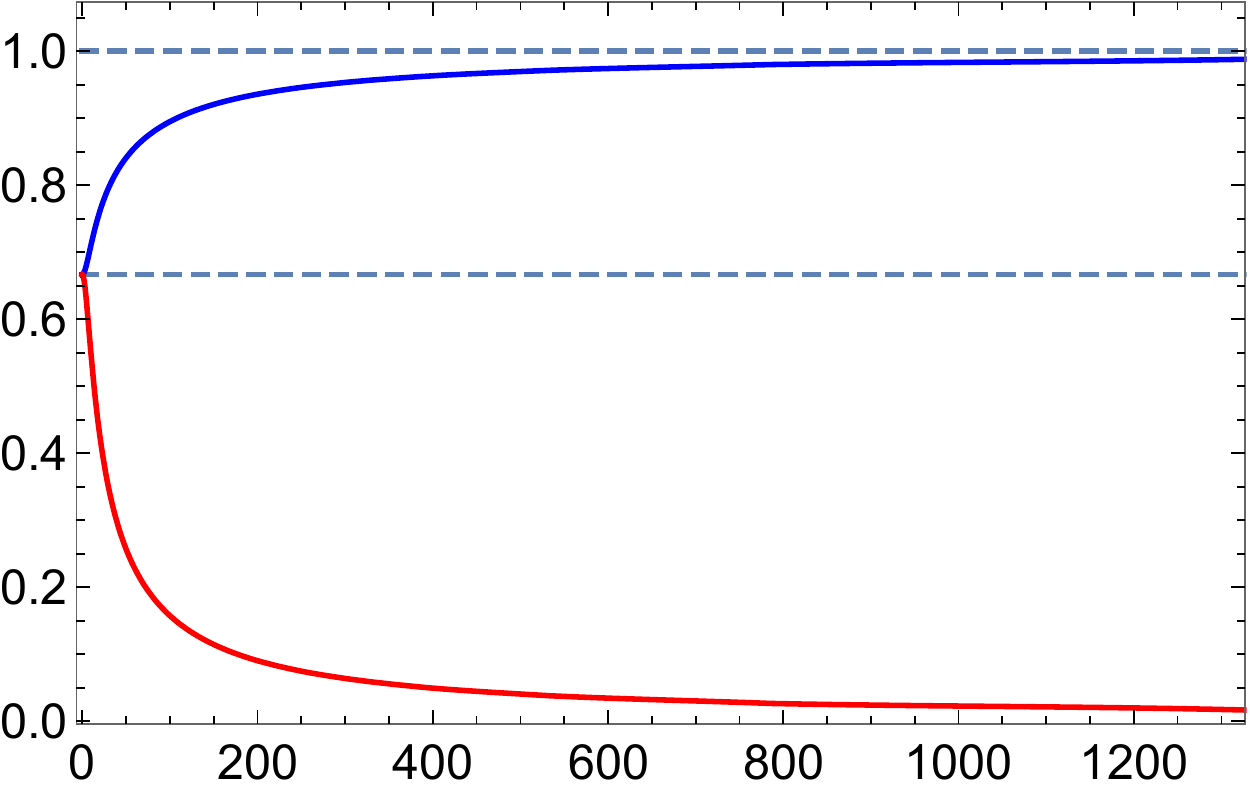} 
\put(-5.2,-.5){\large $\mathcal{B}/T^2$}
\put(-11.3,+3.8){\large $v_{B}^2$}
\put(-8.8,1.5){\large $v_{B,\perp}^2$}
\put(-8.8,+5.3){\large $v_{B,\parallel}^2$}
\put(-6.8,+4.6){\large $v_{B}^2=2/3$}

\end{center}
\caption{ \small
Butterfly velocity squared $v_\mt{B}^2$ versus the dimensionless parameter $\mathcal{B}/T^2$. The blue curve represents the butterfly velocity along the direction of the magnetic field, while the red curve stands for the butterfly velocity along any direction perpendicular to the magnetic field. The horizontal lines represent either the conformal result $v_\mt{B}^2=2/3$ or the the speed of light.}
\label{fig-VB2}
\end{figure}

As mentioned before, for intermediate values of $\mathcal{B}/T^2$ the metric functions can only be obtained numerically, so in figure \ref{fig-VB2} we show the numerical results for the square of $v_{B,\parallel}$ and $v_{B,\perp}$ as function of $\mathcal{B}/T^2$. We can see that, as we increase the value of $\mathcal{B}/T^2$, the butterfly velocities smoothly interpolate between the UV result, given in equation (\ref{eq-vB-UV}), and the IR results, given in equations (\ref{eq-vBz-IR}) and (\ref{eq-vBxy-IR}).

\section{Two-sided mutual Information} \label{sec-mutual}

In this section we compute the two-sided mutual information for strip-like regions in the boundary theory and show how this quantity drops to zero in shock wave geometries. For simplicity, we only consider the case of homogeneous shock waves, in which the shock wave parameter is given by  $\alpha = \text{const} \times e^{2\pi t_0/\beta}$.

\begin{figure}[H]
\begin{center}
\begin{tabular}{cc}
\setlength{\unitlength}{1cm}
\hspace{-0.9cm}
\includegraphics[width=7cm]{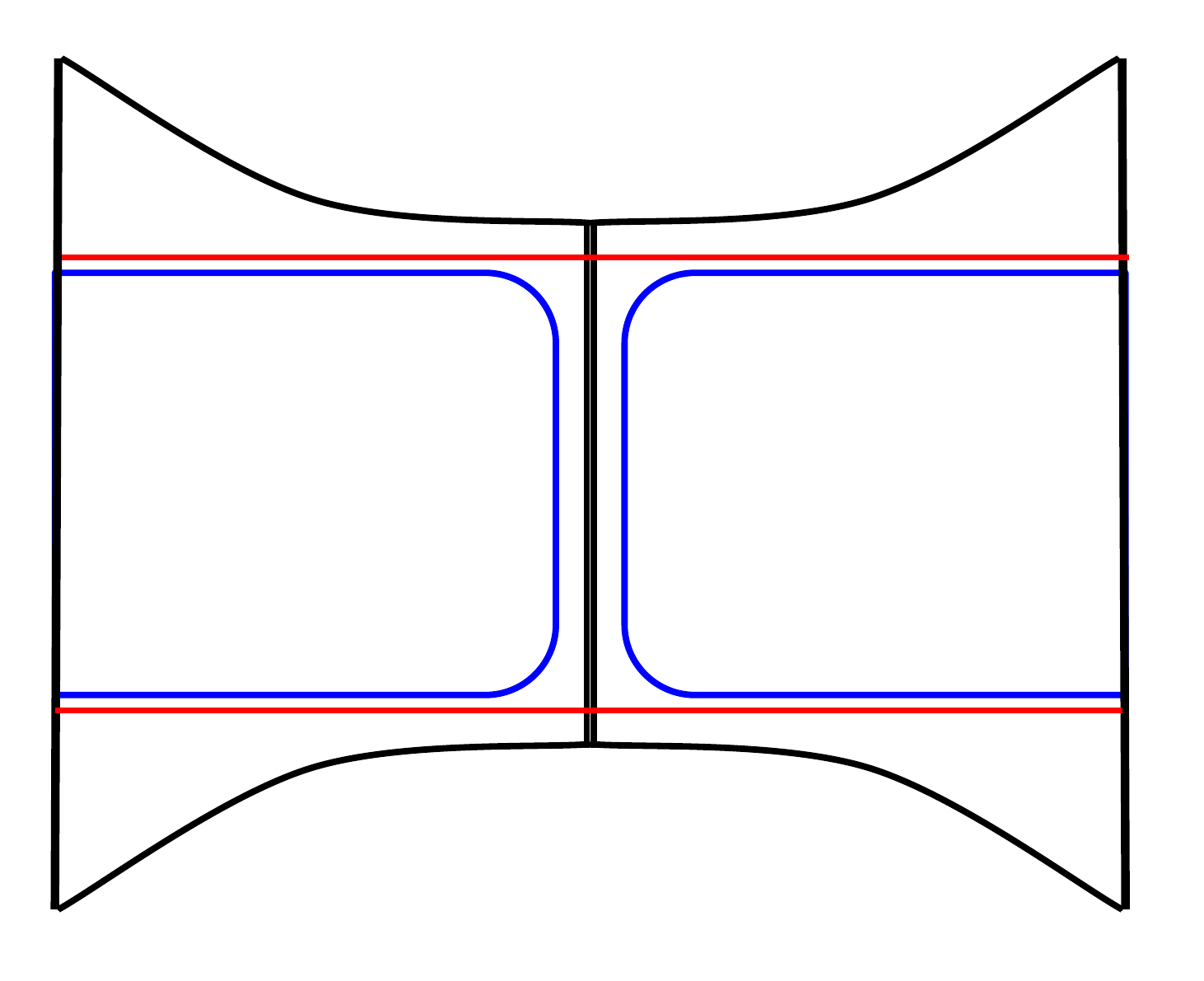} 
\qquad\qquad & 
\includegraphics[width=7.8cm]{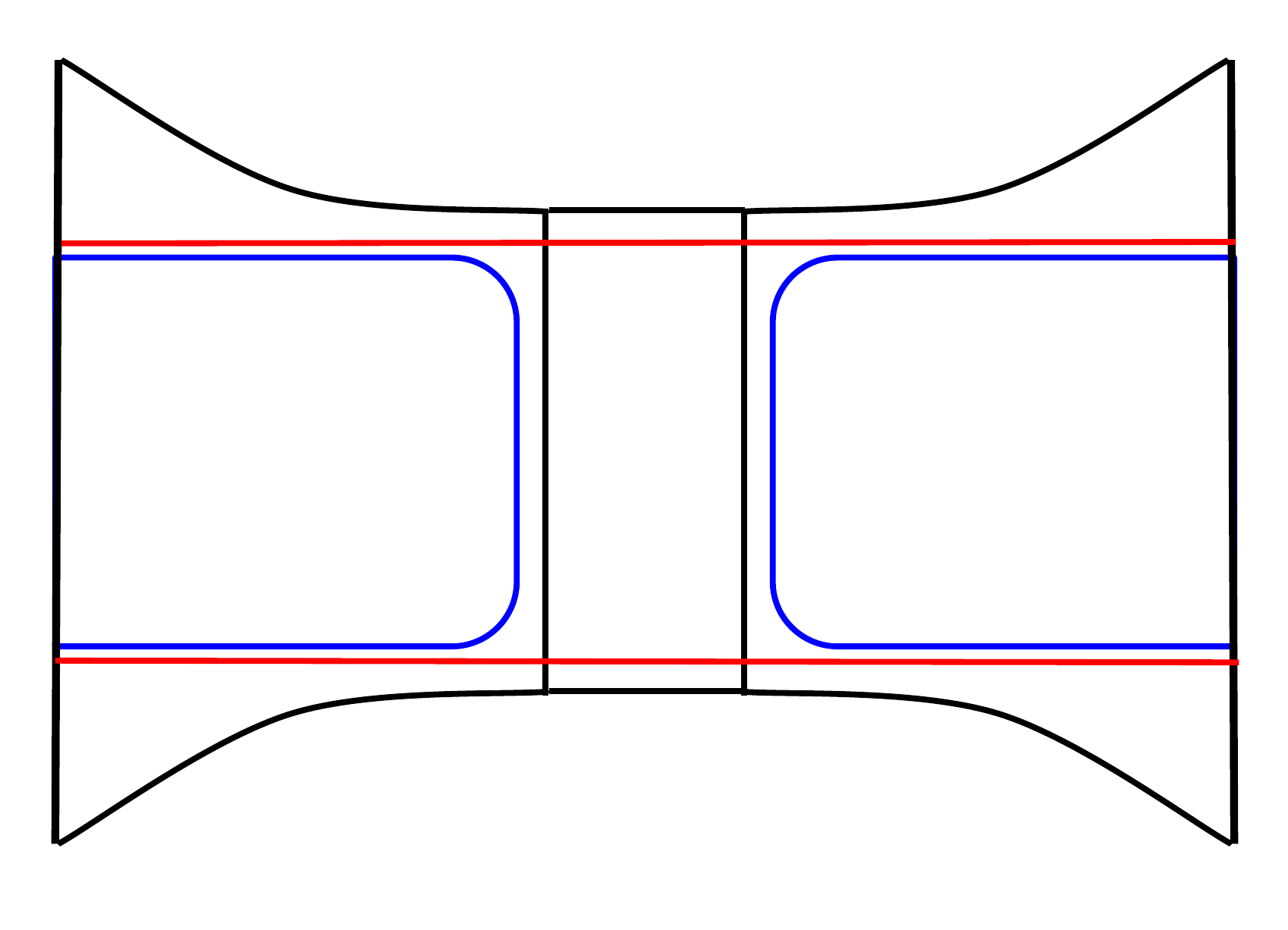}
\qquad
  
         \put(-385,80){$\gamma_A$}
         \put(-355,80){$\gamma_B$}
         
         \put(-150,80){$\gamma_A$}
         \put(-85,80){$\gamma_B$}
        
         \put(-435,125){$\gamma_1$}
         \put(-435,35){$\gamma_2$}
                    
         \put(-200,125){$\gamma_1$}
         \put(-200,35){$\gamma_2$} 
         
         \put(-380,138){$\text{Horizon}$}
         \put(-130,138){$\text{Horizon}$}         
         
         \put(-115,5){$(b)$}
         \put(-370,5){$(a)$}

         \put(-366,138){\rotatebox{-90}{$\rightarrow$}}
         
         \put(-130,138){\rotatebox{-135}{$\rightarrow$}}
         \put(-105,134){\rotatebox{-45}{$\rightarrow$}}
         
         \put(-115,60){\small \rotatebox{90}{$\text{wormhole}$}}
         
         \put(-465,50){\small \rotatebox{90}{$\text{left boundary}$}} 
         \put(-225,50){\small \rotatebox{90}{$\text{left boundary}$}} 
         
         \put(-270,120){\small \rotatebox{-90}{$\text{right boundary}$}}
         \put(-5,120){\small \rotatebox{-90}{$\text{right boundary}$}}

\end{tabular}
\end{center}
\caption{ \small (a) Schematic representation of the $t=0$ slice of the two-sided black brane geometry. (b) Schematic representation of the shock wave geometry, in which the wormhole becomes longer. In both cases the blue curves represent the U-shaped extremal surfaces $\gamma_A$ (in the left side of the geometry) and $\gamma_B$ (in the right side of the geometry). The red curves represent the extremal surfaces $\gamma_1$ and $\gamma_2$ connecting the two sides of the geometry . The extremal surface $\gamma_\text{wormhole}$ defined in the text is given by $\gamma_\text{wormhole}=\gamma_1 \cup \gamma_2$.}
\label{fig-surfaces}
\end{figure}
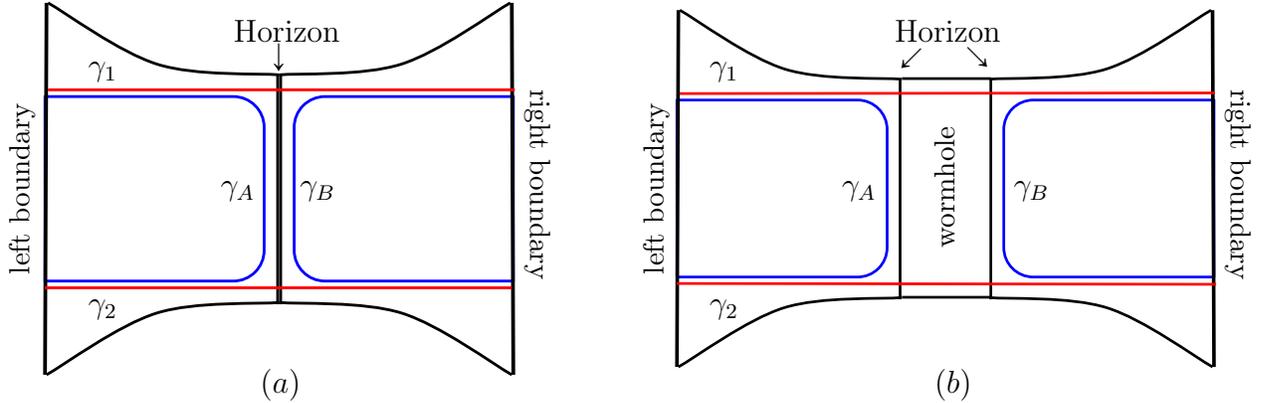

The two-sided mutual information between a region $A$ in the left boundary and a region $B$ in the right boundary is given by
\be
I(A,B)=S_A+S_B-S_{A \cup B}\,,
\ee
where $S_X$ stands for the entanglement entropy of region $X$. This quantity is always positive and provides an upper bound for correlations between $A$ and $B$ \cite{boundIAB}. The atypical entanglement pattern of the thermofield double state at $t=0$ is diagnosed by a positive mutual information between large regions $A$ and $B$. We compute the above entanglement entropies holographically, using the Ryu-Takayanagi prescription \cite{RT,HRT}. For simplicity, we take $A$ and $B$ to be identical strip-like regions at the $t=0$ slice of the geometry. $S_A$($S_B$) is given by $\frac{1}{4G_\mt{N}}$ times the area of the co-dimension 2 extremal surface $\gamma_A$($\gamma_B$) which is homologous to the region $A$($B$). The surface $\gamma_A$ ($\gamma_B$) is a U-shaped surface entirely contained in the left(right) exterior region of the geometry. For the computation of $S_{A \cup B}$ we have two possible options for the extremal surface, and we have to choose the one with minimal area. The first alternative is simply $\gamma_A \cup \gamma_B$, whose area is area($\gamma_A$)+area($\gamma_B$). In this case the mutual information is identically zero $I(A,B)=0$. The other option is the surface $\gamma_\text{wormhole}$ that stretches through the wormhole, connecting the two sides of the geometry. In this case the mutual information is positive
\be
I(A,B)= \frac{1}{4G_\mt{N}}\left( \text{area}(\gamma_A)+\text{area}(\gamma_B)-\text{area}(\gamma_\text{wormhole}) \right) \geq 0\,.
\ee
In figure \ref{fig-surfaces} we make a schematic representation of the surfaces $\gamma_A$, $\gamma_B$ and $\gamma_\text{wormhole}$ in a two-sided black brane geometry with and without a shock wave at the horizon. In section \ref{sec-MIversusL} we will show that for large regions $A$ and $B$, the surface $\gamma_\text{wormhole}$ has the minimal area.

When we perturb the system in the asymptotic past we create a shock wave geometry, in which the left and right exterior regions are not altered, but the wormhole becomes longer. The strength of the shock wave and the length of the wormhole are both controlled by the shock wave parameter $\alpha$. In the shock wave geometry, only the quantities that probe the interior of the black hole are affected by the shock wave. Hence, the U-shaped extremal surfaces $\gamma_A$
and $\gamma_B$ are not affected by the shock wave, while the surface $\gamma_\text{wormhole}$ becomes longer. As a result the entanglement entropy $S_{A \cup B}(\alpha)$ generically depends on the shock wave parameter, and it actually is an increasing function of it. So, in the shock wave geometry, we write the mutual information as $I(A,B;\alpha)=S_A+S_B-S_{A \cup B}(\alpha)$, where we indicate that $S_A$ and $S_B$ do not depend on $\alpha$, while $S_{A \cup B}$ does.

The quantities $S_A, S_B$ and $S_{A \cup B}$ are divergent because they are computed over surfaces that extend all the way to one or two of the asymptotic boundaries in the geometry. None the less, the mutual information is finite, because the divergences in $S_A$ and $S_B$ cancel the divergence of $S_{A \cup B}$.

It is convenient to define a regularized version of $S_{A \cup B}$ as
\be
S^\text{reg}_{A \cup B}(\alpha)=S_{A \cup B}(\alpha)-S_{A \cup B}(\alpha=0)\,, 
\ee
so that by writing
\be
I(A,B;\alpha)=S_A+S_B-\left[ S_{A \cup B}(\alpha)-S_{A \cup B}(0) \right]-S_{A \cup B}(0)=I(A,B;0)-S^\text{reg}_{A \cup B}(\alpha)\,,
\ee
the mutual information not only splits into two finite parts, but also one of them, $I(A,B;0)$, is the mutual information of the unperturbed geometry. As we will show in section \ref{sec-MIversusL}, the value of $I(A,B;0)$ depends on the temperature of the system and on the width of the strip-like regions $A$ and $B$. 

Notice that since $S^\text{reg}_{A \cup B}(\alpha)$ is an increasing function of $\alpha$, the mutual information $I(A,B;\alpha)$ decreases as $\alpha$ gets bigger. Eventually, the area of $\gamma_\text{wormole}$ becomes larger than the area of $\gamma_A \cup \gamma_B$ and the mutual information has a transition to a constant vanishing value. Finally, note that increasing $\alpha$ is equivalent to move the creation of the shock wave to earlier times, leading us to conclude that the two-sided mutual information drops to zero as we move the perturbation further into the past.

%%% extremal surfaces

\subsection{Two-sided mutual information versus strip's width} \label{sec-MIversusL}

In this section we compute the mutual information in the unperturbed geometry as a function of the strip's width. As we are dealing with an anisotropic system, we consider two types of strips: the strips defined by the equation $0 \leq x \leq \ell$, which we call parallel strips, and those defined by the equation $0 \leq z \leq \ell$, which we call orthogonal strips. The above nomenclature is based on the fact that the magnetic field is oriented along the $z$-direction, and rotational invariance in the $xy$-plane implies that no generality is lost when the parallel strips are defined in the the way we just described.

\subsubsection*{Orthogonal strips $0 \leq z \leq \ell$}
This region is delimited by two hyperplanes, one at $z=0$ and the other at $z=\ell$.
The appropriate embedding for this case is $X^m=(0,x,y,z(r),r)$. The components of the induced metric $g_{ab}$ are
\bea
g_{xx}&=& g_{yy}=V(r)\,,\\
g_{rr}&=& \frac{1}{U(r)}+W(r)z'(r)^2\,.
\eea
Let us first compute $S_A$. The corresponding area functional is
\be
\text{area}(\gamma_A)=\int dx\, dy\, dr\, \sqrt{\det(g_{ab})} = V_2 \int dr\, V(r)\,\sqrt{\frac{1}{U(r)}+W(r)z'(r)^2} =V_2 \int dr\,\mathcal{L}(z,z';r)\,,
\ee
where $V_2 = \int dx\,dy$ is the volume of the hyperplanes at $z=0$ and $z=\ell$. The `Lagrangian' $\mathcal{L}(z,z';r)$ does not depend on $z$, and hence there is a conserved quantity associated to translations in $z$
\be
\gamma=\frac{\partial \mathcal{L}}{\partial z'}=\frac{V(r) W(r) z'}{\sqrt{\frac{1}{U}+W z'^2}}= V(r_m)\sqrt{W(r_m)}\,,
\ee
where, in the last equality, we calculated $\gamma$ at the point $r_m$ at which $z' \rightarrow \infty$. The extremal area\footnote{In a slight abuse of language we use the same notation for the area functional and for the extremal area.} can then be calculated as
\be
\text{area}(\gamma_A)=2 V_2 \int_{r_m}^{\infty} dr \frac{V}{\sqrt{U}}\frac{1}{\sqrt{1-\gamma^2 V^{-2}W^{-1}}}\,.
\ee
From the above result we can finally compute $S_A$ as
\be
S_A =\frac{\text{area}(\gamma_A)}{4 G_\mt{N}}=\frac{V_2}{2 G_\mt{N}} \int_{r_m}^{\infty} dr \frac{V}{\sqrt{U}}\frac{1}{\sqrt{1-\gamma^2 V^{-2}W^{-1}}}\,,
\ee
with an identical result for $S_B$. We proceed to the calculation of $S_{A \cup B}$. In this case the surface is the union of two hyperplanes connecting the two sides of the geometry, so that $z'=0$ and the extremal area is given by
\be
\text{area}(\gamma_\text{wormhole})= 4 V_2 \int_{\rh}^{\infty} dr \frac{V}{\sqrt{U}}\,,
\ee
where the overall factor of 4 comes from the two sides of the geometry and from the two hyperplanes. We then find $S_{A \cup B}$ to be
\be
S_{A \cup B}=\frac{V_2}{G_\mt{N}} \int_{\rh}^{\infty} dr \frac{V}{\sqrt{U}}\,.
\label{eq-SAUBunp}
\ee
We finally compute the mutual information as
\be
I(A,B;0)=\frac{V_2}{G_\mt{N}} \left[ \int_{r_m}^{\infty} dr \frac{V}{\sqrt{U}}\frac{1}{\sqrt{1-\gamma^2 V^{-2}W^{-1}}} - \int_{\rh}^{\infty} dr \frac{V}{\sqrt{U}}\right]\,,
\ee
where,as before, the $0$ indicates that $I(A,B;0)$ is computed in the unperturbed geometry. Note that $I(A,B;0)$ depends on the temperature via $\rh$ and on the `turning point' $r_m$. The value of $r_m$ defines the width of the strip as
\be
\ell=\int dz =\int_{r_m}^{\infty} dr\,z'(r)=2\int_{r_m}^{\infty} dr\,\frac{1}{\sqrt{W U}}\frac{1}{\sqrt{\gamma^{-2}V^2 W-1}}\,.
\ee
As both the mutual information and the strip's width depend on $r_m$, we can make a parametric plot of $I(A,B;0)$ versus $\ell$ that we show in figure \ref{fig-MIversusL}(a).

\subsubsection*{Parallel strips $0 \leq x \leq \ell$}
This region is delimited by two hyperplanes, one at $x=0$ and the other at $x=\ell$.
The appropriate embedding for this case is $X^m=(0,x(r),y,z,r)$. The components of the induced metric are
\bea
g_{xx}&=& V(r)\,,\\
g_{zz}&=& W(r)\,,\\
g_{rr}&=& \frac{1}{U(r)}+V(r)x'(r)^2\,.
\eea
Proceeding as before we can compute the mutual information and the strip's width respectively as
\be
I(A,B;0)=\frac{V_2}{G_\mt{N}} \left[ \int_{r_m}^{\infty} dr \frac{\sqrt{V W}}{\sqrt{U}}\frac{1}{\sqrt{1-\gamma^2 V^{-2}W^{-1}}} - \int_{\rh}^{\infty} dr \frac{\sqrt{V W}}{\sqrt{U}}\right]\,,
\ee
and
\be
\ell=\int dx =\int_{r_m}^{\infty} dr\,x'r)=2\int_{r_m}^{\infty} dr\,\frac{1}{\sqrt{V U}}\frac{1}{\sqrt{\gamma^{-2}V^2 W-1}}\,,
\ee
where $\gamma=V(r_m)\sqrt{W(r_m)}$. As before we can make a parametric plot of $I(A,B;0)$ versus $\ell$.

\begin{figure}[H]
\begin{center}
\begin{tabular}{cc}
\setlength{\unitlength}{1cm}
\hspace{-0.9cm}
\includegraphics[width=7.3cm]{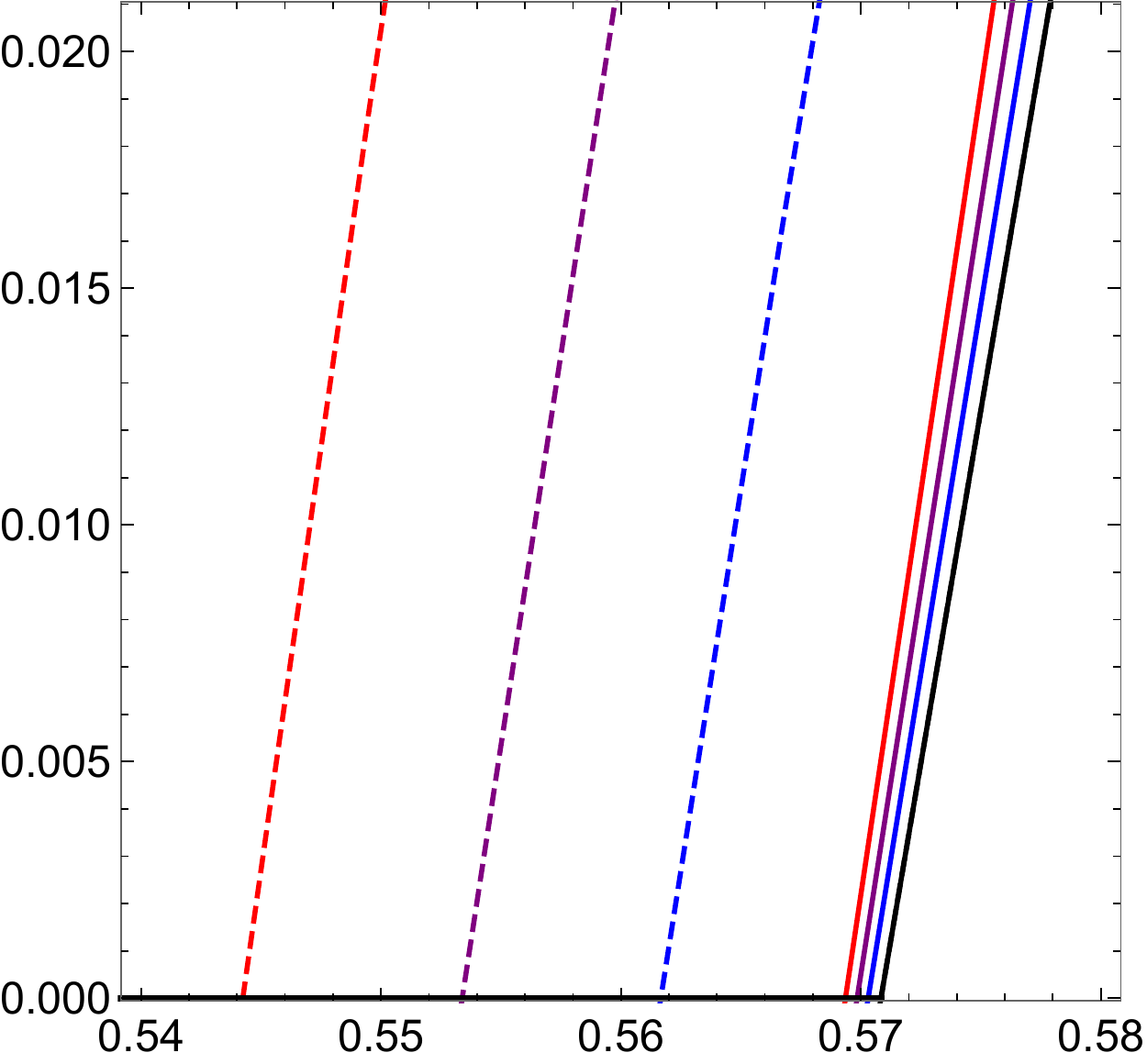} 
\qquad\qquad & 
\includegraphics[width=7cm]{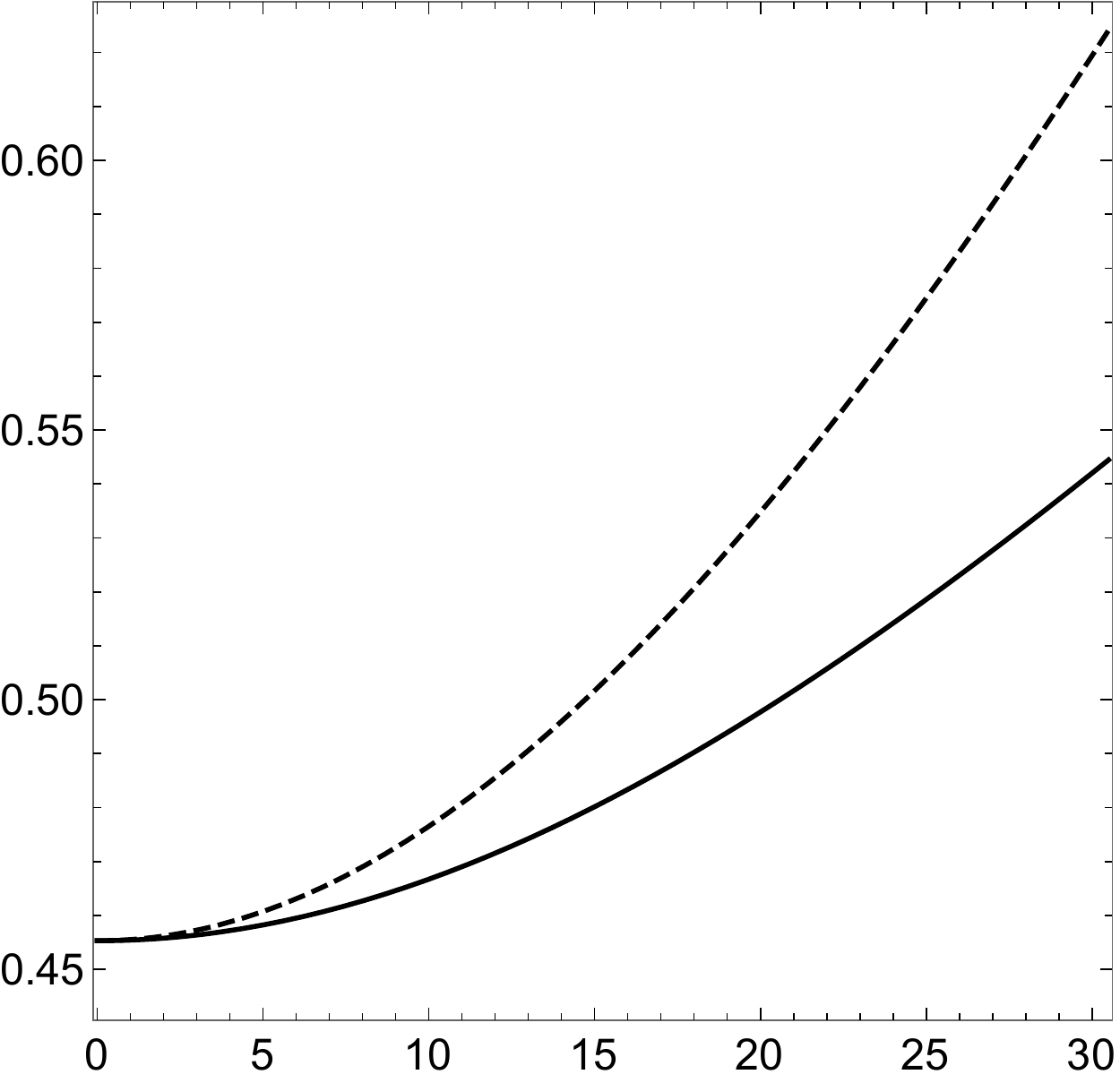}
\qquad
 \put(-355.,-12){\large $\ell/\ell_{AdS}$}
 \put(-110.,-12){\large $\mathcal{B}/T^2$}
 \put(-350.,-30){\large $(a)$}
 \put(-105.,-30){\large $(b)$}
\put(-465,+75){\rotatebox{90}{\large $I(A,B)$}}
\put(-220,+80){\rotatebox{90}{\large $I(A,B)$}}
         
\end{tabular}
\end{center}
\caption{ \small
(a) Mutual Information (in units of $V_2/G_\mt{N}$) as a function of the strip's width $\ell$ for several values of $\mathcal{B}/T^2$. The curves correspond to $\mathcal{B}/T^2 = 0$ (black curve), $\mathcal{B}/T^2 = 14.8$ (blue curves), $\mathcal{B}/T^2 = 21.2$ (purple curves), $\mathcal{B}/T^2 = 27.5$ (red curves). (b) Mutual information (in units of $V_2/G_\mt{N}$) versus $\mathcal{B}/T^2$ for strips of fixed width $\ell/\ell_{AdS} =0.75$. The continuous (dashed) curves correspond to the results for orthogonal (parallel) strips.}
\label{fig-MIversusL}
\end{figure}
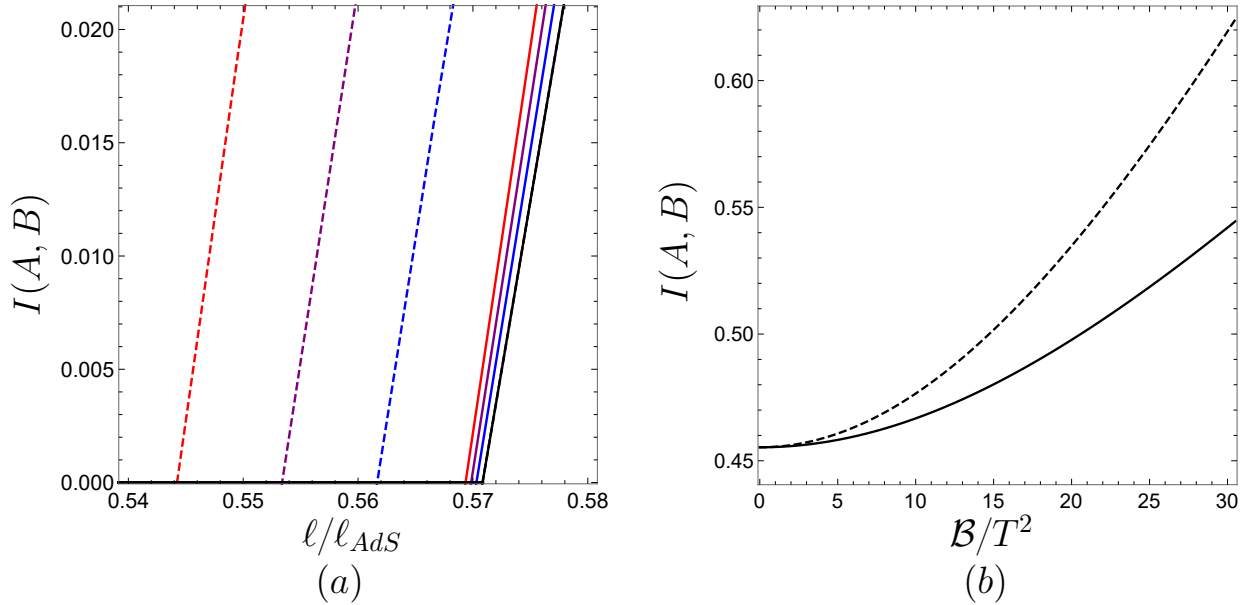

Figure \ref{fig-MIversusL}(a) shows the mutual information as a function of the strip's width for orthogonal and parallel strips for various values of $\mathcal{B}/T^2$. The figure \ref{fig-MIversusL}(b) shows the mutual information as a function of the magnetic field for parallel and orthogonal strips of fixed width.

%\begin{figure}[H]
%\begin{center} 
%\setlength{\unitlength}{1cm}
%\includegraphics[width=0.6\linewidth]{MIvsL.pdf}
%\put(-5.,-.5){\large $\ell$}
%\put(-11.2,+4.6){\rotatebox{90}{\large $I(A,B)$}}
%
%\end{center}
%\caption{ \small
%Mutual Information (in units of $V_2/G_\mt{N}$) as a function of the strip's width $\ell$ for several values of $B/T^2$. The curves correspond to $B/T^2 = 0$ (black curve), $B/T^2 = 14.8$ (blue curves), $B/T^2 = 21.2$ (purple curves), $B/T^2 = 27.5$ (red curves). The continuous (dashed) curves correspond to the results for orthogonal (parallel) strips.}
%\label{fig-MIversusL}
%\end{figure}

\subsection{Disruption of the two-sided mutual information}

\begin{figure}[h!]
\centering
%\captionsetup{justification=centering}

\begin{tikzpicture}[scale=1.5]

%%% singularities
\draw [thick,decorate,decoration={zigzag,segment length=2mm, amplitude=0.5mm}]  (5.5,3) -- (9.5,3);

\draw [thick,decorate,decoration={zigzag,segment length=2mm, amplitude=0.5mm}]  (6.6,0) -- (10.6,0);

%%% bottom left - top right diagonal
\draw [thick] (6.5,0) -- (9.5,3);
\draw [thick] (6.6,0) -- (9.6,3);

% left boundary
\draw [thick] (5.5,3) -- (6.5,0);

% right boundary
\draw [thick] (9.6,3) -- (10.6,0);

% top left - bottom right diagonal
\draw [thick,dashed] (5.5,3) -- (7.5,1);
\draw [thick,dashed] (8.6,2) -- (10.6,0);

%%% extremal surface
\draw[very thick, red] (6,1.5) -- (10.1,1.5);

\draw[thick,red, |-|] (7,1.5) -- (7.5,1.5);

\draw[thick,red, |-|] (7,1.5) -- (8,1.5);

\draw [thick,<->] (8.13,1.4) -- (7.63,.9);

\node [scale=.8] at (8.,1.0) {$\frac{\alpha}{2}$};

\draw [dashed, blue] (5.5,3) to (7.1,1.65) to [out=-35,in=180] (7.5,1.52)
to [out=0,in=-145] (7.9,1.65) to (9.5,3);

\node [scale=.7] at (7.5,1.7) {$r_0$};

\node [scale=.45] at (6.5,1.39) {$I$};
\node [scale=.45] at (7.27,1.39) {$II$};
\node [scale=.45] at (7.73,1.39) {$III$};

%\node [thick,scale=1.5] at (7.1,1.65) {$\cdot$};
%\node [thick,scale=1.5] at (7.9,1.65) {$\cdot$};

%\node [scale=1.2] at (7.7,1.8) {$\alpha$};

\end{tikzpicture}
\vspace{0.1cm}
\caption{ \small Extremal surface (horizontal, red) in the shock wave geometry. We split the left half of the surface into three parts, $I$, $II$ and $III$. The segments $II$ and $III$ have the same area and they are separated by the point $r_0$ at which the constant-$r$ surface (blue, dashed curve, defined by $r=r_0$) intersects the extremal surface.}
\label{fig-surfaceLocation}
\end{figure}
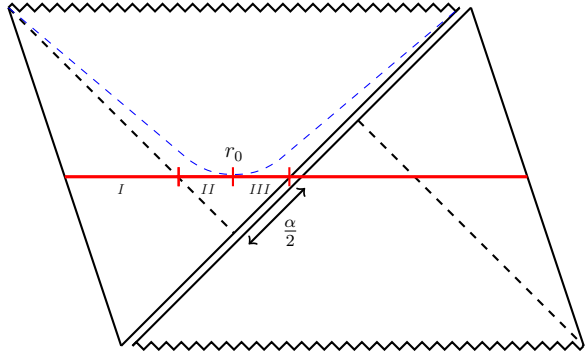

In this section we study how the two-sided mutual information drops to zero in shock wave geometries. In order to simplify the analysis, we first consider the case of semi-infinite strips. The orthogonal strip is defined by $0 \leq z < \infty$, while the parallel strips is defined by $0 \leq x < \infty$. In this case, by symmetry, we now that the extremal surface divides the bulk into two parts, as shown in figure \ref{fig-surfaceLocation}. Once we have the mutual information for a semi-infinite strip we multiply this result by two to obtain the result for a finite strip.

\subsubsection*{Orthogonal strip $0 \leq z < \infty$}
Since $S_A$ and $S_B$ are not affected by the shock wave, we only have to compute $S_{A \cup B}$.
This region is delimited by the hyperplane $z=0$. The appropriate embedding in this case is $X^m=(t,x,y,0,r(t))$. The components of the induced metric are
\bea
g_{xx}&=& g_{yy}=V(r)\,,\\ 
g_{tt}&=&-U+\frac{\dot{r}^2}{U}\,.
\eea
The area functional is then calculated as
\be
\text{area}(\gamma_\text{wormhole})=\int dx\,dy\,dt\,\sqrt{\det(g_{ab})}=V_2 \int dt\,V \sqrt{-U+\frac{\dot{r}^2}{U}}=V_2 \int dt\,\mathcal{L}(r,\dot{r};t)\,.
\ee
The above functional is invariant under $t$-translations and the associated conserved quantity is given by
\be
\gamma_{\perp}=\frac{\partial \mathcal{L}}{\partial \dot{r}}\dot{r}-\mathcal{L}=\frac{V}{\sqrt{-U+\frac{\dot{r}^2}{U}}}
\frac{\dot{r}^2}{U}=-V(r_0)\sqrt{-U(r_0)}\,,
\label{eq-rdotperp}
\ee
where, in the last equality, we computed $\gamma_{\perp}$ at the point $r_0$ at which $\dot{r}=0$. This `turning point' is located inside the horizon, where $U <0$. By solving (\ref{eq-rdotperp}) for $\dot{r}$ we can write the extremal area as
\be
\text{area}(\gamma_\text{wormhole})=V_2 \int dr\, \frac{V}{\sqrt{\gamma_{\perp}^2 V^{-2}+U}}\,.
\ee
We compute the above area in the left side of the geometry, and then we multiply the obtained result by two to account for the two sides of the geometry. As shown in figure \ref{fig-surfaceLocation}, it is convenient to split the left half of the extremal surface into three segments $I$, $II$ and $III$. The segment $I$ goes from the boundary to the horizon (at $v=0$). The segment $II$ starts at the horizon (at $v=0$) and ends at the point $r_0$. The segment $III$ goes from $r_0$ to the horizon at $u=0$. The segments $II$ and $III$ have the same area, so we can split the above integral as $\int_{I \cup II \cup III}=\int_{\rh}^{\infty}+2\int_{r_0}^{\rh}$. Therefore, $S_{A \cup B}$ can be written as
\be
S_{A \cup B}(r_0)= \frac{\text{area}(\gamma_\text{wormhole})}{4G_\mt{N}}=\frac{V_2}{2G_\mt{N}} \left[  \int_{\rh}^{\infty} dr\,  \frac{V}{\sqrt{\gamma_{\perp}^2 V^{-2}+U}}+2\int_{r_0}^{\rh} dr\,  \frac{V}{\sqrt{\gamma_{\perp}^2 V^{-2}+U}} \right],
\ee
where the overall factor of 2 accounts for the two sides of the geometry, and we indicate that $S_{A \cup B}$ depends on the turning point $r_0$. Note that, for $r_0=\rh$ we recover one half of the value given by (\ref{eq-SAUBunp}) for the unperturbed geometry\footnote{We obtain one half of the value given by (\ref{eq-SAUBunp}) because we are considering a semi-infinite strip, while for a finite strip we should multiply the result by two, therefor recovering the result in (\ref{eq-SAUBunp}).}, indicating that $r_0=\rh$ corresponds to the absence of a shock wave. We then define the regularized entanglement entropy of $A \cup B$ as
\be
S^\text{reg}_{A \cup B}=S_{A \cup B}(r_0)-S_{A \cup B}(\rh)=\frac{V_2}{2G_\mt{N}} \left[  \int_{\rh}^{\infty} dr\,  \left( \frac{V}{\sqrt{\gamma_{\perp}^2 V^{-2}+U}}- \frac{V}{\sqrt{U}} \right)+2\int_{r_0}^{\rh} dr\,  \frac{V}{\sqrt{\gamma_{\perp}^2 V^{-2}+U}} \right].
\ee
We would like to express this result in terms of the shock wave parameter $\alpha$, so we write the latter in terms of $r_0$ as\footnote{A detailed derivation of this expression is presented in Appendix A of \cite{Jahnke-2017}.}
\be
\alpha(r_0)=2 e^{K^{\perp}_1(r_0)+K^{\perp}_2(r_0)+K^{\perp}_3(r_0)}
\ee
where
\bea
K^{\perp}_1(r_0)&=&\frac{4\pi}{\beta} \int_{\bar{r}}^{r_0} \frac{dr}{U}\,,\\
K^{\perp}_2(r_0)&=&\frac{2\pi}{\beta} \int_{\rh}^{\infty} \frac{dr}{U}\left(1-\frac{1}{\sqrt{1+\gamma_{\perp}^{-2}V^2 U}} \right)\,,\\
K^{\perp}_3(r_0)&=&\frac{4\pi}{\beta} \int_{r_0}^{\rh} \frac{dr}{U}\left(1-\frac{1}{\sqrt{1+\gamma_{\perp}^{-2}V^2 U}} \right)\,.
\eea
Note that $\alpha(\rh)=0$, corresponding to the absence of a shock wave. $\alpha$ increases as we move $r_0$ deeper into the black hole, and diverges at some critical point $r_0 = r^{\perp}_c$ which is implicitly given by
\be
\frac{(V^2 U)'}{V^2 U}\Big|_{r=r^{\perp}_c}=0,
\ee
where the prime indicates a derivative with respect to $r$. Finally, the mutual information can be simply computed as 
\be
I(A,B;r_0)=I(A,B;\rh)-S^\text{reg}_{A \cup B}(r_0)\,.
\ee
Since $S^\text{reg}_{A \cup B}(r_0)$, $I(A,B;r_0)$ and $\alpha(r_0)$ are functions of the turning point $r_0$, we can make parametric plots of $S^\text{reg}_{A \cup B}$ versus $\log \alpha$ and $I(A,B)$ versus $\log \alpha$. We choose to use $\log \alpha$ because this quantity is proportional to the shock wave time $t_0$. 
\subsubsection*{Parallel strip $0 \leq z < \infty$}
Again, we only need to compute $S_{A \cup B}$. This region is delimited by the hyperplane $x=0$. The appropriate embedding is $X^m=(t,0,y,z,r(t))$ and the components of the induced metric are
\bea
g_{yy}&=&V(r)\,,\\
g_{zz}&=&W(r)\,,\\
g_{tt}&=&-U(r)+\frac{\dot{r}^2}{U(r)}\,.
\eea
Proceeding as before, we can compute the regularized entanglement entropy as
\bea
S^\text{reg}_{A \cup B}&=&S_{A \cup B}(r_0)-S_{A \cup B}(\rh)\\
&=&\frac{V_2}{2G_\mt{N}} \left[  \int_{\rh}^{\infty} dr\,  \left( \frac{\sqrt{V W}}{\sqrt{\gamma_{||}^2 V^{-1}W^{-1}+U}}- \frac{\sqrt{V W}}{\sqrt{U}} \right)+2\int_{r_0}^{\rh} dr\,  \frac{\sqrt{V W}}{\sqrt{\gamma_{||}^2 V^{-1}W^{-1}+U}} \right],\nonumber
\eea
where $\gamma_{||}=-\sqrt{W(r_0)V(r_0)}\sqrt{-U(r_0)}$. As before, the mutual information can be calculated as $I(A,B;r_0)=I(A,B;\rh)-S^\text{reg}_{A \cup B}(r_0)$. The shock wave parameter can be written as a function of $r_0$ as
\be
\alpha(r_0)=2 e^{K^{||}_1(r_0)+K^{||}_2(r_0)+K^{||}_3(r_0)}
\ee
where
\bea
K^{||}_1(r_0)&=&\frac{4\pi}{\beta} \int_{\bar{r}}^{r_0} \frac{dr}{U}\,,\\
K^{||}_2(r_0)&=&\frac{2\pi}{\beta} \int_{\rh}^{\infty} \frac{dr}{U}\left(1-\frac{1}{\sqrt{1+\gamma_{||}^{-2}V W U}} \right)\,,\\
K^{||}_3(r_0)&=&\frac{4\pi}{\beta} \int_{r_0}^{\rh} \frac{dr}{U}\left(1-\frac{1}{\sqrt{1+\gamma_{||}^{-2}V W U}} \right)\,.
\eea
Note that $\alpha(\rh)=0$ again, indicating the absence of a shock wave. $\alpha$ increases as we move $r_0$ deeper into the black hole, and diverges at some critical point $r_0 = r^{||}_c$ which is implicitly given by
\be
\frac{(V W U)'}{V W U}\Big|_{r=r^{||}_c}=0,
\ee
where the prime indicates a derivative with respect to $r$.

Once more, since $S^\text{reg}_{A \cup B}(r_0)$, $I(A,B;r_0)$ and $\alpha(r_0)$ are functions of the turning point $r_0$, we can make parametric plots of $S^\text{reg}_{A \cup B}$ versus $\log \alpha$ and $I(A,B)$ versus $\log \alpha$.

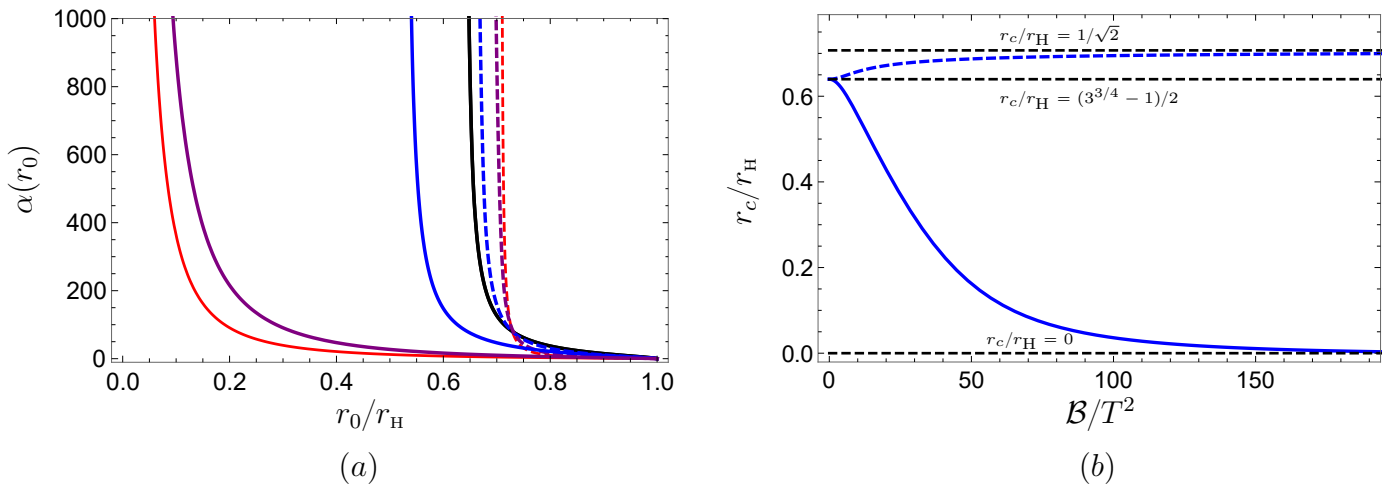
\begin{figure}[H]
\begin{center}
\begin{tabular}{cc}
\setlength{\unitlength}{1cm}
\hspace{-0.9cm}
\includegraphics[width=8.3cm]{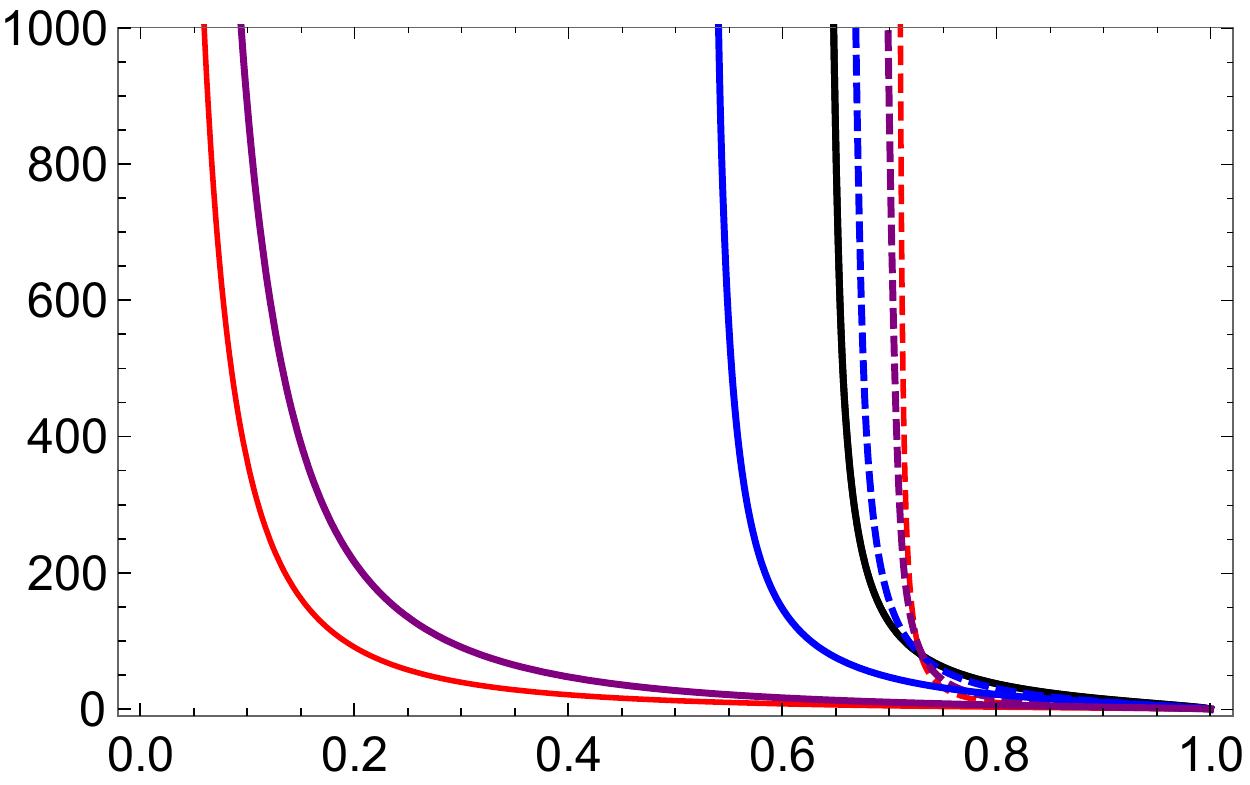} 
\qquad\qquad & 
\includegraphics[width=8cm]{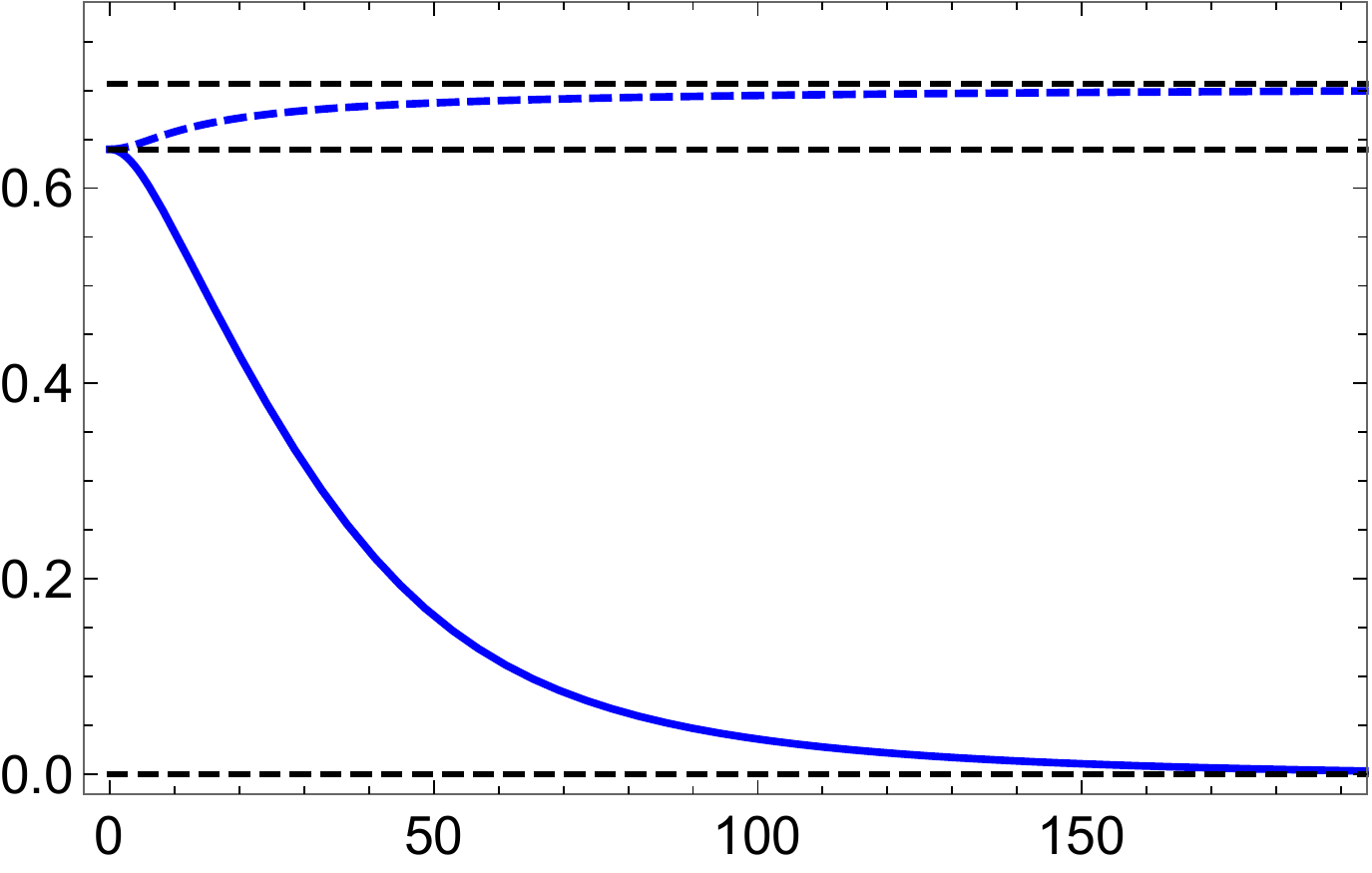}
\qquad
  \put(-520,65){\rotatebox{90}{ $\alpha(r_0)$}}
         \put(-400,-10){ $r_0/\rh$}
         \put(-248,65){\rotatebox{90}{$r_c/\rh$}}
         \put(-120,-10){$\mathcal{B}/T^2$}
         \put(-115,-30){$(b)$}
         \put(-395,-30){$(a)$}
         
         \put(-145,135){\tiny $r_{c}/\rh=1/\sqrt{2}$}
         \put(-150,20){\tiny $r_{c}/\rh=0$}
         \put(-145,111){\tiny $r_{c}/\rh=(3^{3/4}-1)/2$}
         
\end{tabular}
\end{center}
\caption{ \small (a) Shock wave parameter $\alpha$ as a function of the turning point $r_0$ normalized by $\rh$ for some values of $\mathcal{B}/T^2$. The continuous (dashed) curves represent the results for orthogonal (parallel) strips. The black curve represents the result at the UV fixed point ($\mathcal{B}/T^2=0$), while the red curves represent the results at the IR fixed point ($\mathcal{B}/T^2>>1$). The blue curves represent the results for $\mathcal{B}/T^2=12.3$, while the purple curves represent the results for $\mathcal{B}/T^2=133$. (b) Critical point $r_0=r_c$ (at which $\alpha(r_0)$ diverges) versus $\mathcal{B}/T^2$. The continuous (dashed) blue curve represents the result for an orthogonal (parallel) strip. The dashed horizontal line in the middle indicates the result at the UV fixed point, $r_c/\rh=(3^{3/4}-1)/2$, while the top and bottom ones show the results at the IR fixed point, $r_c^{||}/\rh=1/\sqrt{2}$ and $r_c^{\perp}/\rh=0$, respectively.}
\label{fig-rc-alpha}
\end{figure}

Figure \ref{fig-rc-alpha} (a) shows the shock wave parameter as a function of the turning point $r_0$ for several values of the magnetic field. Figure \ref{fig-rc-alpha} (b) shows how the critical points $r^{\perp}_c$ and $r^{||}_c$ vary as a function of $\mathcal{B}/T^2$. Both quantities start at the UV value $r_{c}/\rh=(3^{3/4}-1)/2$ when $\mathcal{B}=0$. As we increase the magnetic field both $r^{\perp}_c$ and $r^{||}_c$ approach their corresponding IR values, which are $r^{\perp}_c=0$ and $r^{||}_c/\rh=1/\sqrt{2}$. 

%%% rc versus b and alpha versus r0

%%%%%%%%%%%%%%%%%%%%%%%%%%%%%%%%%%%%
%%%%%%% SAUBreg and I(A,B) bersus log alpha 

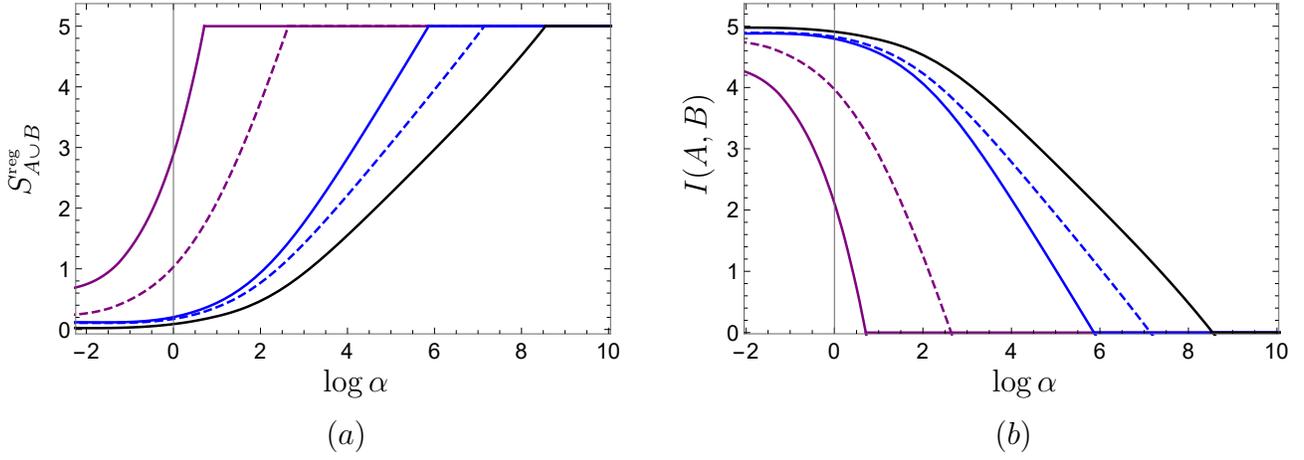
\begin{figure}[H]
\begin{center}
\begin{tabular}{cc}
\setlength{\unitlength}{1cm}
\hspace{-0.9cm}
\includegraphics[width=7.5cm]{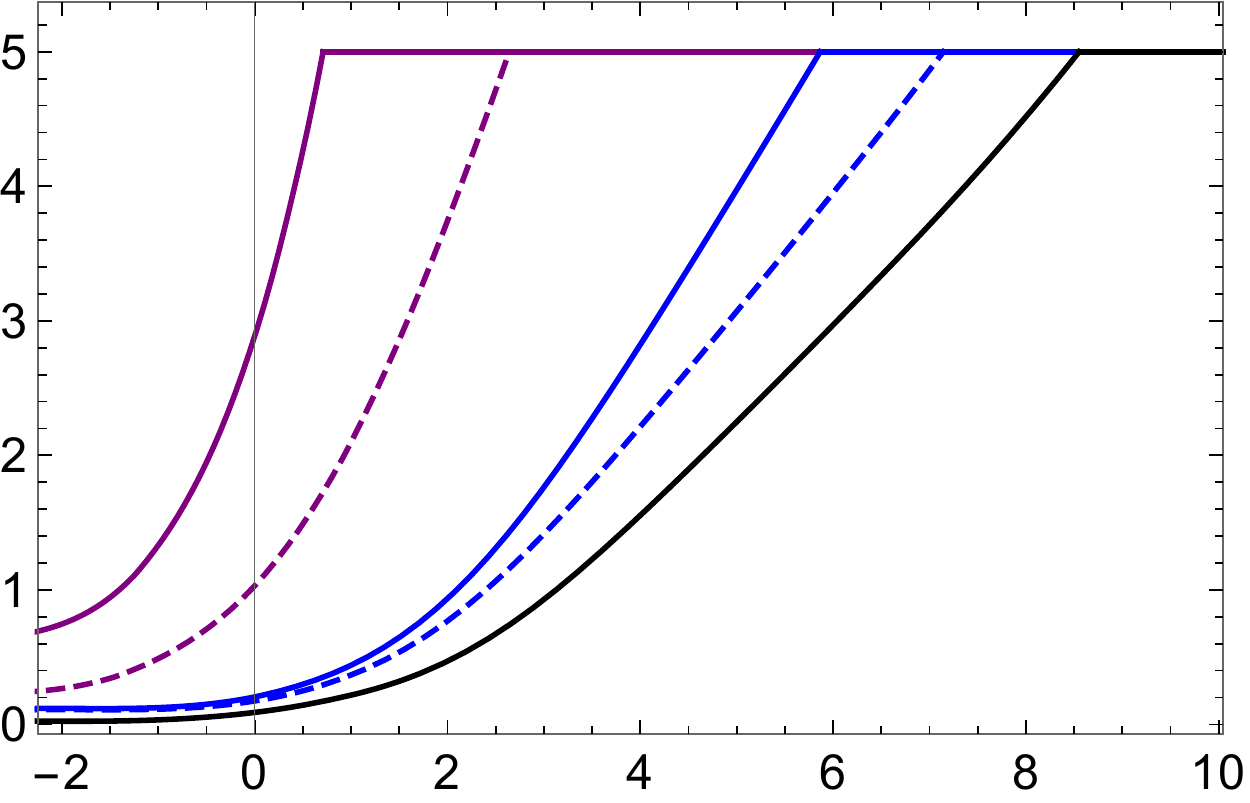} 
\qquad\qquad & 
\includegraphics[width=7.5cm]{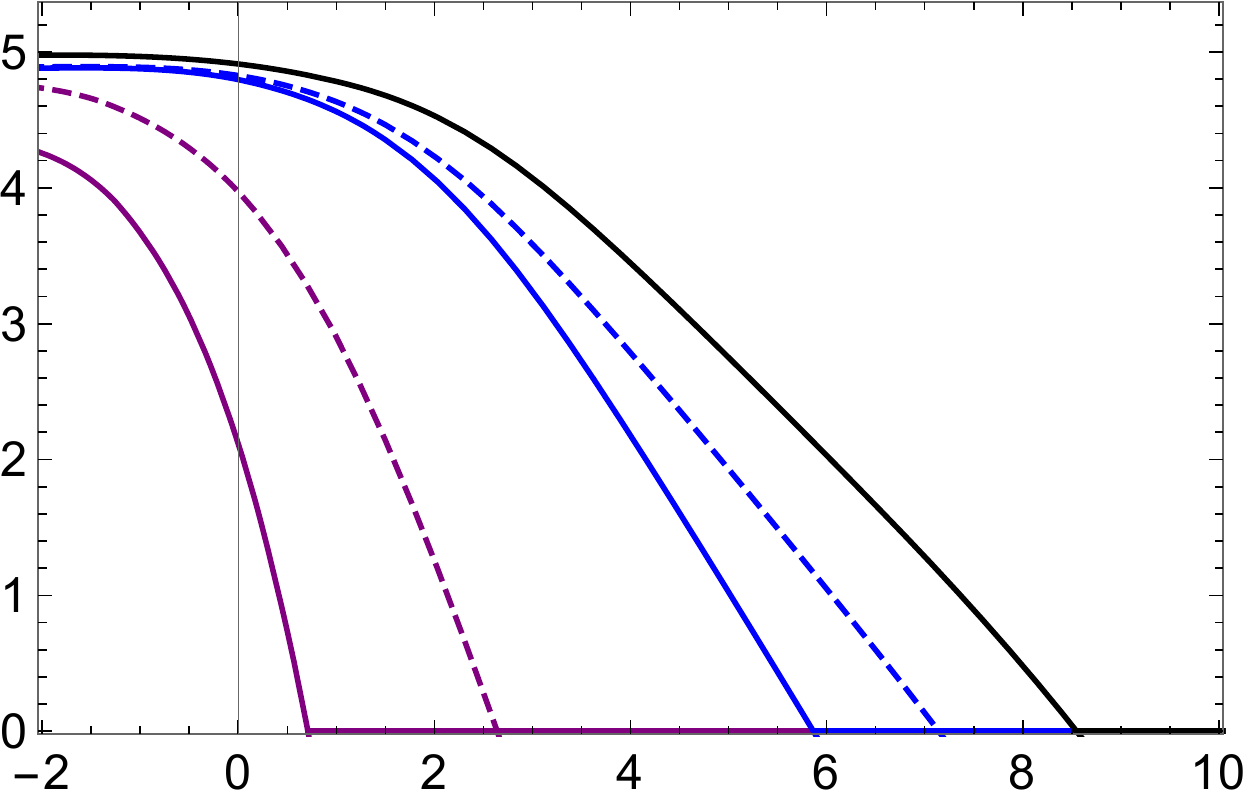}
\qquad
  \put(-485,60){\rotatebox{90}{ $S_{A \cup B}^\mt{reg}$}}
         \put(-370,-10){ $\log \alpha$}
         \put(-230,60){\rotatebox{90}{ $I(A,B)$}}
         \put(-117,-10){ $\log \alpha$}
         \put(-112,-30){$(b)$}
         \put(-365,-30){$(a)$}
         
\end{tabular}
\end{center}
\caption{ \small (a) Regularized entanglement entropy $S_{A \cup B}^\mt{reg}$ (in units of $V_2/G_\mt{N}$) as a function of $\log \alpha$. (b) Mutual information $I(A,B)$ (in units of $V_2/G_N$) as a function of $ \log \alpha$. All the curves have the same mutual information $I(A,B;0)=5$  at $\alpha =0$. In both (a) and (b) the curves correspond (from right to left) to $\mathcal{B}/T^2= 0$ (black curves), $\mathcal{B}/T^2= 12.3$ (blue curves), $\mathcal{B}/T^2=133$ (purple curves).}
\label{fig-spreading}
\end{figure}

Figure \ref{fig-spreading} (a) shows how the regularized entanglement entropy (in units of $V_2/G_\mt{N}$) grows as a function of $\log \alpha$. Figure \ref{fig-spreading} (b) shows how the mutual information (in units of $V_2/G_\mt{N}$) drops to zero as we increase $\log \alpha$. Here we take $I(A,B;0)=5$ at $\alpha =0$. Note that both $S^\text{reg}_{A \cup B}$ and $I(A,B;\alpha)$ have a sharp transition to a constant value for some value of the shock wave parameter $\alpha = \alpha_*$. This happens when the area of $\gamma_\text{wormhole}$ becomes larger than the area of $\gamma_A \cup \gamma_B$, in which case $S_{A \cup B}$ has to be computed from the latter. Given that $\gamma_A$ and $\gamma_B$ stay in the exterior region of the geometry, they are not affected by the shock wave at the horizon and hence $S_{A \cup B}$ does not depend on $\alpha$ whenever this parameter reaches or surpasses $\alpha_*$. As a consequence the mutual information becomes constant and identically zero for $\alpha \geq \alpha_*$.

%%%%%%%%%%%%%%%%%%%%%%%%%%%%%%%%%%%%%%%%

\subsection{Spreading of entanglement} \label{sec-VE}
In this section we show that the disruption of the mutual information is controlled by the so-called entanglement velocity $v_E$. This quantity plays an important role in the spreading of entanglement after a global quench \cite{HM,tsunami1,tsunami2,Mezei-2016v2}. We show that the dependence of $S^\text{reg}_{A \cup B}$ with the shock wave time $t_0$ is very similar to the time behavior of entanglement entropy after global quenches. This shows that the gravitational set up of shock waves in a two-sided black hole provides an additional example of a holographic quench protocol.

\begin{figure}[H]
\begin{center}
\setlength{\unitlength}{1cm}
\includegraphics[width=0.6\linewidth]{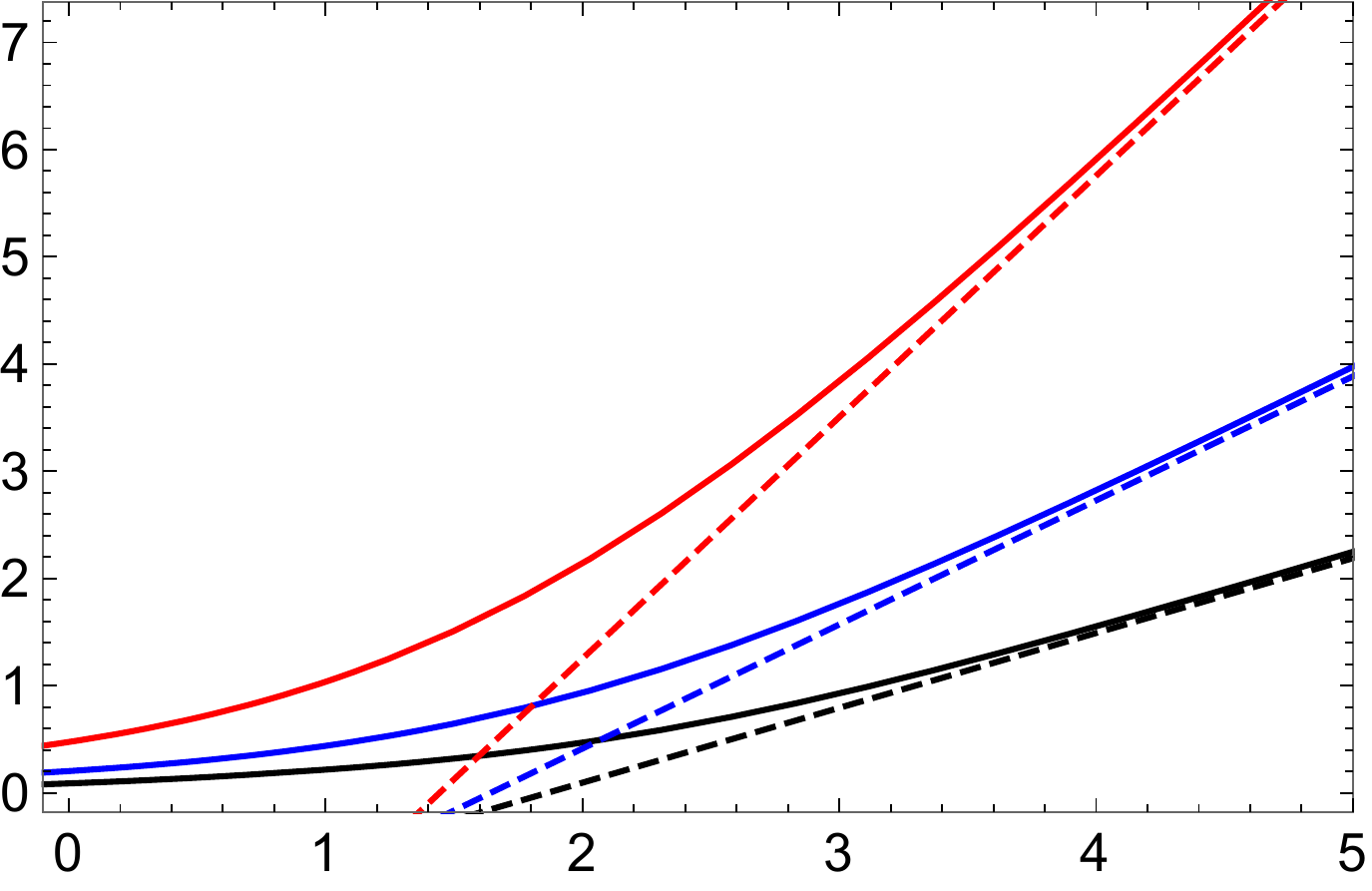} 
\put(-5.5,-.4){\large $\log \alpha$}
\put(-11.2,+3.1){\rotatebox{90}{\large $S_{A \cup B}^\mt{reg}$}}

\end{center}
\caption{ \small
Regularized entanglement entropy of orthogonal strips $S_{A \cup B}^\mt{reg}$ (in units of $V_2/G_\mt{N}$) as a function of $\log \alpha$  . The curves correspond (from bottom to top) to $\mathcal{B}/T^2= 0$ (black curves), $\mathcal{B}/T^2= 12.3$ (blue curves) and $\mathcal{B}/T^2=29.1 $ (red curves). The dashed lines have an angular coefficient given by the equation (\ref{eq-SAUBlog}). We are considering strips of very large width, such that the transition to a constant value (not shown in the figure) occurs at very large $\alpha$.}
\label{fig-SAUBlinear}
\end{figure}

Let us first consider the case of finite orthogonal strips. In the vicinity of $r_0 =r^{\perp}_c$ one can show that\footnote{The equivalent equation for parallel strips is $S^\text{reg}_{A \cup B} \approx \frac{V_2}{G_\mt{N}} \frac{\sqrt{V(r^{||}_c)W(r^{||}_c)}\sqrt{-U(r^{||}_c)}}{2 \pi T} \log \alpha\,$.}
\be
S^\text{reg}_{A \cup B} \approx \frac{V_2}{G_\mt{N}} \frac{V(r^{\perp}_c)\sqrt{-U(r^{\perp}_c)}}{2 \pi T} \log \alpha\,.
\label{eq-SAUBlog}
\ee
In principle, as this approximation requires $r_0$ to be very close to $r^{\perp}_c$ and $\alpha$ diverges at this point, one would expect the approximation to be valid only for large $\alpha$. However, the results of figure \ref{fig-SAUBlinear} actually show that (for large enough regions) the linear approximation is valid within the range $1 \lesssim \alpha \leq \alpha_*$, where $\alpha_*$ is the value of $\alpha$ where $S^\text{reg}_{A \cup B}$ has a sharp transition to a constant value.  

Since $\alpha = \text{const} \times e^{\frac{2 \pi}{\beta}t_0}$, equation (\ref{eq-SAUBlog}) implies that $S^\text{reg}_{A \cup B}$ grows linearly with the shock wave time $t_0$, and therefore
\be
\frac{d S^\text{reg}_{A \cup B}}{d t_0}= \frac{V_2}{G_\mt{N}} V(r^{\perp}_c)\sqrt{-U(r^{\perp}_c)}\,.
\ee
By using the thermal entropy density $s_\mt{th}=\sqrt{V^2(\rh)W(\rh)}/(4 G_\mt{N})$ we can eliminate $G_\mt{N}$ from the above equation that reduces to
\be
\frac{d S^\text{reg}_{A \cup B}}{d t_0}= 4\,V_2\,s_\mt{th}\, v_E^{\perp}\,,
\ee
where
\be
v_E^{\perp}=\frac{V(r^{\perp}_c)\sqrt{-U(r^{\perp}_c)}}{\sqrt{V^2(\rh)W(\rh)}},
\ee
is the entanglement velocity for orthogonal strips. Likewise we can define the entanglement velocity for parallel strips as
\be
v_E^{||}=\frac{\sqrt{V(r^{||}_c) W(r^{||}_c)}\sqrt{-U(r^{||}_c)}}{\sqrt{V^2(\rh)W(\rh)}}\,.
\ee

\subsubsection*{$v_E$ at the UV fixed point}
At the UV fixed point the system is isotropic, and so is the entanglement velocity which in any direction is given by
\be
v_E^\mt{UV} =\frac{\sqrt{2}}{3^{3/4}},
\ee
coinciding with the result for a $d$-dimensional CFT as reported, for instance, in equation 2.11 of \cite{HM}.
\subsubsection*{$v_E$ at the IR fixed point}
At the IR fixed point the entanglement velocity for orthogonal strips is given by
\be
v_{E,\perp}^\mt{IR}=1,
\ee
in agreement now with the result for a BTZ black hole. The entanglement velocity for parallel strips is dictated by the expression
\be
v_{E,||}^\mt{IR}=\frac{\pi}{3^{1/4}}\frac{T}{\sqrt{\mathcal{B}}}\,,
\ee
that vanishes in the IR limit as $\mathcal{B}/T^2 \rightarrow\infty$ at fixed temperature.

\subsubsection*{$v_E$ at intermediate values of $\mathcal{B}/T^2$.}
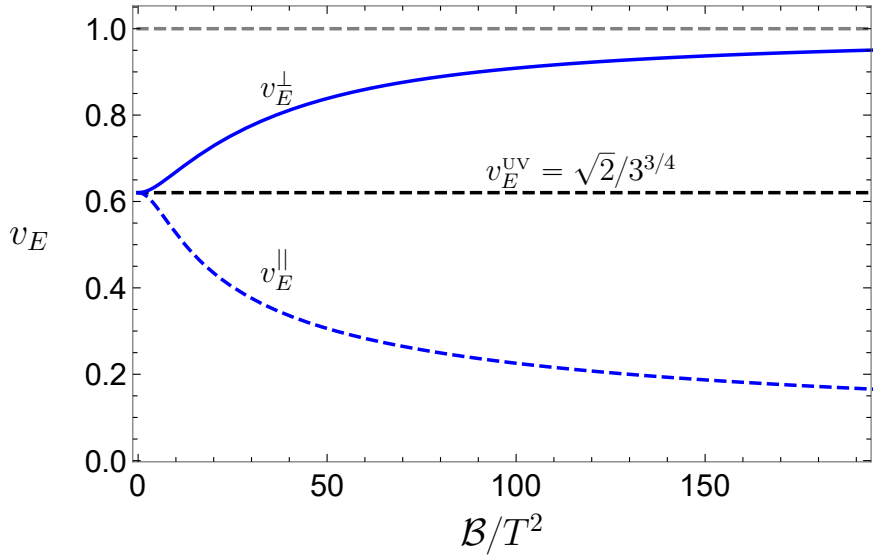
\begin{figure}[H]
\begin{center}
\setlength{\unitlength}{1cm}
\includegraphics[width=0.6\linewidth]{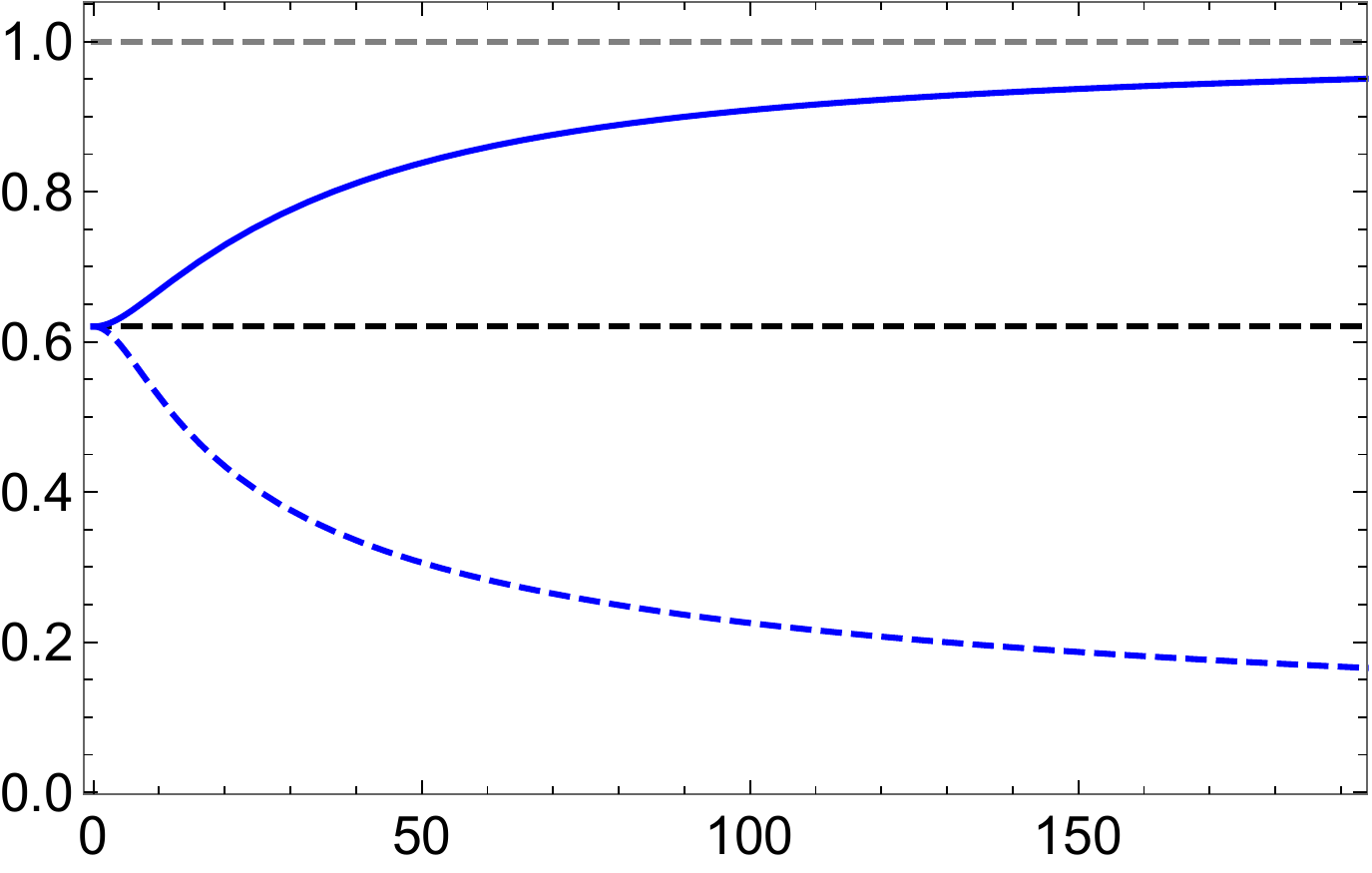} 
\put(-5.5,-.5){\large $\mathcal{B}/T^2$}
\put(-11.5,+3.5){\large $v_{E}$}
\put(-8.3,+5.45){ $v_{E}^{\perp}$}
\put(-8.3,+3.0){ $v_{E}^{||}$}
\put(-5.3,+4.35){ $v_{E}^\mt{UV}=\sqrt{2}/3^{3/4}$}
\end{center}
\caption{ \small
Entanglement velocity $v_{E}$ versus $\mathcal{B}/T^2$. The continuous (dashed) blue curve represents the entanglement velocity for an orthogonal (parallel) strip, while the bottom horizontal black line represent the entanglement velocity at the UV fixed point, which is equal to the conformal result $v_{E}^\mt{UV}=\sqrt{2}/3^{3/4}$. The top horizontal gray line is the speed of light.}
\label{fig-VE}
\end{figure}

As made clear previously, at intermediate values of $\mathcal{B}/T^2$ the metric functions can only be determined numerically, and so is the entanglement velocity. In figure \ref{fig-VE}  we plot $v_E$ for parallel and orthogonal strips as a function of $\mathcal{B}/T^2$. We show that, as we increase the value of $\mathcal{B}/T^2$, these quantities smoothly interpolate between the UV and IR values.

\section{Chaos and diffusivity} \label{sec-diffusion}

In this section we study the relationship between the chaos parameters and diffusion phenomena. For isotropic theories and in the absence of thermoelectric conductivity, the thermal and electric diffusion constants along a given direction can be calculated with the Einstein relations 
\begin{equation}
D_c=\frac{\sigma}{\chi}, \qquad D_T=\frac{\kappa}{C_\rho}.
\label{Diff}
\end{equation}
The different objects in (\ref{Diff}) are the electric susceptibility $\chi$, the specific heat $C_{\rho}$ at fixed charge density $\rho$, the electric conductivity $\sigma$ at vanishing thermal current, and the thermal conductivity $\kappa$ at vanishing electric current. The object of attention in this work is an anisotropic systems with non-vanishing thermoelectric conductivity, and for general systems of this kind, each component of the driving electric field $\vec{E}$ and temperature gradient $\vec{\zeta}\equiv(\vec{\nabla}T)/T$ can in principle induce a heat or electric current in any direction. The general coupled equations relating these driving forces and currents are given by
\begin{eqnarray}
&& \vec{J}=\bar{\sigma}\vec{E}-T\vartheta\vec{\zeta},
\cr
&& \vec{Q}=T\vartheta\vec{E}-T\bar{\kappa}\vec{\zeta},
\label{conductivity_matrix}
\end{eqnarray}
where, for a $d$ dimensional system, $\bar{\sigma}$ is the $d\times d$ electric conductivity matrix, $\bar{\kappa}$ is the corresponding thermal conductivity matrix, and $\vartheta$ is the thermoelectric mixing matrix, which presence implies that a thermal gradient can create an electric current and, conversely, an electric field can cause a thermal current. We see then, that given the anisotropic nature of our system, the expressions in (\ref{Diff}) have to become a pair of matrix equations in terms of some $d\times d$ matrices $\sigma$ and $\kappa$. Since we began defining $\sigma$ and $\kappa$ respectively as the electric conductivity at vanishing thermal current and the thermal conductivity at vanishing electric current, their matrix extension has to be done by finding the combination of electric field and temperature gradient that either cause an electric current without a thermal one, or the other way around, hence (\ref{conductivity_matrix}) dictates
\begin{equation}
\sigma=\bar{\sigma}-T\vartheta\bar{\kappa}^{-1}\vartheta, \qquad \kappa=\bar{\kappa}-T\vartheta\bar{\sigma}^{-1}\vartheta.
\end{equation}
It should be noted that the above expressions are well defined only if $\bar{\kappa}$ and $\bar{\sigma}$ are invertible. If this matrices are singular, that as we will see is our case, the analysis is more subtle and one has to rely on the physical definitions of $\sigma$ and $\kappa$ given before.

Given that $D_c$ and $D_T$ in (\ref{Diff}) bear important information about the system that they are associated with, their relationship with relevant velocities in chaotic systems, like the butterfly velocity, has been explored \cite{Blake1,Blake2}. Similarly, the components of the matrix extension of $D_c$ and $D_T$ in (\ref{Diff}) contain relevant information about the system we are studding, and we will compare some of that information with the butterfly velocity, but it is important to mention that these are not the constants that appear in the coupled diffusion equations. These latter constants constitute the diffusivity matrix described in appendix \ref{AppEigen}. Decoupling the diffusion equations is equivalent to diagonalizing the diffusivity matrix, and the relationships held by its eigenvalues with the electric, thermal, and thermoelectric conductivities, generalize the Einstein relations (\ref{Diff}) \cite{Hartnoll-2014}. Only under very particular circumstances the eigenvalues of the diffusivity matrix reduce to $D_c$ and $D_T$ in (\ref{Diff}) while in general they encode complementary information, making it interesting to also compare them with the butterfly velocity. In our case this proves unfruitful, since, as can be seen in appendix \ref{AppEigen}, the eigenvalues of the diffusivity matrix do not provide meaningful bounds for the chaotic velocities.

There are many holographic methods to compute all the elements of the conductivity matrices \cite{Donos-2015, perturbations_B, Kim-2015, Hartnoll-2007a, Hartnoll-2007b, Davison-2014, perturbations}. However, as remarked in the references just cited, in order to obtain a finite result it is necessary to either break the translation invariance of the theory or introduce a mechanism that does not conserve momentum. Our model, even if anisotropic, is translationally invariant, since, on the one hand, there are not gauge independent quantities that depend on the position, and on the other, there are no fields to which the gauge potential, which is indeed position dependent, should be minimally coupled to. Now, despite Lorentz force not conserving momentum in the directions perpendicular to the magnetic field, it does in the direction parallel to it, so inconsistencies arise when applying the cited methods directly to our system as a whole\footnote{Actually, it is possible to apply the method described in \cite{perturbations,perturbations_B} directly to our system by studying only the directions perpendicular to the magnetic field. The results obtained by this procedure agree with the ones presented here, as it should be. The details of this calculation can be found in appendix \ref{AppD}. }.

To properly carry the calculation it would be necessary to consider a modification of our theory that breaks translation invariance or does not conserve momentum in any direction, compute the conductivity matrix, and then take the appropriate limit to restore translation invariance and momentum conservation in the way our system does. Fortunately the better part of this work is done, since the conductivity matrix in a more general system, that can be reduce to ours, has already been studied in \cite{Donos-2015}, where the authors break translation invariance by adding position dependent axions\footnote{We thank Jerome Gauntlett for pointing this reference to us.}. Thus, in order to obtain the conductivity matrix for our system, we just evaluate (6.10) of \cite{Donos-2015} with the appropriate substitutions and in the right limit. The result is
\begin{eqnarray}
&& \bar{\sigma}=\begin{pmatrix}
0 & 0 & 0 \\
0 & 0 & 0 \\
0 & 0 & \frac{4V(\rh)}{\sqrt{W(\rh)}}
\end{pmatrix}
, \qquad
\vartheta =\begin{pmatrix}
0 & 4\pi\frac{V(\rh)\sqrt{W(\rh)}}{\mathcal{B}} & 0 \\
-4\pi\frac{V(\rh)\sqrt{W(\rh)}}{\mathcal{B}} & 0 & 0 \\
0 & 0 & 0
\end{pmatrix}
{\mathrm{, and}} 
\cr
&& \bar{\kappa}=\begin{pmatrix}
4\pi^{2}T\frac{V(\rh)^{2}\sqrt{W(\rh)}}{\mathcal{B}^{2}} & 0 & 0 \\
0 & 4\pi^{2}T\frac{V(\rh)^{2}\sqrt{W(\rh)}}{\mathcal{B}^{2}} & 0 \\
0 & 0 & \infty
\end{pmatrix},
\label{conduMat}
\end{eqnarray}
from which, in principle, we could compute $\sigma$ and $\kappa$, and where the divergent thermal conductivity in the $z$ direction is recovered as expected in the translational invariant and momentum conserving limit.

We would like to take a moment to list a few benchmarks that indicate the expressions in \eqref{conduMat} to be correct and consistent with previous results. It is reassuring to see that $\bar{\sigma}_{xx}=\bar{\sigma}_{yy}=0$ in \eqref{conduMat}, which is the result previously obtained in \cite{Arciniega:2013dqa,Donos-2015} by yet another two different methods, where $\bar{\sigma}_{zz}$ also coincides with \eqref{conduMat}. In \cite{perturbations_B,Kim-2015} the authors study the effects of a magnetic field on transport in 2+1 dimensional systems at finite charge density $\rho$ and broken translational invariance. We verified that, when evaluated at $\rho=0$ and in the translationally invariant limit, the results in \cite{perturbations_B,Kim-2015} for $\bar{\sigma}$, $\vartheta$ and $\bar{\kappa}$ are consistent with ours in the the directions they study, namely, $x$ and $y$. Moreover, the Hall effect is zero in our system, which is consistent with the results of \cite{Hartnoll-2007a} at zero charge density. As a final check, we note that our thermoelectric coefficients in the $x-y$ plane are consistent with the ones presented in equation (3.39) of \cite{Hartnoll-2007b} at $\omega=0$ and at zero charge density\footnote{To compare our results with the ones presented in \cite{Hartnoll-2007b}, one should notice that $\omega_c \sim \rho =0$ and that $\gamma \sim B^2$. We thank Sean Hartnoll for suggesting these comparison.}.

Before presenting the results for the actual conductivity matrices $\sigma$ and $\kappa$, there are a few remarks we would like to make about what we should expect to find.

In a system with translational invariance, a driving force would lead to an infinite current if a net free density of the charge that it acts upon is present. In our case the driving force $E$ acts on the electric charge while the temperature gradient acts on any matter. Our system is a neutral strongly coupled plasma at finite temperature subject to an external magnetic field, so there is a uniform translational invariant matter density, but a vanishing net electric charge. We would then expect an infinite thermal conductivity $\kappa$ in any direction in which momentum is conserved, but a finite electric conductivity $\sigma$. Despite $\rho=0$ we do not expect $\sigma$ to vanish entirely since, as pointed out in \cite{perturbations}, for a neutral system constituted by charged particle-hole pairs, a current is expected to appear in reaction to $E$ as particles and holes are driven to flow in opposite directions. In this scenario, momentum dissipation will occur as constituents with different charges are dragged with respect to each other. In our system there are no quasi-particles, but the former observation still applies since, even for neutral strongly coupled plasmas, there are exited degrees of freedom with opposite electric charge.

We can now proceed to compute the $\sigma$ and $\kappa$ matrices. The discussion of the previous paragraph is reflected in our results \eqref{conduMat}, but we need to pay particular attention to some of the components, since, if we turn on a thermal gradient perpendicular to the magnetic field to generate a thermal current in this direction, $\vartheta_{xy}=-\vartheta_{yx}\neq 0$ implies that an electric current will be induced in a direction that is also perpendicular to the magnetic field, that, given $\bar{\sigma}_{xx}=\bar{\sigma}_{yy}=0$, will not be possible to stop by applying an electric field. Conversely, if we turn on an electric driving force perpendicular to the magnetic field, $\vartheta_{xy}=-\vartheta_{yx}\neq 0$ implies that a thermal current is generated, while $\bar{\sigma}_{xx}=\bar{\sigma}_{yy}=0$ shows that this happens at vanishing electric current, even if the thermal driving force is zero. The conclusion is that the only combination of driving forces that will lead to a thermal current in the $x-y$ plane at vanishing electric current does not involve a thermal component, leading to an ill-defined $\kappa$ along these directions. 

The previous analysis shows that in our case $D_{T}$ is not the right quantity to compare to the butterfly velocity, because it either diverges in the direction parallel to the magnetic field or is not well defined in the directions perpendicular to it.

In contrast, $\sigma$ is indeed a well defined diagonal matrix. From (\ref{conduMat}) we see that the only current that an electric field in the $z$ direction generates is electric and parallel to it, so we have $\sigma_{zz}=\bar{\sigma}_{zz}$. To compute $\sigma_{xx}$ and $\sigma_{yy}$ it is necessary to determine the combination of driving electric field and thermal gradient that generates an electric current without a thermal one, which can be accomplished because $\vartheta_{xy}=-\vartheta_{yx}\neq 0$. The final result is  
\begin{equation}
\sigma_{xx}=\sigma_{yy}=4\sqrt{W(\rh)}, \qquad \sigma_{zz}=\frac{4V(\rh)}{\sqrt{W(\rh)}}.
\label{conductivity}
\end{equation} 
Note that the explicit dependence on the magnetic field has been eliminated, and the effect of $\mathcal{B}$ only appears indirectly through the metric functions $V(r)$ and $W(r)$, of which the only information we need is at the horizon. Also, $\sigma$ has a smooth limit for any value of $\mathcal{B}/T^{2}$, unlike $\bar{\sigma}$ which is discontinuous in the limit $\mathcal{B}/T^{2}\rightarrow 0$. In order to evaluate \eqref{conductivity} for any value of $\mathcal{B}/T^{2}$ it is necessary to extract $V(\rh)$ and $W(\rh)$ from the numerical solutions.

To compute the electric diffusivity we also need the susceptibility
\begin{equation}
\chi=\left(\frac{\partial\rho}{\partial\mu}\right)_{\mathcal{B},T},
\end{equation}
where $\rho$ is the charge density and $\mu$ is the chemical potential. Given that in our theory both $\rho$ and $\mu$ are zero, the differential, obtained by adding these quantities perturbatively, is evaluated at $\mu=0$. The details of the calculation are contained in Appendix \ref{AppD}, with the final result given by
\begin{equation}
\chi^{-1}=\int_{\rh}^{\infty}\frac{dr}{4V(r)\sqrt{W(r)}},
\label{susceptibility}
\end{equation}
that also needs to be evaluated numerically for arbitrary values of $\mathcal{B}/T^{2}$.

\begin{figure}[H]
\begin{center}
\setlength{\unitlength}{1cm}
\includegraphics[width=0.6\linewidth]{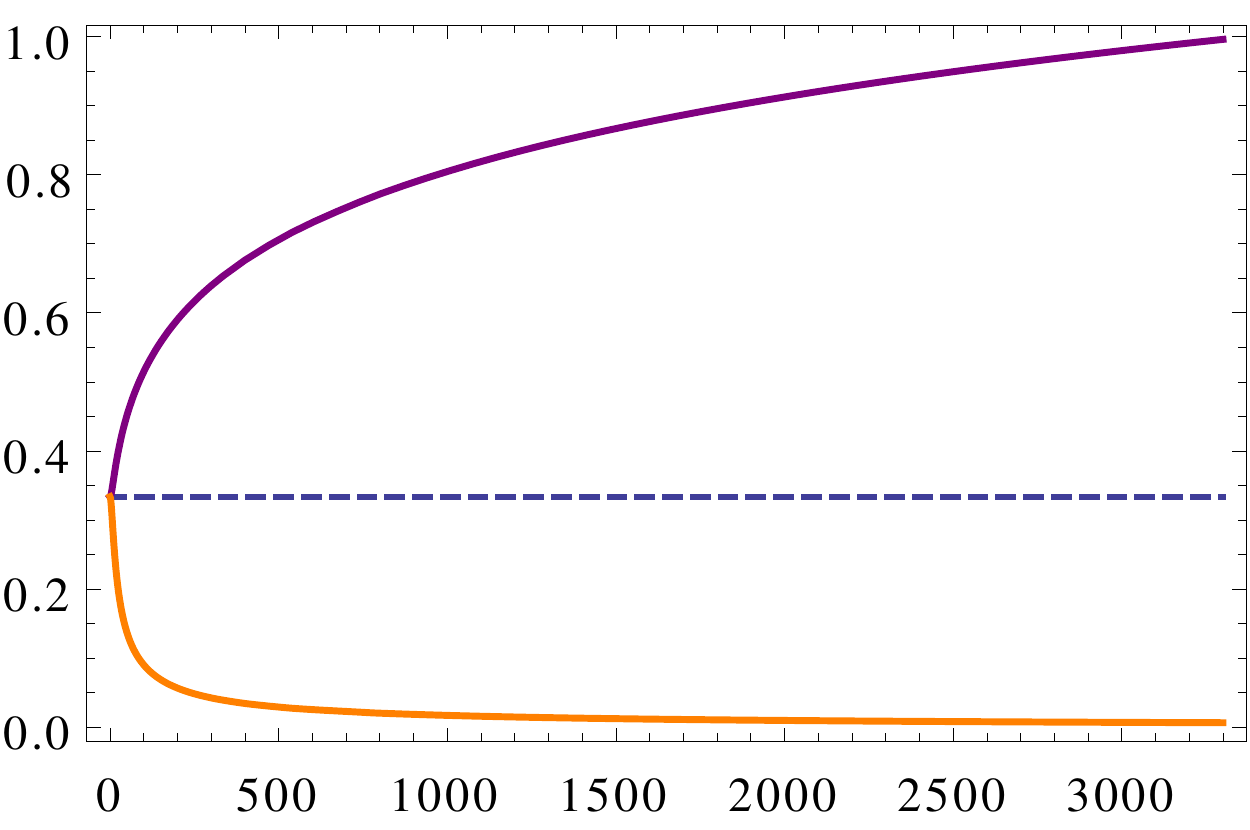} 
\put(-5.5,-.5){\large $\mathcal{B}/T^2$}
\put(-11.3,+3.8){\large $D_{c}$}
\put(-8.8,1.3){\large $D_{c}^{\perp}$}
\put(-8.8,+4.1){\large $D_{c}^{\parallel}$}
\put(-5.5,+3.0){\large $D_{c}=(2\pi T)^{-1}$}
\end{center}
\caption{ \small
Electric diffusivity $D_{c}$ versus the dimensionless parameter $\mathcal{B}/T^2$. The purple (top) curve represents the electric diffusivity along the direction of the magnetic field, while the orange (bottom) one stands for the electric diffusivity along any direction perpendicular to the magnetic field. The horizontal line is at the value of the conformal result $D_{c}=(2\pi T)^{-1}$.}
\label{Dc_BT}
\end{figure}

\begin{figure}[H]
\begin{center}
\setlength{\unitlength}{1cm}
\includegraphics[width=0.6\linewidth]{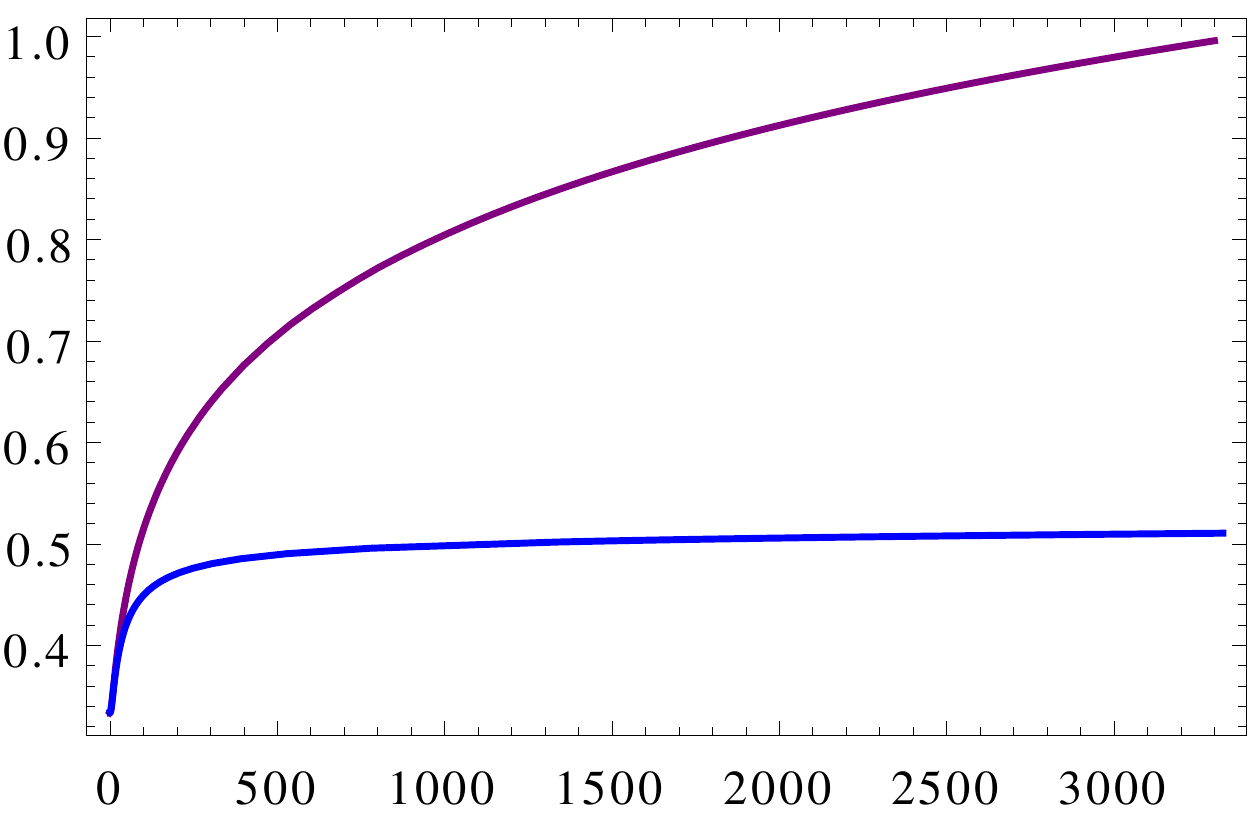} 
\put(-5.5,-.5){\large $\mathcal{B}/T^2$}
\put(-6.8,1.5){\large $\frac{3}{2}v_{B,\parallel}^{2}\tau_{L}$}
\put(-7.8,+3.3){\large $D_{c}^{\parallel}$}
\end{center}
\caption{ \small
Electric diffusivity $D^{\parallel}_{c}$ (purple top curve) and $\frac{3}{2}v_{B,\parallel}^{2}\tau_{L}$ (blue bottom curve) versus the dimensionless parameter $\mathcal{B}/T^{2}$. For any value of $\mathcal{B}/T^{2}$ the inequality $D^{\parallel}_{c}\geq\frac{3}{2}v_{B,\parallel}^{2}\tau_{L}$ holds.}
\label{Dc_vB_par}
\end{figure}

\begin{figure}[H]
\begin{center}
\setlength{\unitlength}{1cm}
\includegraphics[width=0.6\linewidth]{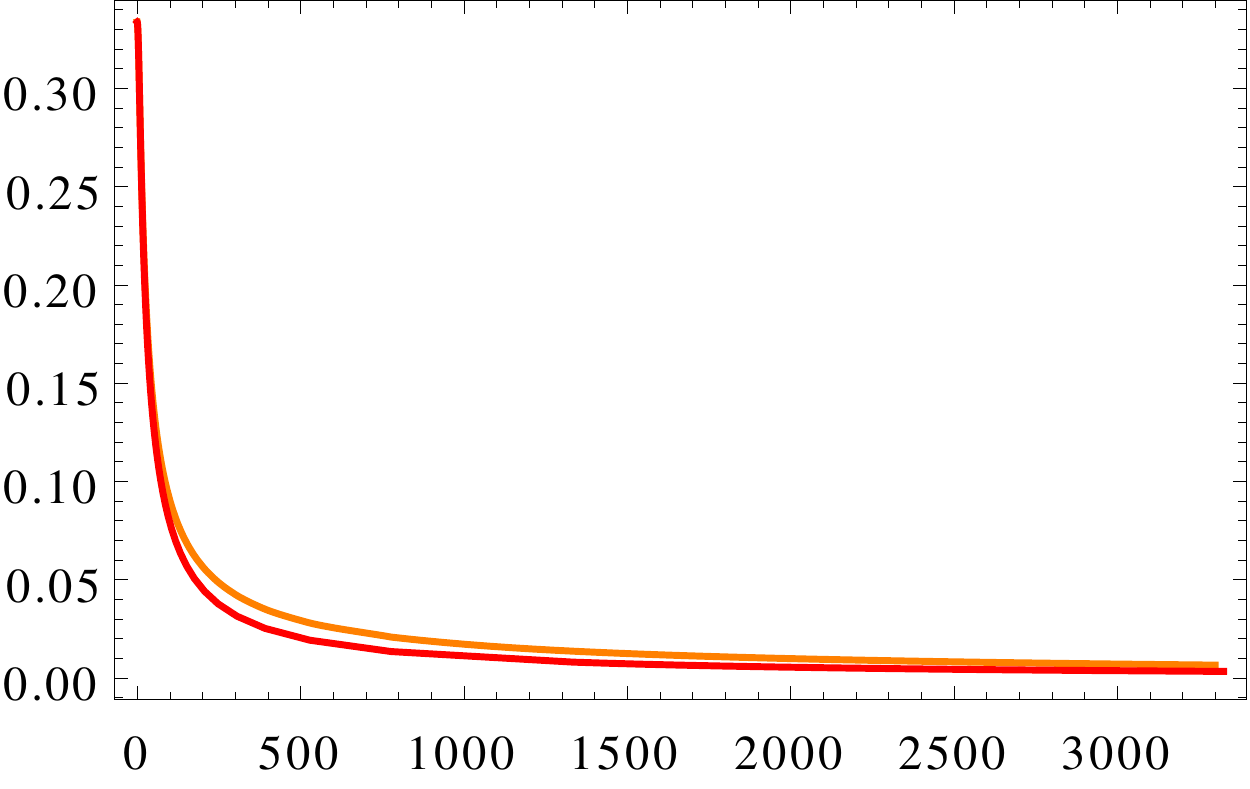} 
\put(-9.3,1.0){\large $\frac{3}{2}v_{B,\perp}^{2}\tau_{L}$}
\put(-8.3,+2.3){\large $D_{c}^{\perp}$}
\put(-5.2,-.5){\large $\mathcal{B}/T^2$}
\end{center}
\caption{ \small
Electric diffusivity $D^{\perp}_{c}$ (orange top curve) and $\frac{3}{2}v_{B,\perp}^{2}\tau_{L}$ (red bottom curve) versus the dimensionless parameter $\mathcal{B}/T^{2}$. For any value of $\mathcal{B}/T^{2}$ the inequality $D^{\perp}_{c}\geq\frac{3}{2}v_{B,\perp}^{2}\tau_{L}$ holds.}
\label{Dc_vB_perp}
\end{figure}

The electric diffusivity in any direction can be calculated by using \eqref{conductivity} and \eqref{susceptibility} in \eqref{Diff}, which gives
\begin{equation}
D^{\perp}_{c}=\sqrt{W(\rh)}\int_{\rh}^{\infty}\frac{dr}{V(r)\sqrt{W(r)}}, \qquad D^{\parallel}_{c}=\frac{V(\rh)}{\sqrt{W(\rh)}}\int_{\rh}^{\infty}\frac{dr}{V(r)\sqrt{W(r)}},
\label{Diffusivity}
\end{equation}
where we use the superscript $\perp$ to denote the diffusivity along any direction perpendicular to the magnetic field and $\parallel$ for the direction parallel to it. In figure \ref{Dc_BT} we show the diffusivity \eqref{Diffusivity} as a function of $\mathcal{B}/T^{2}$. In the limit $\mathcal{B}/T^{2}\rightarrow 0$ the electric diffusivity is the same along any direction and equal to the well known result $D_{c}=\frac{1}{2\pi T}$, thus in this limit the relation between the electric diffusivity and the chaos parameters is
\begin{equation}
D_{c}=\frac{3}{2}v_{B}^{2}\tau_{L},
\end{equation}  
which is consistent with the results from \cite{Blake1} for a $d$ dimensional CFT, $D_{c}=\frac{d}{\Delta_{\chi}}v_{B}^{2}\tau_{L}$, where $\Delta_{\chi}$ is the scaling dimension of the susceptibility. Figures \ref{Dc_vB_par} and \ref{Dc_vB_perp} show that for an arbitrary value for $\mathcal{B}/T^{2}$ the inequality 
\begin{equation}
D_{c}\geq\frac{3}{2}v_{B}^{2}\tau_{L},
\label{inequality}
\end{equation}
is indeed satisfied along any direction.

A simple calculation shows that both $D^{\perp}_{c}$ and $D^{\parallel}_{c}$ diverge when computed a the IR limit $T^{2}/\mathcal{B} \rightarrow 0$. Their ratio, however, is well defined, and one can show that $D^{\perp}_{c} / D^{\parallel}_{c} \sim T^{2}/\mathcal{B}$ as the infrared limit is approached. This is consistent with the results of figure \ref{Dc_BT} and with the fact that $\left( v_{B,\perp}/v_{B,\parallel}\right)^2 \sim T^{2}/\mathcal{B}$ also in this limit.

\section{Discussion} \label{sec-discussion}

We have used holographic methods to study chaos, diffusion and spreading of entanglement of a super Yang-Mills theory at temperature $T$ in the presence of a background magnetic field of constant strength $\mathcal{B}$. The dual geometry can be viewed as a renormalization group flow from an AdS geometry in the ultraviolet to a BTZ-like geometry in the infrared and the parameter controlling this transition is the dimensionless ratio $\mathcal{B}/T^2$, which is very small(large) close to the UV(IR) fixed point.
As explained in more detail below, all of our results can be explained on the basis of the aforementioned  RG flow and the apparent strengthening of the internal interaction of the system due to the presence of the magnetic field.

\subsubsection*{Chaotic properties of the boundary theory}
In section \ref{sec-vb1} we study localized shock waves in the background (\ref{5dmet}) and extract the chaotic properties of the boundary theory from the shock wave profile $\alpha(t,\vec{x})$. We find that the Lyapunov exponent is maximal $\lambda_L=2\pi/\beta$ and the leading order contribution to the scrambling time is controlled by the number of degrees of freedom of the system $t_* \sim \beta \log N^2$, as expected on general grounds.

The results for the butterfly velocity are shown in figure \ref{fig-VB2}. Due to the presence of the magnetic field the butterfly velocity is anisotropic in the $z$-direction, but it still displays rotational symmetry in the $xy$-plane. For simplicity, we only compute the butterfly velocity parallel to the magnetic field $v_{B,\parallel}$ and perpendicular to it $v_{B,\perp}$.

At zero magnetic field we have $v_{B,\parallel}=v_{B,\perp}=\sqrt{2/3}$, which is the value of $v_B$ at the UV fixed point, at which the system is isotropic. As we increase the intensity of the magnetic field $v_{B,\parallel}^2$ increases and approaches the speed of light, while $v_{B,\perp}^2$ decreases and is highly suppressed at large values of $\mathcal{B}/T^2$. This is consistent with the IR results $v_{B,\parallel}^\mt{IR}=1$ and $v_{B,\perp}^\mt{IR}= \frac{2 \pi}{3^{1/4}} \frac{T}{\sqrt{\mathcal{B}}} << 1$, obtained using the formulas (\ref{eq-VBz}) and (\ref{eq-VBxy}) with the metric functions given by (\ref{eq-metric-IR}).

%Figure \ref{fig-VB2} shows that $v_{B,z}^2$ increases as we increase $B/T^2$, and approaches the speed of light for large values of $B/T^2$, while the butterfly velocity along the $xy$ plane $v_{B,xy}^2$ decreases as we increase $B/T^2$ and vanishes as $B/T^2\rightarrow\infty$ at fixed temperature. This is consistent with the IR results $v_{B,z}^\mt{IR}=1$ and $v_{B,xy}^\mt{IR}= \frac{2 \pi}{3^{1/4}} \frac{T}{\sqrt{B}} << 1$.

In \cite{Mezei-2016v2} it was shown that, for $(d+1)$-dimensional isotropic black branes, the null energy condition implies an upper bound for the butterfly velocity, which is given by the conformal result $v_B^2 \leq \frac{d}{2(d-1)}$, of which the right hand sides in a 5-dimensional solution is 2/3. Figure \ref{fig-VB2} shows that as $\mathcal{B}/T^2$ increases, $v_{B,\parallel}^2$ surpasses this bound, while  $v_{B,\perp}^2$ stays below it. This does not contradicts \cite{Mezei-2016v2}, since as seen in \cite{DHoker:2016ncv,Martinez-y-Romero:2017awl}, the theory undergoes a dimensional reduction in the IR fixed point, and the limiting values in the plots are consistent with the IR theory\footnote{The IR fixed point is a CFT that lives in 1+1 dimensions. In this case, the upper bound proposed in \cite{Mezei-2016v2} reads $v_B^2 \leq 1$, which is consistent with the results obtained for $v_{B,\parallel} ^2$.}.

%It is also interesting to note that, for large values of $B/T^2$, the magnetic brane solutions that we consider is effectively described by a 3-dimensional BTZ geometry and the butterfly velocity for a BTZ black hole is equal to the speed of light. Note that this is consistent with the results obtained for $v_{B,z}^2$ and $v_{B,xy}^2$ at large values of $B/T^2$.

As proved in \cite{Qi-2017}, the butterfly velocity should be bounded by the speed of light in asymptotically AdS geometries. This is consistent with our results. In  \cite{Qi-2017} it was also proved that, for isotropic systems, the null energy condition implies that $v_B$ should decrease at the infrared. This is what happens for $v_{B,\perp}$, but our results show that $v_{B,\parallel}$ increases at the infrared. This does not contradicts \cite{Qi-2017} because of the aforementioned dimensional reduction.

Note that, although our results for $v_{B,\parallel}$ violate the upper bound proposed in \cite{Mezei-2016v2}, they remain bounded by their corresponding values at the infrared effective theory, as suggested in \cite{Jahnke-2017}. This only happens because $v_B$ is bounded by the speed of light in asymptotically AdS geometries. If the UV geometry is not asymptotically AdS we do not expect $v_B$ to be bounded by the speed of light. This indeed happens, for instance, in theories defined in non-commutative geometries \cite{VJ-2018}.

\subsubsection*{Mutual information versus strip's width}
The unperturbed two-sided black brane solution has a very particular entanglement pattern between the left and the right side of the geometry, which can be characterized by a positive mutual information between large regions in the left and right boundaries of the geometry. For simplicity, we calculate the two-sided mutual information for strip-like regions.
Figure \ref{fig-MIversusL}(a) shows how the two-sided mutual information in the unperturbed geometry varies as a function of the strip's width $\ell$. If we define the critical width $\ell_c$ as the value of $\ell$ below which the mutual information is zero, this quantity measures how large the strips should be so that the system can have two-sided correlations at $t=0$. Note that $\ell_c$ decreases with the intensity of the magnetic field and this effect is more pronounced for parallel strips than for the orthogonal ones, but in general, the magnetic field permits for smaller regions in our system to share mutual information.

To more explicitly notice the impact of the magnetic field, in figure \ref{fig-MIversusL}(b) we plot the mutual information against ${\cal{B}}/T^2$ at fixed $\ell$, and we see that $I(A,B)$ is a monotonically increasing function of ${\cal{B}}/T^2$ growing faster for parallel strips than for perpendicular ones. The different behavior for the two orientations can be understood by realizing that the increment on $I(A,B)$ has two contributions. On the one hand, the presence of the magnetic field, could have a direct impact on the mutual information between two regions due to a physical process, but on the other hand, increasing ${\cal{B}}/T^2$ makes it so that a separation in the $x$ or $y$ directions in the UV fixed point corresponds to a larger distance for energy scales closer to the IR theory. For parallel strips the width $\ell$ lies on the $x-y$ plane, so, on top of any physical impact of the magnetic field, they are subject to the geometric effect just described. For orthogonal strips $\ell$ lies along the $z$ direction while their extension in the $x-y$ planes is infinite, so their geometry is not modified by ${\cal{B}}/T^2$, leaving them only exposed to the physical impact that the magnetic field could have on their mutual information. From the fact that even for orthogonal strips the mutual information increases with ${\cal{B}}/T^2$, we infer that the magnetic field indeed contributes for the correlation between regions to become stronger by increasing the left-right entanglement of the thermofield double state at $t=0$.

\subsubsection*{Disruption of the two-sided mutual information}

By considering homogeneous shock waves, for which $\alpha = \text{constant} \times e^{\frac{2\pi}{\beta}t_0}$, we study how the two-sided mutual information drops to zero when the system is (homogeneously) perturbed far in the past.

In this case it turned out to be convenient to write the shock wave parameter in terms of a point inside the horizon $r_0$, which also characterizes the area of the extremal surfaces relevant for the computation of $I(A,B)$. Figure \ref{fig-rc-alpha} (a) shows the shock wave parameter as a function of the turning point $r_0$. Note that $r_0=\rh$ gives $\alpha(\rh)=0$, which corresponds to the absence of a shock wave. Moreover, $\alpha$ increases as we move $r_0$ deeper into the black hole, and diverges at some critical point $r_0 = r_c$.

Figure \ref{fig-rc-alpha} (b) shows the critical point $r_0=r_c$ versus $\mathcal{B}/T^2$. When $\mathcal{B}/T^2=0$ both $r_c^{\perp}$ and $r_c^{||}$ have the UV value $r_c/\rh=(3^{3/4}-1)/2$. As we increase the value of $\mathcal{B}/T^2$ both quantities approach their corresponding IR values, which are given by  $r^{\perp}_c=0$ and $r^{||}_c/\rh=1/\sqrt{2}$.

Note that, for orthogonal strips, we can probe a larger region inside the black hole as we increase $\mathcal{B}/T^2$. Indeed, for high values of $\mathcal{B}/T^2$ we can probe a region arbitrarily close to the singularity at $r = 0$. The opposite happens for parallel strips. In this case, as we increase the value of $\mathcal{B}/T^2$, the value of $r_c^{||}$ increases, becoming closer to the horizon. This means that the extremal surface probes a smaller region inside the horizon, as compared to the $\mathcal{B}/T^2=0$ case.

Figure \ref{fig-spreading} (a) shows $S^\text{reg}_{A \cup B}$ versus $\log \alpha$ for orthogonal and parallel strips and for several values of $\mathcal{B}/T^2$. The physical interpretation of these results will be done together with those of the mutual information in section \ref{sec-conclusions}, but for the moment we just notice that the regularized entanglement entropy grows faster as we increase the magnetic field and, for fixed $\alpha$, the result for a orthogonal strip is larger than the corresponding result for a parallel strip.  At some value of the shock wave parameter $\alpha=\alpha_*$, this quantity has a sharp transition to a constant value.  This happens when the area of $\gamma_\text{wormhole}$ becomes larger than the area of $\gamma_A \cup \gamma_B$, in which case $S^\text{reg}_{A \cup B}$ has to be computed from the area of $\gamma_A \cup \gamma_B$. Since $\gamma_A$ and $\gamma_B$ stay in the exterior region of the geometry, they are not affected by the shock wave at the horizon and hence $S^\text{reg}_{A \cup B}$ does not depend on $\alpha$ whenever $\alpha \geq \alpha_*$.
The saturation value of $S^\text{reg}_{A \cup B}$ depends on the width $\ell$ of the strips defining the regions $A$ and $B$. We choose $\ell$ such that, at $\alpha=0$ we have $I(A,B;0)=5$ (in units of $V_2/G_\mt{N}$). Note that, for a fixed temperature (or fixed $\rh$), the mutual information in the unperturbed geometry only depends on $\ell$. As we have fixed $\rh=1$ in our calculations, the mutual information  $I(A,B;0)$ only depends on $\ell$.

Figure \ref{fig-spreading} (b) shows how the mutual information $I(A,B;\alpha)$ drops to zero as we increase the value of $\log \alpha$, what is equivalent to move the perturbation that created the shock wave further into the past. Given that $I(A,B;\alpha)=I(A,B;0)-S^\text{reg}_{A \cup B}$, the information of this figure is basically the same as the information of figure \ref{fig-spreading} (a). Notice that the mutual information drops to zero faster as we increase the ratio $\mathcal{B}/T^2$, and the mutual information for orthogonal strips drops to zero faster than the corresponding results for a parallel strip. So, the magnetic field increases the two-sided correlations in the unperturbed system, but it makes them drop to zero faster when the system is perturbed. This behavior was also observed in another anisotropic systems \cite{Jahnke-2017,VJ-2018}.

\subsubsection*{Spreading of entanglement}
Figure \ref{fig-SAUBlinear} shows that the linear approximation given by equation (\ref{fig-SAUBlinear}) is indeed correct whenever $\alpha  \gtrsim \mathcal{O}(1)$. The linear behavior persists up the saturation (not shown in the figure) of $S^\text{reg}_{A \cup B}$ to a constant value. As explained in section \ref{sec-VE}, the linear behavior is controlled by the entanglement velocities associated to the orthogonal and parallel strips. Note that the magnetic field delays the start of the linear behavior of $S^\text{reg}_{A \cup B}$ with $\log \alpha$.

As pointed out in \cite{Jahnke-2017}, the behavior of $S^\text{reg}_{A \cup B}$ with the shock wave time $t_0$ is very similar to the time behavior of entanglement entropy of subregions in the context of global quenches \cite{HM,tsunami1,tsunami2,Mezei-2016v2}. This indicates that the gravitational setup used in this paper provides an additional example of a quench protocol. Note that the quench effectively starts after a scrambling time $\alpha  \gtrsim 1$, so maybe this setup can be thought of as a holographic model for a slow quench.

Figure \ref{fig-VE} shows how the entanglement velocities $v_E^{||}$ and $v_{E}^{\perp}$ vary as a function of $\mathcal{B}/T^2$. When $\mathcal{B}/T^2=0$, both velocities are equal to the UV result $v_E^\mt{UV} =\frac{\sqrt{2}}{3^{3/4}}$. The entanglement velocity for orthogonal strips $v_{E}^{\perp}$ increases as we increase $\mathcal{B}/T^2$, and approaches the speed of light for large values of $\mathcal{B}/T^2$, while the entanglement velocity for parallel strips $v_E^{||}$ decreases as we increase $\mathcal{B}/T^2$ and it is highly suppressed for large values of $\mathcal{B}/T^2$. This is consistent with the IR results $v_{E,\perp}^\mt{IR}=1$ and $v_{E,||}^\mt{IR}=\frac{\pi}{3^{1/4}}\frac{T}{\sqrt{\mathcal{B}}} << 1$.

As well as the butterfly velocity, the entanglement velocity of isotropic systems was also shown to be bounded by its corresponding value for a Schwarzschild black hole $v_E \leq v_E^\mt{Sch}$ \cite{Mezei-2016v2}. For a 5-dimensional black brane, this upper bound is equal to $\sqrt{2}/3^{3/4}$. Note that the entanglement velocity for perpendicular strips $v_{E,\perp}$ violate this bound, but remain bounded by the speed of light. This does not contradicts \cite{Mezei-2016v2} because, at the IR, the system flows to a CFT that lives in 1+1 dimensions and, in this case, the upper bound is given by $v_E \leq 1$.

Unfortunately, our numerical solution for the metric functions inside the horizon does not have  enough precision to calculate $v_E$ for larger values of the ratio $\mathcal{B}/T^2$ (we consider $\mathcal{B}/T^2$ up to 200). However, our numerical results strongly suggests that $v_E^{\perp}$ approaches the speed of light for very large $\mathcal{B}/T^2$.

%{\bf NOTE:} The entanglement velocity is determined from a region inside the horizon and it interpolates between its corresponding values at IR and UV fixed points. Again, it would be interesting to understand the geometrical realization of the RG flow from the point of view of the region inside the black hole.

\subsubsection*{Chaos and diffusivity}

Previous work \cite{Lucas-2016,Davison-2016,Baggioli-2016,Kim-2017,Blake3,Ahn-2017} considered theories with a non-zero charge density. This couples the charge and momentum transport, which means that the thermoelectric conductivity $\vartheta$ in \eqref{conductivity_matrix} is non-zero. However, because they also break translational symmetry by adding position dependent axion fields, the thermal conductivity $\kappa$ is finite.

In our case, the magnetic field $\mathcal{B}$ also couples the charge and momentum transport, as is reflected in a non-zero $\vartheta$ matrix. However, as previously mentioned, since translational symmetry is not broken, a well define $\kappa$ matrix is not expected. This is also intuitive, since translational invariance along a particular direction implies that shifting the vacuum of the theory along that direction comes at no cost, so any driving thermal force could have an infinite effect.

Figures \ref{Dc_vB_par} and \ref{Dc_vB_perp} shows that the lower bound \eqref{inequality} for the electric diffusivity in terms of the chaos parameters proposed by Blake \cite{Blake1,Blake2} is valid in our case. Along the direction of the magnetic field the bound is saturated only for $\mathcal{B}/T^{2}=0$. As explained by Blake \cite{Blake1,Blake2}, for $\mathcal{B}/T^{2}>0$ the integral \eqref{susceptibility} is dominated by the UV region of the geometry, whereas the chaos parameters are determinate by the IR data. On the other hand, for the directions perpendicular to the magnetic field the bound is saturated for $\mathcal{B}/T^{2}=0$ and $\mathcal{B}/T^{2}\rightarrow\infty$, where both the butterfly velocity and the diffusivity tend to zero. 

Finally, note that our numerical results for  $D^{\perp}_{c}$ and $D^{\parallel}_{c}$ are consistent with their corresponding results at the IR fixed point, in the sense that $D^{\perp}_{c} / D^{\parallel}_{c} \sim \frac{T^2}{\mathcal{B}} \rightarrow 0$ in this case. This is also consistent with the fact that  $\left( v_{B,\perp}/v_{B,\parallel}\right)^2 \sim \frac{T^2}{\mathcal{B}} \rightarrow 0$ in the IR limit.

\section*{Conclusions and future directions}\label{sec-conclusions}

One of the things that we care the most to comment is that the results of the different quantities that we have computed seam to indicate that, loosely speaking, the magnetic field makes our system more rigid, in the sense that it increases the mutual information between regions but also makes the impact of a perturbation to propagate faster by disrupting the entanglement across it. We are uncertain of the mechanism behind this observation, and consider that further investigation is necessary to clarify it. One possibility is that, as shown in \cite{Ayala:2016bbi,Ayala:2015bgv,Ayala:2018wux} and references therein, the internal interaction of the system gets intensified for strengths of the magnetic fields above the square of the temperature of the system, which is certainly the regime that we explore in detail. The resolution we use for magnetic fields smaller than the square of the temperature was not thought to test the effect of inverse magnetic catalysis on the chaotic properties of our system, which is a directions worth exploring in future work.

Our results for the butterfly and entanglement velocity also strongly suggests that both quantities are very useful tools for diagnosing RG flows. It would be interesting to investigate the behavior of these quantities under other examples of RG flows.

Another interesting extension of this work would be to consider shock waves in two-sided black holes as an holographic quench protocol and investigate further the connections between chaos and spreading of entanglement, following the ideas of \cite{Mezei-2016}.

\subsection*{Acknowledgements}
It is a pleasure to thank Keun-Young Kim, Hyun-Sik Jeong, Yongjun Ahn, and Leopoldo Pando-Zayas for helpful discussions and insightful comments on the draft. We also thank Sean Hartnoll and Jerome Gauntlett for useful correspondence.

VJ was supported by Mexico's National Council of Science and Technology (CONACyT) grant CB-2014/238734. DA and LP were partially supported by PAPIIT grant IN113618.

\appendix
\section{Interior extension of the background}\label{InteriorNumeric}

\begin{figure}
\begin{center}
\begin{tabular}{cc}
\setlength{\unitlength}{1cm}
\hspace{-0.9cm}
\includegraphics[width=7cm]{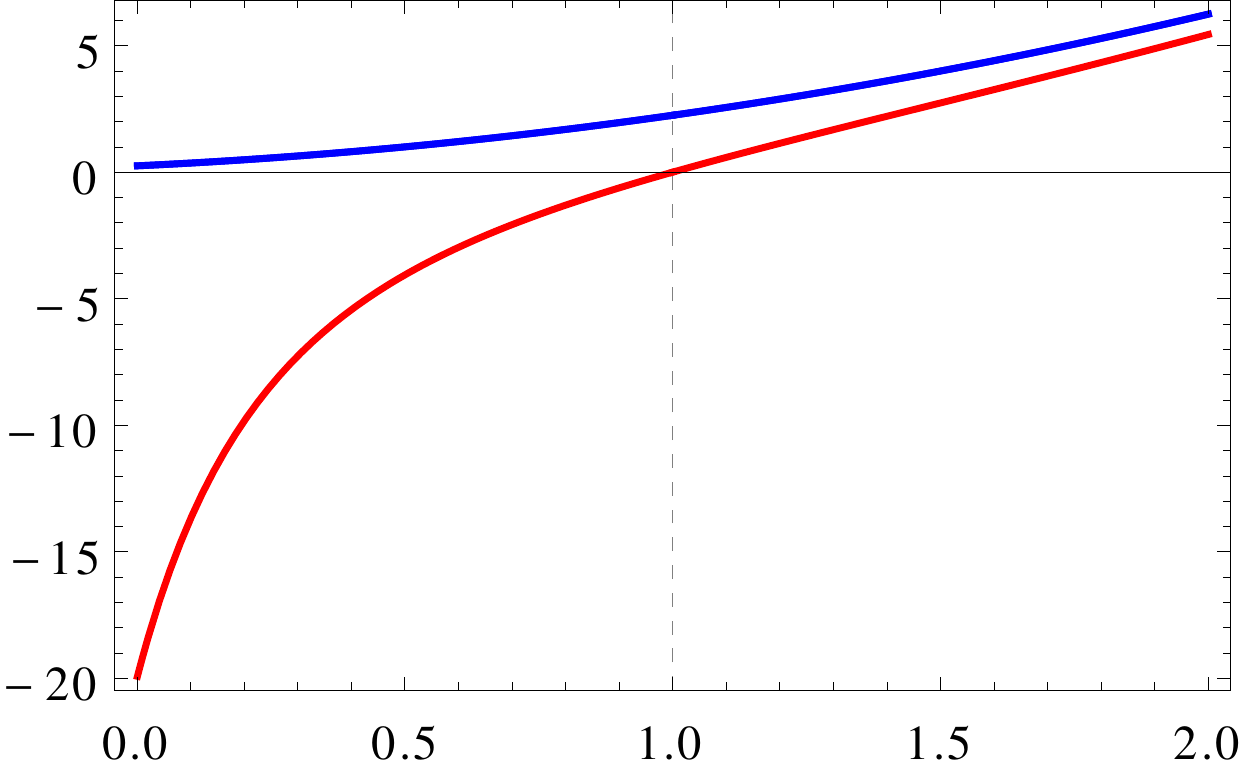} 
\qquad\qquad & 
\includegraphics[width=7cm]{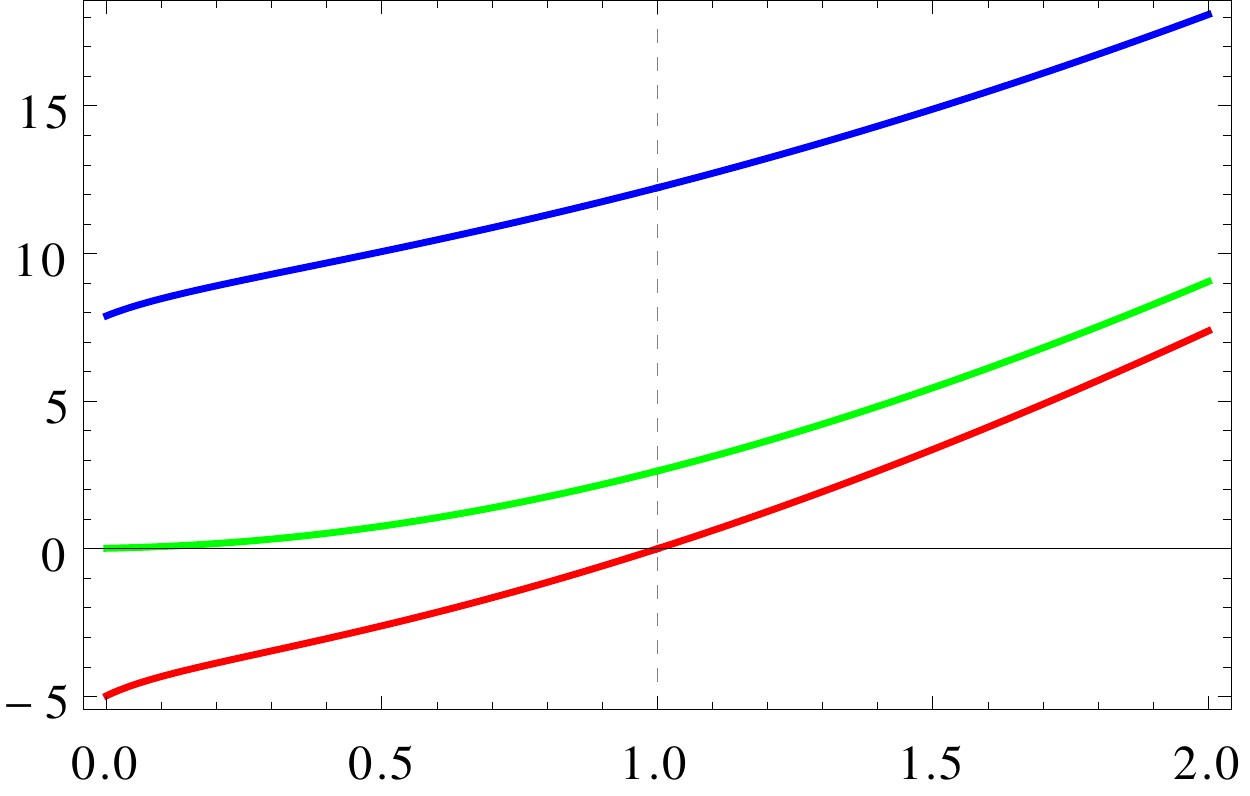}
\qquad
   \put(-250,-10){$\frac{r}{\rh}$}
   \put(-400,105){V,W}
   \put(-400,70){U}
   \put(-18,-10){$\frac{r}{\rh}$}
   \put(-50,40){U}
   \put(-130,50){W}
   \put(-130,92){V}
 \\
 (a) & (b) 
\end{tabular}
\end{center}
\caption{\small 
(a) Metric functions $U(r)$ (red) and $V(r)=W(r)$ (blue) for $\mathcal{B}/T^{2}=0$. (b) Metric functions $U(r)$ (red), $V(r)$ (blue) and $W(r)$ (green) for $\mathcal{B}/T^{2}\approx 78.006$. In both cases the dashed vertical line corresponds to the horizon.}
\label{MetricInt}
\end{figure}
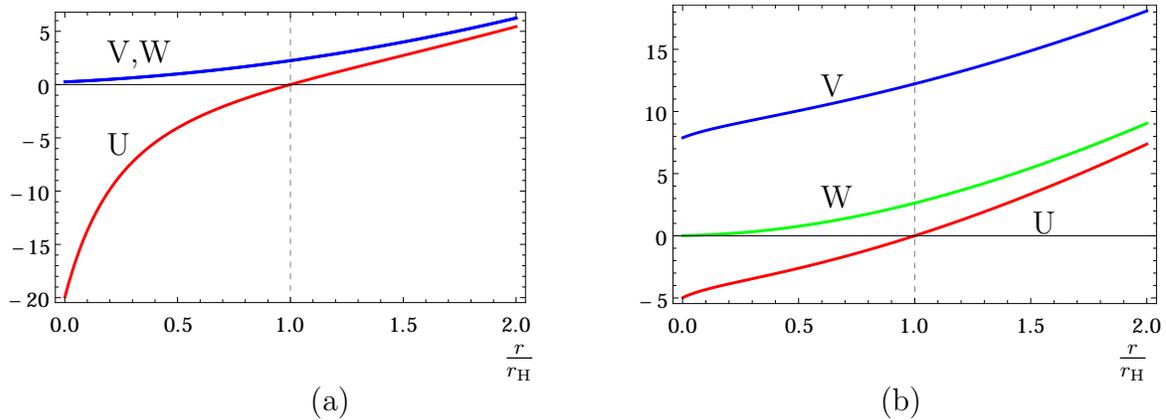

Since the equations of motion for the background are degenerated at $\rh$, in the past \cite{Arean:2016het} we have used a small distance from the horizon as an expansion parameter $\epsilon$ to perturbatively solve the equations of motion near $\rh$. The perturbative solution is then evaluated at $r_p=\rh+\epsilon$ to provide boundary conditions for the numeric integration that is to be done towards the boundary. To extend this solution to the interior of the horizon in such a way that it connects smoothly across it, all that needs to be done is evaluate the same perturbative solution at $r_p=\rh-\epsilon$, providing then boundary conditions for the numeric integration that now will be performed towards the singularity at $r=0$. Using the same perturbative solution, evaluated in the corresponding side of the horizon, to generate the interior and exterior boundary conditions guaranties that the metric functions are smooth across the horizon, see figure \ref{MetricInt}, just as long as the same value for $\mathcal{B}$ is used in both cases.

There is a subtlety that is relevant to mention here. The generic solution obtained in the way just described has an asymptotic behavior for large $r$ given by $U_{r\rightarrow\infty}\rightarrow r^2$, $V_{r\rightarrow\infty}\rightarrow v_\infty r^2$ and $W_{r\rightarrow\infty}\rightarrow w_\infty r^2$, where $v_\infty$ and $w_\infty$ are constants that differ from the unit and therefore some scaling has to be done to attain a geometry that approaches $AdS_5$. As can be seen in \cite{Arean:2016het}, the scaling allowed by the equations of motion is given by $\tilde{V}(r)=V(r)/v_\infty, \tilde{\mathcal{B}}=\mathcal{B}/v_\infty$ and $\tilde{W}(r)=W(r)/w_\infty$, so the background that actually approaches $AdS_5$ has an intensity for the magnetic field given by $\tilde{\mathcal{B}}$. Once $v_\infty$ and $w_\infty$ have been numerically obtained from the exterior solution, the interior solution has to be scale accordingly and we need to keep in mind that the intensity of the magnetic field is given by $\tilde{\mathcal{B}}$. To keep the notation simple, we will refer to this normalized quantity simply as $\mathcal{B}$, since it is the actual intensity of the field in the gauge theory.

As mentioned in the body of this work, the interior geometry transits from BTZ$\times R^2$ close to the horizon, to a black brane solution as we get closer to the singularity. This is the case because for $0\leq \mathcal{B}/T^2<\infty$, the black brane metric is an attractor as either $r\rightarrow\infty$ or $r\rightarrow 0$ in the equations of motion coming from (\ref{EMact}), so any solution will approach this geometry in both limiting regions\footnote{This can also be confirmed by directly solving the equations of motions close to the boundary and close to the singularity.}.

We will now studying the behavior of the metric functions to show that the transition, from the BTZ$\times R^2$ near horizon geometry to the black brane solution, occurs at a radial position that gets closer to the singularity as $\mathcal{B}/T^{2}$ increases.

The BTZ$\times R^2$ geometry is given by
\begin{equation} \label{btz}
U_{BTZ}(r)=3(r^2-r_h^2),\;\;\; V_{BTZ}(r)=\frac{\mathcal{B}}{\sqrt{3}}\;\;\; \mathrm{and}\;\;\; W_{BTZ}(r)=3r^2,
\end{equation}
while the metric functions for the black brane solutions are given by
\begin{align}
\label{bb}
U_{BB}(r) &=(r+\frac{r_h}{2})^2(1-\frac{(\frac{3}{2}r_h)^4}{(r+\frac{r_h}{2})^4}),\nonumber\\
V_{BB}(r) &=\frac{4V_0}{9r_h^2}(r+\frac{r_h}{2})^2,\\
W_{BB}(r) &=\frac{4}{3}(r+\frac{r_h}{2})^2. \nonumber
\end{align}

The coordinate $r$ in (\ref{bb}) has been shifted so that its Hawking temperature is the same as that of (\ref{btz}).

To study the near horizon behavior we can write (\ref{btz}) exactly as
\bea
U_{BTZ}(r)&=&6 r_h (r-r_h)+3(r-r_h)^2, \nonumber \\
V_{BTZ}(r)&=&\frac{ \mathcal{B} }{\sqrt{3}}, \label{BTZhor}\\
W_{BTZ}(r)&=&3 {r_h}^2+6 r_h (r-r_h)+3(r-r_h)^2, \nonumber
\eea
and approximate (\ref{bb}) by
\bea
U_{BB}(r)&=&6 r_h (r-r_h)-2 (r-r_h)^2+\frac{8 }{3 r_h}(r-r_h)^3+{\cal O}(4) \nonumber \\
V_{BB}(r)&=&V_0+\frac{4 V_0 }{3 r_h}(r-r_h)+\frac{4 V_0 }{9 r_h^2}(r-r_h)^2, \label{BBP}\\
W_{BB}(r)&=&3 r_h^2+4 r_h (r-r_h)+\frac{4}{3} (r-r_h)^2, \nonumber
\eea
where the only series that needs higher order corrections is that of $U_{BB}$. The fact that the two solutions share the same temperature is now apparent.

\begin{figure}
\begin{center}
\includegraphics[width=.9\linewidth]{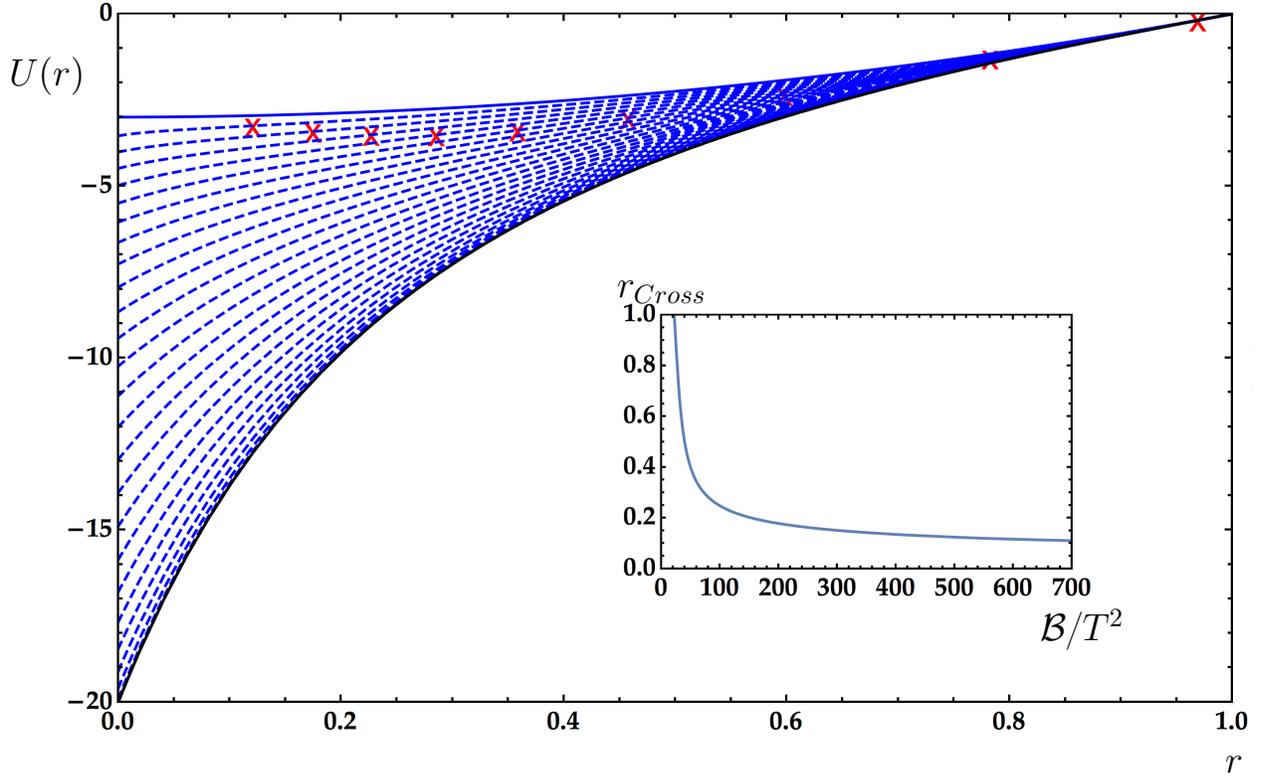} 
\put(-470,280){\large $U(r)$}
\put(-10,20){\large $r$}
\put(-240,200){\large $r_{Cross}$}
\put(-80,70){\large $\mathcal{B}/T^{2}$}
\end{center}
\caption{\small The numerical solutions for the metric function $U$ inside the horizon for values of $\mathcal{B}/T^2\approx$0, 0.68, 1.4, 2.1, 2.8, 3.6, 4.5, 5.4, 6.4, 7.5, 8.7, 10, 12, 14, 16, 19, 23, 28, 34, 43, 57, 78, 118, 207, 534, are shown in dashed lines from bottom to top. For each solution $U$ we indicated the place where its second derivative, with respect to $r$, changes sign, and in the inset we depict the radius $r_{Cross}$ where this change takes place as a function of $\mathcal{B}/T^2$. The bottom solid line is the black brane solutions and the top solid line is the BTZ$\times R^2$.}
\label{Uint}
\end{figure}

Lets start by analyzing $U$, about which, from the last two sets of equations, we notice that the leading term for $U_{BTZ}$ and $U_{BB}$ is the same when expanded around the horizon. Nonetheless, the second term is not only different, but is contrary in sign. The second derivative of $U_{BTZ}$ is positive at all points between the singularity and the horizon, so an indicator of a radial position at which the geometry has already departed from BTZ$\times R^2$ is the place, that we will call $r_{Cross}$, at which the second derivative becomes negative. In figure \ref{Uint} we have plotted our numerical solutions for a a number of values of $\mathcal{B}/T^2$ ranging from 0 to 500, along with (\ref{btz}) and (\ref{bb}). We have marked the points in which the second derivative changes signs for each solution, showing that $r_{Cross}$ is indeed smaller for plots with larger $\mathcal{B}/T^2$. In the inset we have plotted $r_{Cross}$ as a function of $\mathcal{B}/T^2$ to make explicit the decreasing nature of this radius with respect to this parameter.

\begin{figure}
\begin{center}
\includegraphics[width=.9\linewidth]{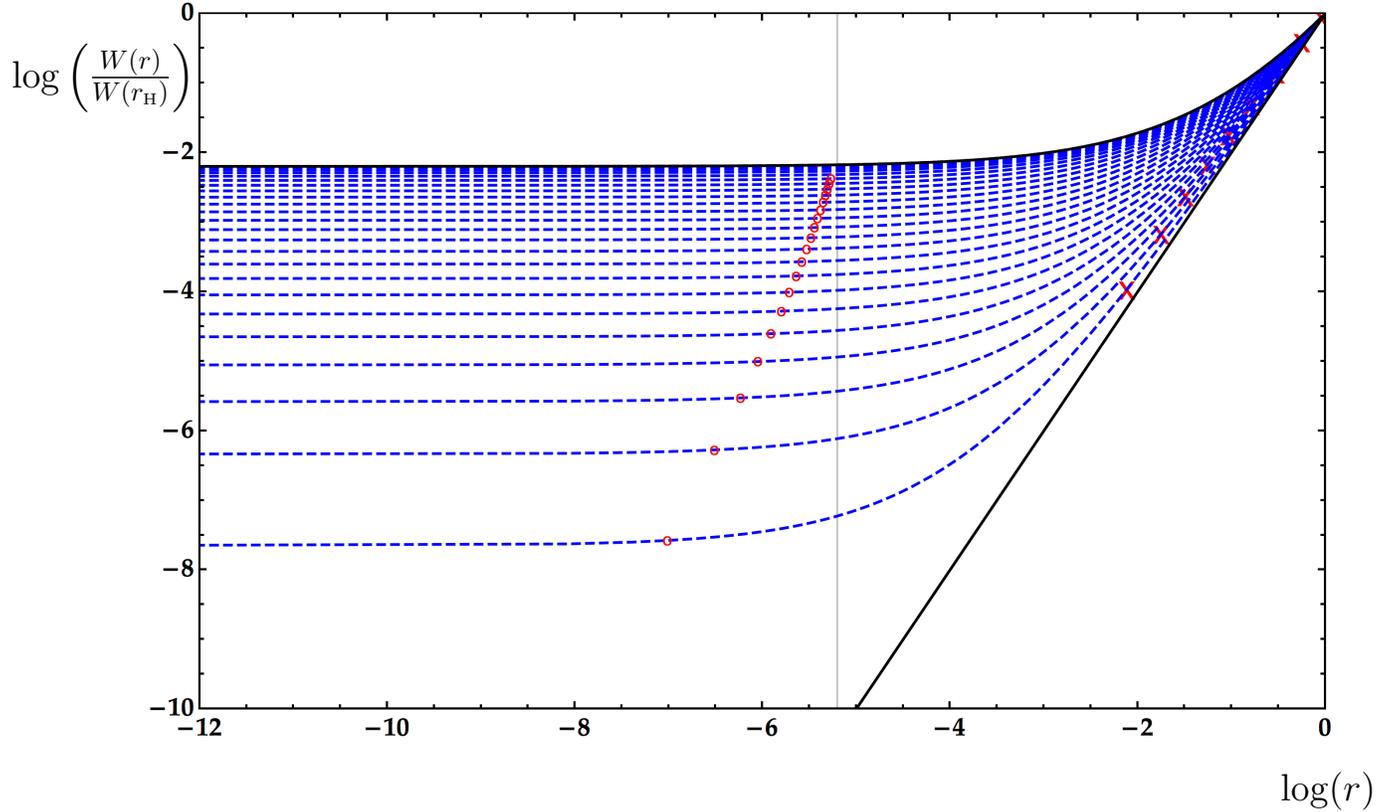} 
\put(-500,280){\large $\log\left(\frac{W(r)}{W(\rh)}\right)$}
\put(-20,10){\large $\log(r)$}
\end{center}
\caption{\small The numerical solutions for the metric function $W$ inside the horizon for values of $\mathcal{B}/T^2\approx$0, 0.68, 1.4, 2.1, 2.8, 3.6, 4.5, 5.4, 6.4, 7.5, 8.7, 10, 12, 14, 16, 19, 23, 28, 34, 43, 57, 78, 118, 207, 534, are shown in dashed lines from top to bottom. For each solution $W$ we indicated with a circle the place up to which, when moving away from the singularity, it remains constant up to one part in a hundred, and the vertical line indicates where the black brane solution fails to satisfy this condition. Whit an ``x" we mark the evaluation of each function $W$ at the $r_{Cross}$ obtained from its corresponding $U$ in figure (\ref{Uint}). The top solid line is the black brane solutions and the diagonal solid line on the right is the BTZ$\times R^2$.}
\label{Wint}
\end{figure}

The behavior of $W$ is better analyzed closer to the singularity. From (\ref{btz}) we see that $W_{BTZ}$ remains as $3r^2$ everywhere, approaching zero at $r=0$, but, since the singularity was shifted in (\ref{bb}), $W_{BB}$ approaches $\rh^2/3$. In a logarithmic plot this two behaviors are clearly separated, so in figure \ref{Wint} we plot  $\log(W(r)/W(\rh))$ versus $\log(r)$, where the metric functions was normalized by its value at the horizon to facilitate the comparison. We can see that as we get closer to the singularity the plots become horizontal, indicating that the solution is approaching the black brane solution. In this case we proceed conversely, and determine how far from the singularity the black brane behavior extends, taking as an arbitrary indicator the point where the solution remains a constant up to one part in a hundred. By marking this point in each plot of figure \ref{Wint} we confirm that as $\mathcal{B}/T^2$ gets smaller, the black brane solution extends closer to the horizon, pushing the transition region along with it. To bond the transition region from above, we also indicate the evaluation of each $W$ at the $r_{Cross}$ determined from the behavior of its corresponding $U$. This analysis also indicates that the BTZ$\times R^2$ geometry penetrates further inside the horizon as $\mathcal{B}/T^2$ increases, moving the transition to the black brane behavior closer to the singularity.

\section{Diffusion perpendicular to the magnetic field}
\label{AppD}

All the elements of the conductivity matrix can be computed holographically in a background solution $\{{G^{BG}}_{mn},F^{BG}\}$ by introducing small perturbations of the metric and the Maxwell field 
\begin{equation}
F=F^{BG}+\epsilon dA, \qquad G_{mn}={G^{BG}}_{mn}+\epsilon g_{mn},
\end{equation}
and solving the equations of motion for $dA$ and $g_{mn}$ to first order in $\epsilon$. The electric and thermal currents appearing in \eqref{conductivity_matrix} are then related to first integrals of the equations of motion of the perturbations. The procedure outlined in \cite{perturbations_B,perturbations} indicates that to this end it suffice to consider a perturbation given by
\begin{eqnarray}
&& A_{i}=t\delta f_{i}(r)+\delta a_{i}(r),
\cr
&& g_{ti}=t\delta h_{i}(r)+\delta g_{ti}(r),
\cr
&& g_{ri}=\delta g_{ri}(r),
\label{general_perturbationsB}
\end{eqnarray}
that can be consistently studied in on our background \eqref{5dmet} to first order in $\epsilon$ without the need to perturb any other component of the metric or the gauge potential. In fact, the perturbations along the direction of the magnetic field and the perturbations along the directions perpendicular to the magnetic field decouple. Thus, here we will study only the directions perpendicular to the magnetic field, being the $\{x,y\}$ plane in this particular choice of coordinates.

The resulting equations of motion further decouple in two groups. The first group couples $A_{x}$, $g_{ry}$ and $g_{ty}$. Focusing on $A_{x}$, Maxwell equations reads
\begin{equation}
\partial_{r}\left(-U\sqrt{W}\left(t\delta f_{x}'+\delta a_{x}'+\frac{\mathcal{B}}{V}\delta g_{ry}\right)\right)+\frac{\mathcal{B}\sqrt{W}}{UV}\delta h_{y}=0.
\label{Maxwell_x}
\end{equation}
The only way the time dependence drops from this equation is to fix $\delta f_{x}$ to a constant value which, from the definition of $A_{z}$, gives the intensity of the driving electric field in the $x$-direction
\begin{equation}
\delta f_{x}=-E_{x},
\end{equation}
After this choice \eqref{Maxwell_x} can be immediately integrated, defining a constant quantity which we identify as the electric current along the $x$-direction
\begin{equation}
J^{x}=-4U\sqrt{W}\left(\delta a_{x}'+\frac{\mathcal{B}}{V}\delta g_{ry}\right)+4\int_{\rh}^{r} \frac{\mathcal{B}\sqrt{W}}{UV}\delta h_{y}.
\label{Jx1}
\end{equation}
On the other hand, Einstein equations are
\begin{eqnarray}
&& \delta a_{x}'+\frac{\mathcal{B}}{V}\delta g_{ry}=\frac{V^{2}}{4\mathcal{B}U}\partial_{r}\left(\frac{\delta h_{y}}{V}\right),
\cr
&& \partial_{r}\left(\frac{U^{2}\sqrt{W}}{4}\partial_{r}\left(t\frac{\delta h_{y}}{U}+\frac{\delta g_{ty}}{U}\right)\right)+\frac{\mathcal{B}\sqrt{W}}{V}E_{x}=0.
\label{Einstein_y}
\end{eqnarray}
The time dependence drops out completely only if $\delta h_{y}$ is proportional to $U$. The proportionality constant gives the intensity of the driving thermal gradient in the $y$-direction
\begin{equation}
\delta h_{y}=-\zeta_{y}U,
\label{hy}
\end{equation}
given that $\delta h_{y}$ appears on the definition of $g_{ty}$. Then the first equation in \eqref{Einstein_y} completely determinates $\delta g_{ry}$ in terms of $\delta a_{x}'$ 
\begin{equation}
\delta g_{ry}=-\zeta_{y}\frac{V^{3}}{4\mathcal{B}^{2}U}\partial_{r}\left(\frac{U}{V}\right)-\frac{V}{\mathcal{B}}\delta a_{x}'.
\label{gry}
\end{equation}
Remarkably, \eqref{hy} and \eqref{gry} solve \eqref{Maxwell_x}. After this choices the equation for $g_{ty}$ can be integrated, from which we obtain the thermal current in the $y$-direction. The final result is
\begin{eqnarray}
&& J^{x}=\zeta_{y}\left(\frac{V^{2}\sqrt{W}}{\mathcal{B}}\partial_{r}\left(\frac{U}{V}\right)-\mathcal{M}(r)\right),
\cr
&& Q^{y}=-U^{2}\sqrt{W}\partial_{r}\left(\frac{\delta g_{ty}}{U}\right)-E_{x}\mathcal{M}(r),
\end{eqnarray}
where 
\begin{equation}
\mathcal{M}(r)=4\int_{\rh}^{r}\frac{\mathcal{B}\sqrt{W}}{V},
\end{equation}
corresponds to the total magnetisation density of the boundary theory as $r\rightarrow\infty$ \cite{perturbations_B}.

The second group, which involves $A_{y}$, $g_{tx}$ and $g_{rx}$, is analogous to the previous one. That is, time dependence drops out completely if
\begin{equation}
\delta f_{y}=-E_{y}, \qquad \delta h_{x}=-\zeta_{x}U,
\end{equation}
which fixes
\begin{equation}
\delta g_{rx}=\frac{V}{\mathcal{B}}\delta a_{x}'-\zeta_{x}\frac{V^{3}}{4\mathcal{B}^{2}U}\partial_{r}\left(\frac{U}{V}\right),
\end{equation}
and thus the currents are given by
\begin{eqnarray}
&& J^{y}=-\zeta_{x}\left(\frac{V^{2}\sqrt{W}}{\mathcal{B}}\partial_{r}\left(\frac{U}{V}\right)+\mathcal{M}(r)\right),
\cr
&& Q^{x}=-U^{2}\sqrt{W}\partial_{r}\left(\frac{\delta g_{tx}}{U}\right)+E_{y}\mathcal{M}(r).
\end{eqnarray}

After gathering all our results we are left then with the set of perturbations given by
\begin{eqnarray}
&& A_{x}=-E_{x}t+\delta a_{x}(r),
\cr
&& A_{y}=-E_{y}t+\delta a_{y}(r),
\cr
&& g_{tx}=-\zeta_{x}t U(r)+\delta g_{tx}(r),
\cr
&& g_{ty}=-\zeta_{y}t U(r)+\delta g_{ty}(r),
\cr
&& g_{rx}=\delta g_{rx}(r),
\cr
&& g_{ry}=\delta g_{ry}(r),\label{pertcons}
\end{eqnarray}
which components need to satisfy infalling boundary conditions at the horizon
\begin{equation}
A_{x}=-E_{x}\nu, \qquad A_{y}=-E_{y}\nu, \qquad g_{tx}=U(r)(g_{rx}-\zeta_{x}\nu), \qquad g_{ty}=U(r)(g_{ry}-\zeta_{y}\nu),
\label{infalling1}
\end{equation}
where $\nu$ is the infalling Eddington-Finklestein coordinate given by
\begin{equation}
\nu=t+\frac{1}{4\pi T}\log(r-\rh)+\mathcal{O}(r-\rh).
\end{equation}
This conditions in turn imply that
\begin{eqnarray}
&& \delta a_{x}(r)=-\frac{E_{x}}{4\pi T}\log(r-\rh)+\mathcal{O}(r-\rh),
\cr
&& \delta a_{y}(r)=-\frac{E_{y}}{4\pi T}\log(r-\rh)+\mathcal{O}(r-\rh),
\cr
&& \delta g_{tx}(r)=U(r)\delta g_{rx}(r) -\frac{\zeta_{x}U(r)}{4\pi T}\log(r-\rh)+\mathcal{O}(r-\rh),
\cr
&& \delta g_{ty}(r)=U(r)\delta g_{ry}(r)-\frac{\zeta_{y}U(r)}{4\pi T}\log(r-\rh)+\mathcal{O}(r-\rh),
\cr
&& \delta g_{tz}(r)=\mathcal{O}(r-\rh)^{0}.
\label{infalling2}
\end{eqnarray}

The electric and thermal currents can then be expressed in terms of $E_{i}$ and $\zeta_{i}$ after using \eqref{infalling2} to perform some evaluations at the horizon (also using the fact that $V(r)$ and $W(r)$ are regular at $\rh$ and that $U(r)=4\pi T(r-\rh)$ near the horizon). The final result is
\begin{eqnarray}
&& J^{x}=\zeta_{y}T\left(4\pi\frac{V(\rh)}{\mathcal{B}}\sqrt{W(\rh)}\right),
\cr
&& J^{y}=-\zeta_{x}T\left(4\pi\frac{V(\rh)}{\mathcal{B}}\sqrt{W(\rh)}\right),
\cr
&& Q^{x}=-E_{y}T\left(4\pi\frac{V(\rh)}{\mathcal{B}}\sqrt{W(\rh)}\right)-\zeta_{x}T\left(4\pi^{2}T\frac{V(\rh)^{2}}{\mathcal{B}^{2}}\sqrt{W(\rh)}\right),
\cr
&& Q^{y}=E_{x}T\left(4\pi\frac{V(\rh)}{\mathcal{B}}\sqrt{W(\rh)}\right)-\zeta_{y}T\left(4\pi^{2}T\frac{V(\rh)^{2}}{\mathcal{B}^{2}}\sqrt{W(\rh)}\right),
\label{currents_rH}
\end{eqnarray}

Another necessary quantity to calculate the diffusion constants is the electric susceptibility, and to compute it we need to perturbatively add
\begin{equation}
A_{t}=\delta a_{t}(r).
\label{At}
\end{equation}
It is important to note that even if we decide to turn on \eqref{general_perturbationsB} and \eqref{At} simultaneously, both decouple allowing us to analyse them separately. The equation of motion for $A_{t}$ then reads
\begin{equation}
\partial_{r}(-V\sqrt{W}\delta a_{t}')=0,
\end{equation}
which defines a constant quantity that we identify as the charge density
\begin{equation}
\rho=-4V\sqrt{W}\delta a_{t}'.
\label{charge}
\end{equation}
The holographic dictionary relates the chemical potential $\mu$ to value of $\delta a_{t}$ at the boundary, thus integration of \eqref{charge} gives
\begin{equation}
\mu=\rho\int_{\rh}^{\infty}\frac{dr}{4V\sqrt{W}},
\label{charge_density}
\end{equation}
from where \eqref{susceptibility} is obtained.

\section{Eigenvalues of the diffusivity matrix}
\label{AppEigen}

Following \cite{Hartnoll-2014} and references therein, the diffusion of charge and energy in a strongly coupled system can be described by the conservation equations
\begin{equation}
\partial_{t}\delta\rho=-\partial_{i}J_{i}, \qquad \partial_{t}\delta s=-\partial_{i}\frac{Q_{i}}{T}.
\label{continuity}
\end{equation}
Here $J_{i}$ and $Q_{i}$ are the components of the electric and thermal currents described in the main text, while $\delta\rho$ and $\delta s$ are perturbations in the charge and entropy density respectively. To obtain the coupled diffusion equations for $\delta\rho$ and $\delta s$ it is necessary to consider the constitutive relations
\begin{equation}
J_{i}=-\bar{\sigma}_{ij}\partial_{j}\mu-\vartheta_{ij}\partial_{j}T, \qquad \frac{Q_{i}}{T}=-\vartheta_{ij}\partial_{j}\mu-\frac{1}{T}\bar{\kappa}_{ij}\partial_{j}T,
\label{constitutive}
\end{equation}
that give the electric and thermal currents generated by a given temperature and chemical potential gradient (in the absence of an external electric field). From \eqref{constitutive} in \eqref{continuity} we obtain
\begin{equation}
\partial_{t}\delta\rho=\bar{\sigma}_{ij}\partial_{i}\partial_{j}\mu+ \vartheta_{ij}\partial_{i}\partial_{j}T, \qquad \partial_{t}\delta s=\vartheta_{ij}\partial_{i}\partial_{j}\mu+\frac{1}{T}\bar{\kappa}_{ij}\partial_{i}\partial_{j}T.
\label{d1}
\end{equation}
The crucial point here is that if the system features a constant magnetic field, which is the case of interest in this work, then the conductivity matrices in the directions perpendicular to such field can be decomposed into longitudinal and Hall components in the particular form $\bar{\sigma}_{ij}=\bar{\sigma}_{L}\delta_{ij}+\bar{\sigma}_{H}\epsilon_{ij}$. The antisymmetric nature of $\epsilon_{ij}$ makes it so that the contribution of the Hall components will cancel completely after index contraction with $\partial_{i}\partial_{j}$ in \eqref{d1}. The next step is to express the changes of the chemical potential and temperature in terms of the perturbations of charge and entropy density. This is achieved with the thermodynamic identity
\be
\begin{pmatrix}
\delta\mu\\
\delta T
\end{pmatrix}=\begin{pmatrix}
\chi & \xi \\
\xi & \frac{C_{\mu}}{T}
\end{pmatrix}^{-1}\begin{pmatrix}
\delta\rho\\
\delta s
\end{pmatrix}.
\label{thermo}
\ee

After direct substitution of \eqref{thermo} in \eqref{d1} we obtain the diffusion equations for $\delta\rho$ and $\delta s$
\be
\partial_{t}\begin{pmatrix}
\delta\rho\\
\delta s
\end{pmatrix}=D\cdot\nabla^{2}\begin{pmatrix}
\delta\rho\\
\delta s
\end{pmatrix},
\ee
where $D$ is the diffusion matrix for our system, that using \eqref{conduMat} can be explicitly written as
\begin{equation}
D=\begin{pmatrix}
0 & 0 & 0 & 0 & 0 & 0 \\
0 & 0 & 0 & 0 & 0 & 0 \\
0 & 0 & \frac{4V(\rh)}{\chi\sqrt{W(\rh)}} & 0 & 0 & 0 \\
0 & 0 & 0 & 4\pi^{2}\frac{V(\rh)^{2}\sqrt{W(\rh)}}{C_{\mu}\mathcal{B}^{2}} & 0 & 0 \\
0 & 0 & 0 & 0 & 4\pi^{2}\frac{V(\rh)^{2}\sqrt{W(\rh)}}{C_{\mu}\mathcal{B}^{2}} & 0 \\
0 & 0 & 0 & 0 & 0 & \infty
\end{pmatrix}.\label{DM}
\end{equation}
Note in particular that given that the longitudinal components of our thermoelectric conductivity matrix are identically zero, $\vartheta$ doesn't enter the diffusion equations at all. As a result, the diffusion equations are already decoupled and we can read the eigenvalues of $D$ directly from \eqref{DM}.

Of the six eigenvalues of $D$, only four are independent, and are associated to electric and thermal diffusion in directions parallel and perpendicular to the magnetic field. We see that the eigenvalue corresponding to the perpendicular electric diffusion vanishes, so no electric diffusion occurs in this plane as a result of $\bar{\sigma}_{xx}=\bar{\sigma}_{yy}=0$. Also, the divergence of one of the eigenvalue associated to thermal diffusion is a consequence of momentum conservation along the direction of the magnetic field, coming from the fact that $\bar{\kappa}_{zz}=\infty$. 

The other two eigenvalues, associated to electric and thermal diffusion in parallel and perpendicular directions respectively, need to be evaluated numerically. One of them is directly $D_{c}^{\parallel}$, which relation with the butterfly velocity has already been studied in the main text. The only quantity that we have not yet computed and is necessary to evaluate the remaining eigenvalue
\begin{equation}
D_T^\perp=4\pi^{2}\frac{V(\rh)^{2}\sqrt{W(\rh)}}{C_{\mu}\mathcal{B}^{2}},
\end{equation}
is $C_\mu$, given by
\begin{equation}
C_{\mu}=T\left(\frac{\partial s}{\partial T}\right)_{\mu,\mathcal{B}},
\end{equation}
where as usual the entropy density $s$ is related to the area of the horizon  
\begin{equation}
s=4\pi V(\rh)\sqrt{W(\rh)}.
\label{entropy_density}
\end{equation}
To compute $C_\mu$ we need the derivative of the entropy density with respect to the temperature at fixed chemical potential and magnetic field. Given that in our solutions $\mu=0$ the former is immediately achieved, while for the latter it is convenient to note that the entropy density is related to a function $H$ that only depends on the dimensionless ratio $\frac{\mathcal{B}}{T^{2}}$ by
\begin{equation}
\frac{s}{\mathcal{B}^{\frac{3}{2}}}=H\left(\frac{\mathcal{B}}{T^{2}}\right).
\end{equation}
The function $H$ can be computed numerically using \eqref{entropy_density}. After a little algebra one can show that the derivative of $H$ is related to the specific heat $C_{\mu}$ by
\begin{equation}
C_{\mu}=-\frac{2\mathcal{B}^{\frac{5}{2}}}{T^{2}}H'\left(\frac{\mathcal{B}}{T^{2}}\right).
\end{equation}

\begin{figure}
\begin{center}
\begin{tabular}{cc}
\setlength{\unitlength}{1cm}
\hspace{-0.9cm}
\includegraphics[width=7cm]{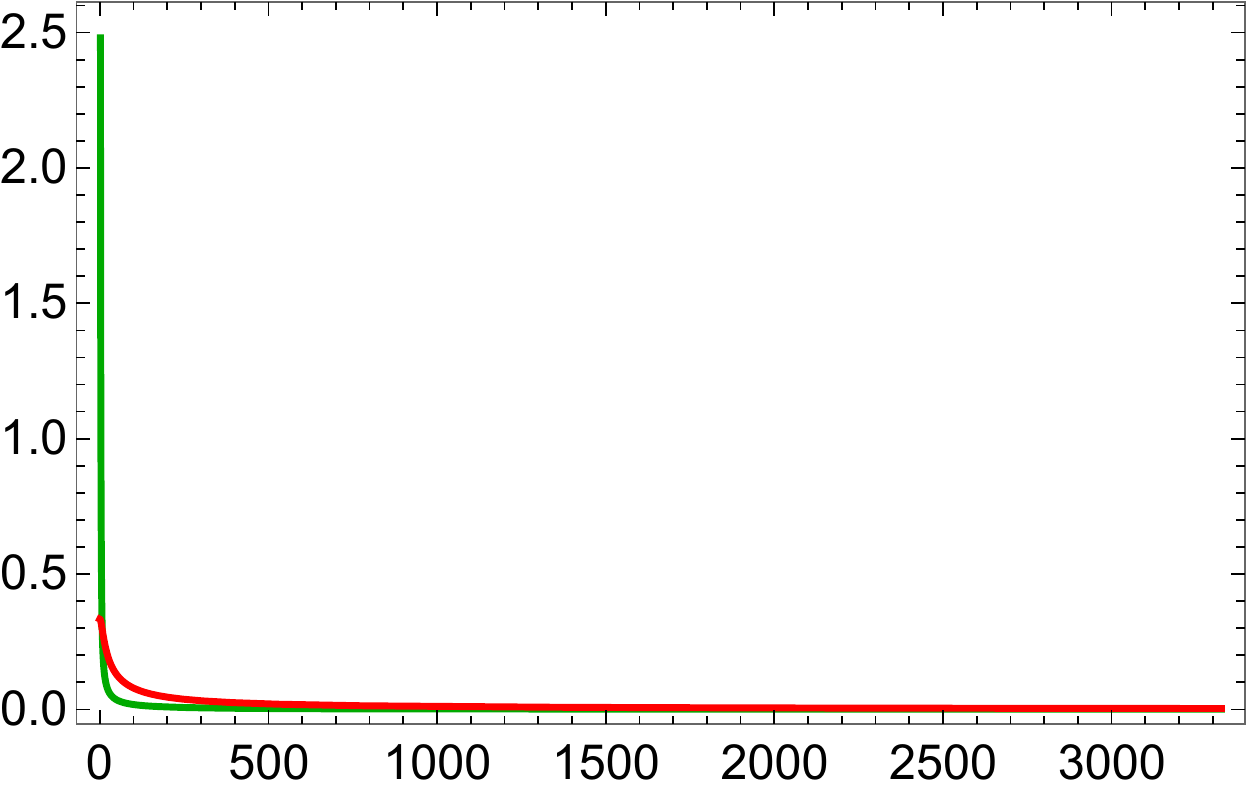} 
\qquad\qquad & 
\includegraphics[width=7cm]{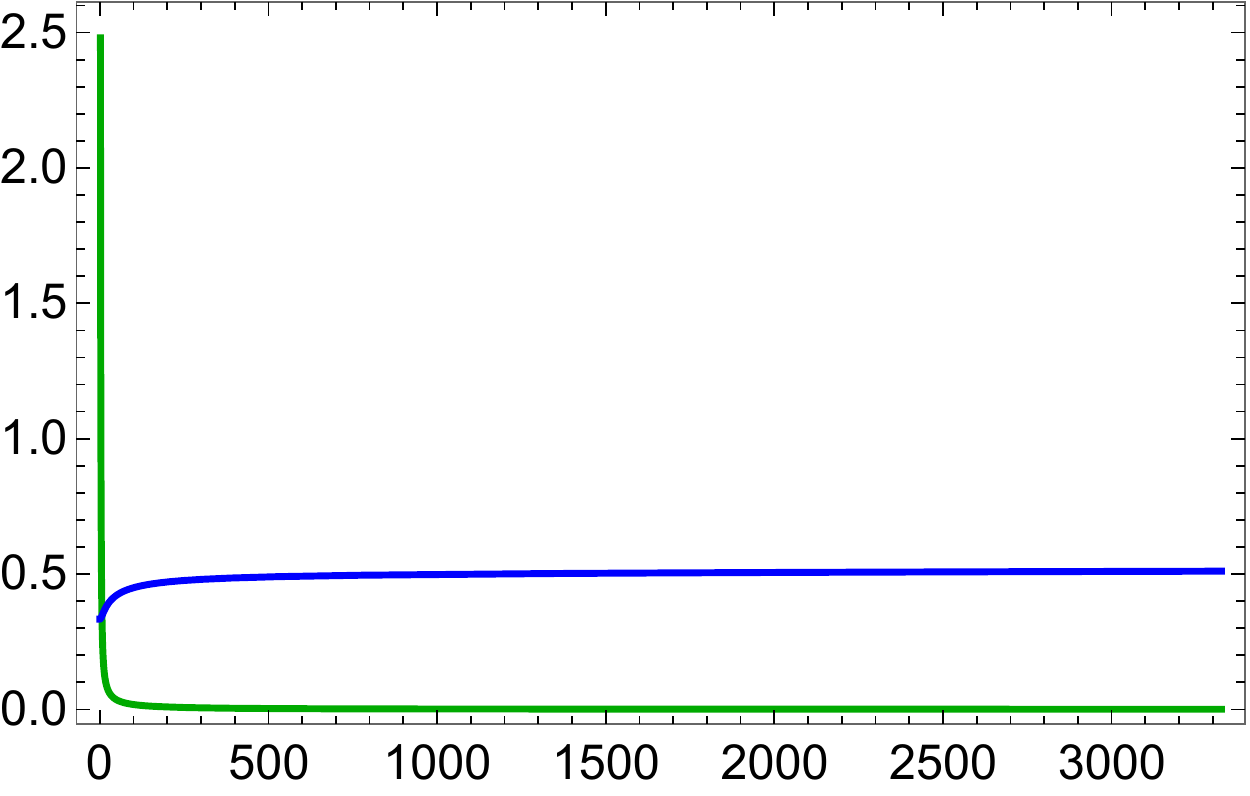}
\qquad
   \put(-250,-10){$\mathcal{B}/T^{2}$}
   \put(-300,20){$\frac{3}{2}v_{B,\perp}^{2}\tau_{L}$}
   \put(-420,110){$D_T^\perp$}
   \put(-18,-10){$\mathcal{B}/T^{2}$}
   \put(-50,50){$\frac{3}{2}v_{B,\parallel}^{2}\tau_{L}$}
   \put(-180,110){$D_T^\perp$}
 \\
 (a) & (b) 
\end{tabular}
\end{center}
\caption{\small 
(a) Eigenvalue $D_T^\perp$ of the diffusion matrix (green curve) and $\frac{3}{2}v_{B,\perp}^{2}\tau_{L}$ (red curve). (b) Eigenvalue $D_T^\perp$ of the diffusion matrix and $\frac{3}{2}v_{B,\perp}^{2}\tau_{L}$ (blue curve).}
\label{EigenPlot}
\end{figure}
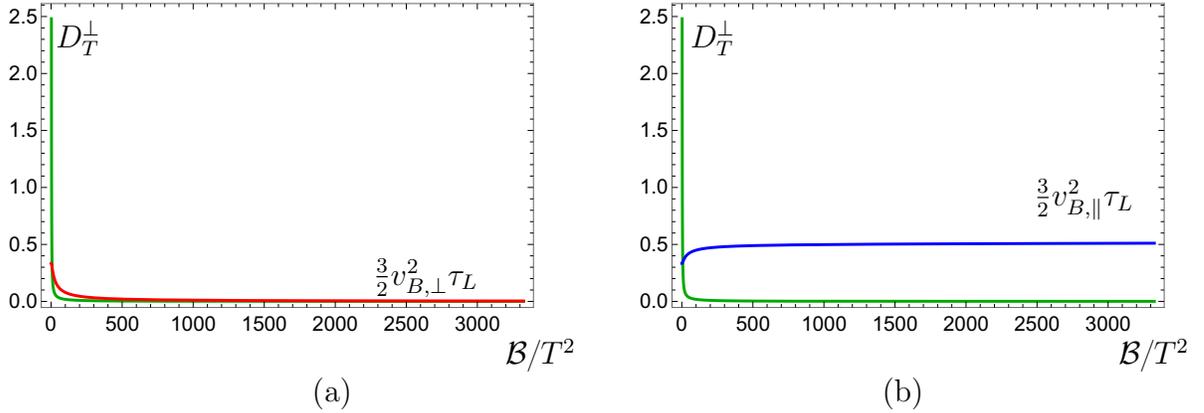

In figure 7 we show $D_T^\perp$ as a function of the dimensionless quantity $\mathcal{B}/T^{2}$, and compare it to the butterfly velocity. We conclude that for generic values of $\mathcal{B}/T^{2}$ this diffusion eigenvalue is not a good bound for the butterfly velocity along any direction. It is possible that this is because this eigenvalue has information about both thermal and electric effects while $D_{c}$, as mentioned in the main text, contains information about charge currents only.

%%%%%%%%%%%%%%%%%%%%%%%%%%%%%%%%%%%%%%%%%%%%%%%%%%%%%%%%%%%%%%%%%%%%%%%%%

\end{document}